\documentclass[acmsmall,nonacm]{acmart}
\AtBeginDocument{%
  \providecommand\BibTeX{{%
    \normalfont B\kern-0.5em{\scshape i\kern-0.25em b}\kern-0.8em\TeX}}}

\usepackage{booktabs}
\usepackage{amsthm}
\usepackage{mathtools}
\usepackage{dsfont}

\usepackage[labelformat=simple]{subcaption}

\usepackage{multirow}
\usepackage{bm}
\usepackage{colortbl}
\usepackage{color}
\usepackage[ruled]{algorithm2e} 

\def \qed{$\square$}
\def \mc{\mathcal}
\def \bo{\boldsymbol}
\def \tm{\textrm}

\def \P{\mathcal{P}}

\def \A{\mathcal{A}}

\def \D{\mathcal{D}}

\def \Q{\mathbf{Q}}

\def \E{\textrm{E}}

\def \po{\textrm{po}}
\def \cmin{\textrm{cmin}}
\def \dom{\textbf{\textup{dom}}}

\def \all{\textrm{all}}

\def \1{ \mathds{1}}

\def \mR{\mathbb{R}}

\def \b{\mathbf}
\def \h{\hat}
\def \q{\mathbf{q}}

\def \x{\mathbf{x}}
\def \y{\mathbf{y}}
\def \z{\mathbf{z}}
\def \e{\mathbf{e}}

\def \0{\mathbf{0}}
\def \d{\mathbf{d}}
\def \p{\mathbf{p}}
\def \sim {\textrm{sim}}

\def \tb{\textcolor{blue}}

\def \udl{\underline}

\def \bm{\mathbf}

\newtheorem{assumption}{Assumption}
\newtheorem{remark}{Remark}

\usepackage{siunitx}
\begin{document}

\title{Price Interpretability of Prediction Markets: A Convergence Analysis}

\author{Dian Yu}
\email{deaniiyu@outlook.com}

\affiliation{%
  \institution{Industrial Bank., Co., Ltd.}
  \city{Fuzhou}
  \country{China}
}

\author{Jianjun Gao}
\affiliation{%
  \institution{School of Information Management, 
Shanghai University of Finance and Economics}
  \city{Shanghai}
  \country{China}}
\email{gao.jianjun@shufe.edu.cn}

\author{Weiping Wu}
\affiliation{%
  \institution{Fuzhou University}
  \city{Fuzhou}
  \country{China}
}
\email{wu.weiping@fzu.edu.cn}

\author{Zizhuo Wang}
\affiliation{%
 \institution{School of Data Science, The Chinese University of Hong Kong, Shenzhen}
 \city{Shenzhen}
 \country{China}}
\email{wangzizhuo@cuhk.edu.cn}


\begin{abstract}
\tb{Prediction markets are long known for prediction accuracy. This study systematically explores the fundamental properties of prediction markets, addressing questions about their information aggregation process and the factors contributing to their remarkable efficacy. We propose a novel multivariate utility (MU) based mechanism that unifies several existing automated market-making schemes. Using this mechanism, we establish the convergence results for markets comprised of risk-averse traders who have heterogeneous beliefs and repeatedly interact with the market maker. We demonstrate that the resulting limiting wealth distribution aligns with the Pareto efficient frontier defined by the utilities of all market participants. With the help of this result, we establish analytical and numerical results for the limiting price in different market models. Specifically, we show that the limiting price converges to the geometric mean of agent beliefs in exponential utility-based markets. In risk-measure-based markets, we construct a family of risk measures that satisfy the convergence criteria and prove that the price can converge to a unique level represented by the weighted power mean of agent beliefs. In broader markets with Constant Relative Risk Aversion (CRRA) utilities, we reveal that the limiting price can be characterized by systems of equations that encapsulate agent beliefs, risk parameters, and wealth. Despite the potential impact of traders' trading sequences on the limiting price, we establish a price invariance result for markets with a large trader population. Using this result, we propose an efficient approximation scheme for the limiting price. Numerical experiments underscore the accuracy of this approximation across various scenarios, outperforming existing approximation methods. Our findings serve to aid market designers to better tailor and adjust the market-making mechanism for more effective opinions elicitation.}
\end{abstract}


\keywords{Dynamic mechanisms, information elicitation, prediction markets}

\maketitle
\section{Introduction}\label{sec:introduction}
Information, often scattered among the crowd, is valuable yet hard to aggregate. Throughout the centuries, researchers and practitioners have invented various tools in an attempt to piece them together. Classical mechanisms like polls, surveys, and brainstorming work fine.\footnote{\tb{According to \cite{berg2008prediction}, polls only accurately predict the outcome in $32\%$ of cases in presidential elections. \cite{Pathak:2015} find that expert forecasts performed worse than polls and \tb{prediction markets}. In the corporate setting, \cite{Cowgill:2015} discover that expert forecasts had a $25\%$ higher mean square error compared to prediction markets.} } However, they lack the commitment to let people put their money where their mouth is. Over the past two decades, the \textit{prediction market}, a mechanism designed specifically for information elicitation
and aggregation, has gained much popularity. For example, the Iowa Electronic Markets (IEM), arguably the pioneer of all existing prediction markets designed primarily for various elections, almost consistently beats professional opinion polls \cite{berg2008prediction}. \tb{Other success stories include corporate consensus pooling, box office forecasting, and online recreational sports betting (e.g., \cite{chen2005-comparison,Cowgill:2009,Cowgill:2015}).} There is theoretical evidence
that prediction markets are robust against price manipulation \cite{hanson2009manipulator}. \tb{Overall, prediction markets have become a popular tool for information aggregation (e.g., \cite{wolfers2004prediction,Arrow:science_2008}).}

\tb{Due to the liquidity issue, a prediction market is usually organized by an automated market maker  \cite{hanson2003combinatorial}.\footnote{\tb{The ``liquidity issue'' refers to a phenomenon where buyers' or sellers' orders face delays or extended waiting times before being matched. In prediction markets, which usually have fewer traders than financial markets, one approach to address this issue is introducing a market maker to provide liquidity for buy and sell orders \cite{wolfers2004prediction}.} } In this market, traders exchange specialized securities with a market maker. These securities offer rewards based on uncertain outcomes, revealed once the true state is known. Traders, guided by their subjective beliefs, trade with the market maker to gain expected utility. Simultaneously, the market maker gathers valuable information through trading to enhance the overall accuracy of predictions. While the prediction market literature has primarily focused on designing effective market-making mechanisms (e.g., \cite{chen2007utility,abernethy2013efficient,Othman:EC2013-sensitive-liquidity,Slamka:IEEE-PM}), there have been relatively few attempts to explore the market's evolutionary behavior. In this paper, we aim to narrowing this gap by addressing the following questions and providing additional insights into the analysis of prediction markets.
}
\begin{enumerate}
\item \tb{Under what conditions can we expect a prediction market to reach a consensus? That is, when does the trading process (market price) converge?}

\item How to interpret the limiting price? What is the relation between such limiting price with each trader's belief, wealth and their risk attitude?
\end{enumerate}

\tb{
In line with the spirit of \cite{sethi2016belief}, we explore a dynamic prediction market model with multiple securities, involving a finite number of myopic, risk-averse traders with heterogeneous beliefs interacting with an automated market maker. We make several contributions that enhance the understanding and effectiveness of prediction markets.}

\tb{
Firstly, we introduce a unified multivariate utility (MU)-based mechanism that incorporates several market making mechanisms \tb{(e.g., \cite{hanson2003combinatorial,hanson2007logarithmic,chen2007utility,agrawal2011unified})} into a consistent framework. Based on this framework, we propose a method for establishing market convergence that subsumes and extends the existing results in \cite{frongillo2015elicitation} and \cite{sethi2016belief}. More importantly, we demonstrate that the limiting wealth distribution is Pareto optimal with respect to the utility of each trader. Additionally, this approach allows us to bypass the challenges of analyzing transient behavior in price dynamics and enables a direct examination of the limiting price.}

\tb{
By leveraging the general convergence result, we study the convergence properties of various utility types. Specifically, we derive the analytical form of price dynamics for the exponential utility-based market, revealing that the limiting price is the geometric mean of agents' beliefs and is independent of the trading sequences. We extend this result to the convex risk-measure-based market. Specifically, we present an explicit method to construct risk measures that satisfy the convergence requirements. For a wide range of convex risk measures, we derive the analytical form of the limiting price, which emerges as a weighted power mean of agent beliefs. 
}

\tb{
Furthermore, we explore the more commonly used market model where all participants adopt the utility with \textit{constant relative risk aversion} (i.e., CRRA utility) to make decisions. We analyze traders' optimal trading decisions and examine the limiting wealth distribution resulting from the aggregate Pareto-optimization problem. It reveals that the limiting price can be characterized through a system of equations including all participants' risk parameters, initial wealth allocation, and subjective beliefs. Crucially, we establish a fundamental result indicating that, even though trading sequences can influence the limiting price, their impact diminishes to insignificance as the trader population increases. Using these results, we introduce a heuristic weight called \textit{Pareto Optimal Induced (POI) weights} and derive the associated approximate price formula that underscores the important role of risk aversion and initial wealth in price determination. To evaluate this approximation scheme, we conduct numerical experiments, which demonstrate that our POI price can closely track the actual price resulting from the trading with high accuracy across various settings. Through further comparison, our approximation scheme consistently outperforms previous attempts in the literature, such as the one presented in \cite{sethi2016belief}. This performance improvement significantly advances the state-of-the-art in our comprehension of market prices.
}

\tb{Our findings hold significant potential for various applications. Firstly, the MU-based mechanism offers extensive flexibility in designing pricing formulations, empowering the market designer to tailor markets for diverse purposes. Secondly, our theoretical results establish a link between the parameters and the limiting price, enhancing the market designer's ability to elicit traders' beliefs effectively. Specifically, this suggests that market maker should exercise caution when selecting risk and wealth parameters to balance liquidity and accuracy. Our theoretical result on price stability in a random trading market suggests attracting more traders to achieve a reliable limiting price. Moreover, in scenarios like artificial prediction markets (\cite{Chakravorti:2023}), where traders are artificially generated supervised learners, our results provide guidance for configuring hyper-parameters for these artificial traders. Lastly, our convergence scheme and the subsequent price analysis can be effortlessly extended to prediction markets without a market maker. As an ancillary outcome, the theoretical proof of market convergence is closely related to the distributional algorithm of weakly-coupled multi-objective optimization. This proof can be extended to address more generalized problems sharing a similar structure.
}


This paper is organized as follows. In the remainder of this section, we review the related literature. In Section \ref{sec:market_model}, we introduce the market model and the multivariate utility (MU)-based pricing mechanism. In Section \ref{sec:convergence}, we present the general convergence result. In Section \ref{sec:expU_Risk}, we study the exponential utility-based market and the risk
measure-based market. \tb{We then explore the CRRA utility-based market in Section \ref{sec:CRRA}. In Section \ref{sec:price-volume-nonstationary}, we discuss the price-volume relationship and the non-stationary market setting.} We conclude the paper in Section \ref{se:conclusion}. All the proofs and related results are provided in the Appendix.

\subsection{Related Research}\label{sse:review}

\tb{
Our work is closely related to the research on automated market-making mechanisms in prediction markets. The logarithm market scoring rule (LMSR) introduced by \cite{hanson2003combinatorial,hanson2007logarithmic}, stands out as one of the most popular mechanisms in internet prediction markets due to its desirable properties. In practice, LMSR can be implemented in the form of a  cost function \cite{chen2010new}. This cost function-based mechanism has been refined in various directions (e.g., \cite{Othman:EC2013-sensitive-liquidity,abernethy2013efficient,abernethy:EC2014-vol-parameter}). Furthermore, \cite{chen2007utility} propose a univariate utility function-based framework, which strongly motivates our adoption of the multivariate utility (MU)-based mechanism. Additionally, \cite{agrawal2011unified} introduce the sequential convex pari-mutuel mechanisms, allowing the market to accept limit orders. Partial equivalence among these mechanisms has been established \cite{chen2007utility,chen2010new,agrawal2011unified}. To deal with the liquidity-adaption problem in practical implementation, \cite{Othman:EC2013-sensitive-liquidity} propose a liquidity-sensitive mechanism. Recent developments have extended these mechanisms in various directions (e.g., \cite{Freeman:2017,Ban:2018,Freeman:2018ICAA}). As for the performance comparison of different mechanisms, \cite{Slamka:IEEE-PM} provide simulation studies.
}

\tb{
In prediction markets, trading-generated prices serve as predictors of future events, reflecting an aggregate estimate of the likelihood of a certain event occurring (\cite{manski2006interpreting,wolfers2006interpreting}). As for the market convergence and price interpretation, there are generally two streams of studies. The first one involves a common prior and heterogeneous information. \cite{ostrovsky2012information} and \cite{iyer2014information} find that when all participants are risk-neutral or risk-averse, prices converge to a common posterior. These findings align with the well-known theorem established by \cite{aumann1976agreeing}, stating that people with a common prior must have a common posterior if all posteriors are common knowledge, or in short, \textit{people cannot agree to disagree}. The second stream of the model is characterized by heterogeneous priors but common information. In this model, several researchers adopt the static equilibrium model, where the price results from the market clearing condition. Following this idea, \cite{gjerstad2005risk} and \cite{manski2006interpreting} find a bias between traders' mean beliefs and the equilibrium price. However, \cite{pennock1999aggregating} and \cite{wolfers2006interpreting} show that the market equilibrium coincides with the wealth-weighted average of trader's beliefs when traders adopt the logarithm utility to make decisions. 
}

\tb{In contrast to the above static model, another line of research explores the dynamic trading model. \cite{othman2010markets} investigate a market with risk-neutral traders with heterogeneous beliefs. In this setting, a finite number of traders interact with the market only once, resulting in a dependence between the trading order and the last posted price. In a related study, \cite{Frongillo:NIPS2015} adopt risk measures to model traders' preferences in a market equipped with a cost function-based mechanism. When the traders trade with the market maker in random order, the dynamic trading process exhibits similarities with a machine learning algorithm that aims to minimize aggregate risk. While they develop some convergence conditions, their approach critically relies on the translation invariance property of risk measures (\cite{Follmer:book2015}), which cannot be directly applied in our general MU-based market. Our work shares more similarities with \cite{sethi2016belief}, who establish price convergence for a binary predication market with risk-averse traders and an LMSR-based market maker. However, our research focuses on a broader market setting, encompassing multiple securities, a general market-making mechanism, and a variety of trading decision paradigms. These complexities make it challenging to apply their approaches effectively to analyze our general model. We also note that \cite{bonnisseau2013discrete} have considered a discrete Walrasian process, which is related to our model. However, the key difference is that their process involves pure exchange without a market maker. In contrast, our model is more intricate due to the externality introduced by the market maker and the pairwise trading restriction. The conditions required for market convergence in this scenario remain unclear, which partially motivates our current research.
}

\tb{
Additionally, several studies have examined prediction markets with specific market structures. For example, \cite{abernethy2014information} analyze the equilibrium properties of a market with traders whose beliefs are drawn from exponential family distributions. \cite{carvalho2017participation} demonstrates that in a binary prediction market with an LMSR-based mechanism and risk-neutral budget-constrained traders, the price converges to the median belief of traders. \cite{tarnaud2019convergence} investigates the asymptotic properties of a simple binary market with an LMSR-based mechanism and two traders. As valuable tools for forecasting and information aggregation, prediction markets also receive attention in the operations research and management science societies. For example, \cite{chen2008modeling} develop a diffusion model for predicting the price of political events in prediction markets, while \cite{berg2009searching} use prediction markets to forecast market capitalization before an initial public offering. \cite{healy2010prediction} investigate different trading mechanisms in small markets with complex environments. \cite{atanasov2017distilling,Pavel_EC_2022} systematically compare prediction markets with other popular information elicitation methods. Through several large-scale experimental tests, they find these two methods have advantages in different situations. \cite{jian2010aggregation} and \cite{choo2022manipulation} study the accuracy of prediction markets when there is price manipulation.
}

\tb{Besides the human-populated markets, artificial prediction markets with artificial agents (bot-traders) are developed as supervised learning tools for probability estimation and the aggregation of weak classifiers (e.g., \cite{storkey2012isoelastic,Chakravorti:2023}). Tuning the hyper-parameters (i.e., wealth, risk parameter, and beliefs of each bot-trader) in these artificial markets is a key challenge, as highlighted by \cite{Chakravorti:2023}. Our convergence result and interpretation of the limiting price provide potential guidance for tuning these hyper-parameters.
}

\textbf{Notations}: \tb{A vector is denoted by a boldface letter, i.e., $\x=(x_1, \ldots, x_n) \in \mathbb{R}^n$ stands for a column vector $\x$. Let $\e$ and $\mathbf{0}$ be the all-one and all-zero vectors in the appropriate dimension, respectively. Given vectors $\y$ and $\x$ in $\mathbb{R}^n$, the notation $\y\geq\x$ means that $y_i\geq x_i$ for all $i$, while $\y>\x$ indicates that $\y\geq \x$ and there exists an $i$ such that $y_i>x_i$. Notations $\mathrm{recc}(\mathcal{A})$ and $\mathrm{recc}(F)$ represent the recession cone of the convex set $\mathcal{A}$ and the convex function $F$, respectively (as defined in \citealt{bertsekas2003convex}). The domain of function $F(\cdot)$ is denoted as $\dom(F)$, and the probability simplex is represented as $\Theta_n \triangleq \{(q_1, \ldots, q_n) \in \mathbb{R}^n_+ | \sum_{i=1}^n q_i=1\}$, where $\mathbb{R}_+$ denotes the set of nonnegative real numbers.}

\section{Market Model and Trading Process}\label{sec:market_model}
In this section, we introduce the market model. We consider a state space $\{1, 2,\ldots, I\}$ with $I \ge 2$, comprising mutually exclusive and exhaustive outcomes. Each state $i \in \{1,2,\ldots, I\}$ corresponds to an Arrow-Debreu security, which pays $1$ dollar if state $i$ occurs. These securities are traded in a prediction market organized by a central market maker. A total of $J>1$ traders engage in repeated transactions with the market maker at discrete time points $t=0,1,2,\ldots$. We represent the market maker's wealth at time $t$ as $W_t \in \mathbb{R}$ and their outstanding securities position (i.e., the number of securities already sold) as $\Q_t\in \mathbb{R}^I$. The market maker's initial values are $W_0>0$ and $\Q_{0} =\mathbf{0}$. For each trader indexed by $j=1,\ldots,J$, we use $w_{j,t} \in \mathbb{R}$ and $\q_{j,t} \in \mathbb{R}^I$ to denote their wealth and security positions at time $t=0,1,2,\ldots$. The initial values are $w_{j,0}>0$ and $\q_{j,0}=\mathbf{0}$.

\tb{At each time $t$, one trader interacts with the market maker (we will specify the order of arrivals of the traders in later discussions) by submitting an order $\Delta \q \in \mR^I$. For instance, in the context of a football match outcome, such an order might be expressed as $\Delta \q=(2, 1, -1)$, representing $2$ shares for a buying order on ``win'', $1$ share for a buying order on ``loss'', and $1$ share for a selling order on ``tie''.} At any time $t$, if the $j$-th trader pays $\Delta w$ to the market marker (if $\Delta w<0$, then it means the trader is paid $-\Delta w$ by the market maker) and buys
$\Delta \q$ share of security, then the market state is updated as
follows:\footnote{In this work, most of the results are irrelevant 
to the condition of the nonnegative wealth restriction (bankruptcy restriction). 
However, we specify the difference when it is necessary.}
\begin{align}
\textrm{Trader $j$:}
\begin{dcases}
    w_{j,t+1} =  w_{j,t} - \Delta w\\
   \q_{j,t+1} = \q_{j,t} + \Delta\q
\end{dcases},
~~~~~\textrm{Market Maker:}
\begin{dcases}
     W_{t+1} =  W_t + \Delta w\\
    \Q_{t+1} = \Q_t + \Delta\q.
\end{dcases}\label{eq:dynamic_w_q}
\end{align}
The other traders' positions keep unchanged, i.e., $w_{k,t+1} = w_{k,t}$ and $\q_{k,t+1}= \q_{k,t}$,
for $k=1,\ldots,J$ and $k\not =j$.

To price an order $\Delta \q \in \mR^I$, the market maker adopts a \textit{multivariate-utility} (MU)-based mechanism described as follows. Suppose the market marker' MU function is $U(\cdot):\mR^I\rightarrow \mR$, which maps the net wealth position $\b{y}_t \triangleq  W_t \cdot \e - \Q_t \in \mR^I$ to some real value. We
call $\y_t$ the \textit{net wealth position} because it represents all the possible wealth level if the final state is revealed at time $t$.\footnote{\tb{For example, let $\y_t=\big( y_{1,t},\ldots,y_{I,t} \big)$ and $Q_{i,t}$ be the $i$-th element of $\Q_t$. Then, the $i$-th element of $\y$ is $y_{i,t}=W_t-Q_{i,t}$ which represents the wealth after paying the obligation if $i$-th event occurs.}} Similar to the utility-based pricing mechanism in
\cite{chen2007utility}, the price of the order $\Delta \q$ is determined by solving the following optimization problem:\footnote{{Note that the domain of function $U(\cdot)$, denoted by $\dom(U)$, may not necessarily be $\mR^I$. As a result, in problem ($\P_{\cmin}$), we explicitly add the constraint $(W_t + \Delta w )\cdot \b{e} - (\Q_t+ \Delta\q)\in \dom(U)$. This constraint can typically be expressed by some linear or nonlinear inequalities. } }
\begin{align}
\P_{\cmin}(\Delta \q):~&~\min_{\Delta w \in \mathbb{R}}~~\Delta w \notag\\
\textrm{Subject to:}~&~ U\big((W_t + \Delta w )\cdot \b{e} - (\Q_t+ \Delta\q)\big)\ge U(W_0\cdot \e),\label{eq:P_cmin_U}\\
~&~(W_t + \Delta w ) \cdot \e - (\Q_t + \Delta\q )\in \dom(U).\notag
\end{align}
The minimizer of problem $\P_{\cmin}(\Delta \q)$ is the price of the order $\Delta \q$. Constraint (\ref{eq:P_cmin_U}) can also be written as $U(\y_t + (\Delta w \cdot \b{e} - \Delta \q))\geq U(W_0\cdot\e)$, which means that the utility of the updated net wealth position is no less than that of the initial state. This is consistent with the utility-preserving property proposed in \cite{chen2007utility}, i.e., the market maker tries to find the lowest possible charge for the order $\Delta \q$ such that the post-trade utility level is no less than the initial level.

To ensure the above pricing problem is well-posed, throughout this paper, we assume $U(\cdot)$ satisfies the following assumptions. 
\begin{assumption} \label{assumption:U_mild}
$U(\cdot)$ satisfies the following conditions: (i) it is
{continuous}, concave and its domain $\dom(U)$ is convex with
$\e \in \mathrm{recc}(\dom(U))$; (ii) it is monotonically increasing.\footnote{In
this paper, we call a multivariate function
$F(\cdot):\mR^I\rightarrow \mR$ {\it monotonically increasing}, if
and only if for any $\y,\hat{\y}\in\dom(F)$, if $\y \ge \hat{\y}$,
then $F(\b{y})\ge F(\hat{\y})$ and inequality holds strictly when
$\y>\h{\y}$. }
\end{assumption}
%

Assumption \ref{assumption:U_mild} is mild and natural for the utility function except for the requirement $\e \in \textrm{recc}(\dom(U))$ which means that the utility is well-defined even if the wealth goes to infinity. \tb{
Under Assumption \ref{assumption:U_mild}, it can be verified that the MU-based pricing rule satisfies the axioms proposed in \cite{abernethy2013efficient}. These axioms serve as general requirements for a reasonable pricing mechanism. Furthermore, the MU-based pricing rule possesses several other desirable properties. For instance, it demonstrates responsiveness to any finite order with a unique price, effectively eliminates arbitrage opportunities for surebets, and exhibits a finite worst-case loss (under mild conditions). Moreover, there is abundant freedom to construct MU functions (see examples in Appendix \ref{apdx:ssec:MU_examples}). The following result establishes the relation between the MU-based pricing rule with the MSR and the cost function-based mechanism.}


\begin{proposition}\label{prop:equivalence_diff_mechanism}
An MU-based mechanism with a utility function satisfying Assumption \ref{assumption:U_mild} is equivalent with a cost function-based mechanism with a corresponding cost function. Furthermore, if an MU-based mechanism has a bounded loss under worst-case scenario, then the underlying utility can induce a proper scoring rule. Conversely, any proper scoring rule can induce an MU-based mechanism that has bounded worst-case loss.
\end{proposition}

\tb{The detailed proof of the above proposition is provided in Appendix \ref{apdx:ssec:equivalence_mechanism}}. In this work, we adopt the MU-based pricing rule in our analysis mainly due to the following reasons. First, as shown in Proposition \ref{prop:equivalence_diff_mechanism}, the MU-based pricing rule is general enough to cover existing classical pricing rules. More importantly, the MU-based mechanism provides a tractable way to study the convergence property of prices generated by a dynamic trading process. Once we have these convergence results, they can be directly applied to other pricing rules. Second, compared with the cost function-based pricing rule \cite{chen2007utility,abernethy2013efficient} and the LMSR \cite{hanson2003combinatorial}, the MU-based mechanism could explicitly incorporate the risk preferences and the heterogeneous beliefs of all participants in a unified framework, which enables us to explore how these heterogeneous factors affect the final prices of these securities. Third, it is more convenient to construct the MU function in practice compared with the other mechanisms. For example, using risk-measure to define the cost function needs it to be convex and \textit{translation invariant}, which means that the cost function should not be \textit{strictly convex}. Indeed, it is usually not an easy task to customize a valid cost function in an analytical form.\footnote{\tb{Although \cite{follmer2011stochastic} show} that any convex risk measure (cost function) can be defined numerically via an \textit{acceptance set} or by the dual representation, it requires solving a constrained optimization problem whenever a new order is submitted.} \tb{In contrast, the MU-based utility only requires the function to be \textit{concave} and \textit{monotone}, and it can be constructed in various ways, including the classic \textit{expected utility theory (EUT)} \cite{chen2007utility} and convex risk measures (e.g., \cite{chen2007utility, follmer2011stochastic, hu2014multi, frongillo2015elicitation}). Moreover, the price of an order can be computed by a simple line search. We provide additional properties of the MU function and some examples in Appendix \ref{apdx:sec:MU_function}.}

On the trader's side, we assume that each trader $j\in\{1,\ldots, J\}$ also adopts an MU function, $V_j(\cdot):\mR^I\rightarrow \mR$, to evaluate his/her net wealth position.\footnote{\tb{The choice of trader's MU function $V_j(\cdot)$ is similar to that of the market maker's MU function $U(\cdot)$. It can be based on conventional expected utility or constructed using a convex risk measure. Typically, in these formulas, the trader's subjective belief is explicitly specified.}} Following the spirit of \cite{sethi2016belief}, the $j$-th trader solves the following problem to decide the amount of securities to trade,
\begin{align}
\bar{\P}_j(w_{j,t},\q_{j,t}):~&~\max_{ \Delta \q \in \mR^I, ~\Delta w \in \mR }~~
V_j\big( (w_{j,t} - \Delta w)\cdot \e + ( \q_{j,t}+ \Delta \q) \big)\notag\\
\textrm{subject to}~&~ \Delta w = \arg \min_{w \in \mR }~\big\{
w~|~U\big((W_t + w) \cdot \e -(\Q_t
                       + \Delta \q)\big)\ge U(W_0\cdot\e),\notag\\
                       ~&~~~~~~~~~~~~~~~(W_t + w) \cdot  \e -(\Q_t + \Delta \q) \in\dom(U),\notag\\
                       ~&~~~~~~~~~~~~~~~(w_{j,t} - w)\cdot \e + ( \q_{j,t}+ \Delta \q) \in \dom(V_j)\big\}.
                       \label{eq:barP_max_constraint}
\end{align}

In this problem, the term $(w_{j,t} - \Delta w)\cdot \e + (\q_{j,t}+ \Delta \q)$ in the objective function represents the $j$-th trader's net wealth position after the current trading. Problem $\bar{\P}_j(w_{j,t},\q_{j,t})$ can be viewed as a bi-level optimization problem. It fully characterizes all the information in one round of trading, i.e., the size of the order $\Delta \q$ is determined by the upper-level problem, and the payment of the order $\Delta \q$, is determined by the subproblem in constraint (\ref{eq:barP_max_constraint}). The lower level subproblem in constraint (\ref{eq:barP_max_constraint}) is nothing but the problem $\P_{\cmin}(\Delta \q)$. To guarantee the whole trading process is
well defined, we impose the following conditions for $V_j(\cdot)$, $j=1,\ldots,J$.
\begin{assumption}\label{assumption:V_j_mild}
Each $V_j(\cdot)$ satisfies the following conditions: (i) it is {continuous} and strictly concave for
non-equivalent wealth vectors\footnote{We call $\y\in\mR^I$ and $\x\in\mR^I$ are equivalent wealth, if and only if
    $\y=\x+t\cdot\e$ for some $t\in\mR$. Strictly concavity of non-equivalent wealth means that, for any $\x, \y \in
    \dom(V_j)$ who are not equivalent wealth vectors, it has $V_j(\alpha\x+(1-\alpha)\y)>\alpha
V_j(\x)+(1-\alpha)V_j(\y)$ for any $\alpha\in(0,1)$.}; (ii) $\dom(V_j)$ is convex with $\mathrm{recc}(V_j)\subseteq
\mR^I_{+}$ and $\e \in \mathrm{recc}(\dom(V_j))$; (iii) it is monotonically increasing.
\end{assumption}

Assumption \ref{assumption:V_j_mild} is similar to Assumption \ref{assumption:U_mild} except that it requires $V_j(\cdot)$ to be strictly concave except along the direction $\e$. This condition is necessary to establish trading process convergence in the later part of this paper. Note that such an assumption is weaker than global strict concavity, and it can also accommodate risk measure-based MU function, which is linear along direction $\e$. 

\tb{
Although the state $\{w_{j,t},\q_{j,t}\}$ fully represents the trading process, it is complicated to study the
convergence of the trading process by solving problem $\bar{\P}_j(w_{j,t},\q_{j,t})$ directly. It is more convenient to use the \textit{net wealth position} to characterize the trading process. Specifically, at time $t$, let $\x_{j,t} \triangleq w_{j,t}\cdot \e + \q_{j,t}$ and $\y_{t}\triangleq W_t \cdot \e - \Q_t $ be the $j$-th trader's and the market maker's net wealth position, respectively. At time $t=0$, it has $\x_{j,0} = w_{j,0}\cdot \e$ and $\y_0= W_0 \cdot \e$. In each round of trading, when an order $\Delta \q$ is placed at price $\Delta w$, the variation of market maker's wealth is denoted as $\z \triangleq  \big(z_1, z_2, \ldots, z_I \big)= -\Delta w \cdot \e + \Delta \q$. Then, the dynamics (\ref{eq:dynamic_w_q}) transform into $\x_{j,t+1} = \x_{j,t} + \z$ if the $j$-th trader interacts with the market at time $t$, and $ \x_{k,t+1} = \x_{k,t}$ for $k \not =j$, along with the market maker's wealth update, $\y_{t+1} = \y_{t}-\z$. It is worth noting that the market total wealth adheres to the following condition:}
\begin{align}
\y_t + \sum_{j=1}^J \x_{j,t} = \sum_{j=1}^J \x_{j,0} + \y_0= {w}^{\all}_0 \cdot \e,\label{def_xysum}
\end{align}
for all $t =0,1,\ldots$, where $ {w}^{\all}_0 \triangleq \sum _{j=1}^J
w_{j,0} + W_0$ is the total wealth of all participants in the market. Eq. (\ref{def_xysum}) suggests that the total
net wealth position in the market maintains at a constant level throughout the trading process.

Using the wealth vectors $\x_{j,t}$ and $\y_t$, the utility preserving property (\ref{eq:P_cmin_U}) can be expressed as
$U\big(\y_{t} -\z) \geq U(W_0\cdot \e)$. This formulation motivates us to consider an equivalent formulation of the $j$-th trader's decision problem:
\begin{align}
\P_j(\x_{j,t}):~~\max_{ \z \in \mR^I}~&~V_j(\x_{j,t+1}) = V_j(\x_{j,t}+\z)\notag\\
\textrm{Subject to:}~&~
U(\y_{t} -\z) \ge U(W_0 \cdot \e),\label{Pj_constrt}\\
~&~\y_t - \z \in \dom(U),~\x_{j,t}+\z \in \dom(V_j),\notag
\end{align}
for given $\x_{j,t}$ and $\y_{j,t}$ satisfying (\ref{def_xysum}).
The following result establishes equivalence between problems
$\P_j(\x_{j,t})$ and $\bar{\P}_j( w_{j,t}, \q_{j,t})$ in the sense
of the optimal solutions.

\begin{proposition}\label{prop:equivalent_form}
Under Assumptions \ref{assumption:U_mild} and \ref{assumption:V_j_mild}, problem $\P_j(\x_{j,t})$ admits a unique
solution $\z^*$. Furthermore, any feasible solution $\{\Delta w^*, \Delta\q^*\}$ of problem $\bar{\P}_j(w_{j,t},\q_{j,t})$ satisfying $\Delta \q^* - \Delta w^* \cdot \e = \z^*$ is an optimal solution of $\bar{\P}_j(w_{j,t},\q_{j,t})$.
\end{proposition}

The above result suggests that $\P_j(\x_{j,t})$ captures all the
essence of market evolution encoded in the original problem
$\bar{\P}_j(w_{j,t},\q_{j,t})$. It greatly simplifies the original
market process characterized by $w_{j,t}$, $\q_{j,t}$, $W_{t}$ and
$\Q_t$ with the pricing problem $\bar{\P}_j(w_{j,t},\q_{j,t})$.
Therefore, in the consequent analysis, we focus on the formulation
$\P_j(\x_{j,t})$.

\begin{remark}[No-bankruptcy Restriction]\label{remark:no-bankruptcy}
\tb{In practical applications, traders often operate under the no-bankruptcy restriction, which means that the constraint $\x_{j,t}\geq 0$ holds for all $t$ and $j=1,\ldots, J$. When the problem $\P_j(\x_{j,t})$ includes such constraints, we denote the corresponding problem as $\P_j^+(\x_{j,t})$.}
\end{remark}

To facilitate analysis, we make an additional assumption.\footnote{
Assumption \ref{assumption:differentiabiliy} is not necessary to define
the MU-based pricing mechanism and not necessary for general convergence result. 
Imposing these assumptions helps simplify the analysis in dynamic trading process.}
\begin{assumption}\label{assumption:differentiabiliy}
$U(\cdot)$ and $V_j(\cdot)$, $~j=1,\ldots,J$ are continuously differentiable functions.
\end{assumption}

Before we analyze the convergence property, we discuss how to compute the instantaneous prices of the securities under the
MU-based mechanism. The instantaneous prices of the securities can be viewed as an integrated forecast of the corresponding event (see, e.g., \citealt{wolfers2004prediction}). Let $\y= (y_1, y_2,\ldots, y_I)$ be the market maker's net wealth posisition. The \textit{instantaneous price} of each security is defined as the marginal cost per share for
purchasing an infinitesimal quantity of this security. The
instantaneous price of the $i$-th security, $i\in\{1,\ldots, I\}$,
can be computed by pricing the order $\Delta \q = \epsilon \cdot \e_i$
where $\epsilon>0$ is a small number and $\e_i$ is the $i$-th
unit vector. If $\Delta w(\epsilon)$ is the solution of problem
$\P_{\cmin}(\epsilon\cdot \e_i)$, then the instantaneous price of the
$i$-th security is $p_i = \lim_{\epsilon \rightarrow 0} \Delta 
w(\epsilon)/\epsilon$. Under some assumptions of $U(\cdot)$, the
price $p_i$ can be explicitly computed.

\begin{proposition}\label{prop:inst_price}
Under Assumptions \ref{assumption:U_mild} and \ref{assumption:differentiabiliy}, given market state $\y$, if one of the following conditions holds: (i) the state variable $\y$ is an interior point of
$\dom(U)$ or (ii) $U(\y)$ approaches $-\infty$ whenever $\y$ approaches the boundary of  $\dom(U)$, then the
instantaneous price of the $i$-th security is
\begin{align}
p_i = \frac{\partial U(\y)/\partial y_i}{\sum_{k=1}^I \partial
U(\y)/\partial y_k},~~i=1,\ldots, I. \label{def_inst_price}
\end{align}
\end{proposition}
If we aggregate all $p_i$ in a vector $\b{p}=(p_1,\ldots, p_I)$, then the price vector is exactly the
normalized gradient of $U(\y)$, i.e., $\p = \nabla U(\y)\big/(\e^{\top}\nabla U(\y))$. Note that, when the assumptions in Proposition \ref{prop:inst_price} fail to hold, the domain boundary of $U(\cdot)$ will affect 
the solution of problem $\P_{\cmin}(\epsilon \cdot \e_i)$. In those cases, one can still compute the price by using the envelope theorem. However, it will be much more complicated as it will involve the form of the boundary of $U(\cdot)$. We omit the detailed discussions for those cases.

\section{Convergence of Market States in General Setting}\label{sec:convergence}

In this section, we study the convergence properties of the market states (wealth and prices) generated by the
sequential interactions between the traders and the market maker. We define the \textit{trading sequence} as an infinite sequence of traders indexed by time, denoted by $\mathcal{S}=\{j_1,j_2,\ldots,j_t,\ldots\}$ with $j_t\in \{1,\ldots, J\}$. We say a sequence $\mathcal{S}$ satisfies the {\it infinite participation} or simply the IP property, if for any $j\in\{1,\ldots,J\}$, the set $\{t|j_t = j, j_t\in \mathcal{S}\}$ has infinite elements. We denote all trading sequences satisfying the IP property by $\Phi$. In other words, $\Phi$ contains all trading sequences such that each trader interacts with the seller for an infinite number of times.\footnote{Note that the IP property is a weaker condition than the assumption made in \cite{sethi2016belief}. In Assumption 1 of \cite{sethi2016belief}, it is assumed that there exists a constant $m$ such that each trader must trade once during any period of length $m$. Obviously, this implies the IP property but not the reverse. Therefore, our assumption allows a more flexible trading pattern than that in \cite{sethi2016belief}.} \tb{Given a sequence $\mathcal{S}$, we can define the trading process as follows (which is referred to as Trading Process \ref{algtrading}):}
\vspace{5pt}
\vspace{10pt}
\begin{algorithm}[H]\label{algtrading}
\SetAlgorithmName{Trading Process} \SetAlgoLined ~Initialization:
set $\b{y}_0= W_0\cdot \e$, $\x_{j,0}=w_{j,0}\e$,  $\forall j=1,\ldots,
J$; $\mathcal{S} =\{j_1,j_2,\ldots, j_t,\ldots\}$ is given; Set $t =
1$.\

\While{True} {
  Select the $j$-th trader at time $t$ as $j = j_t$\;
  Solve $\z$ from problem $\P_j(\x_{j,t})$ (or problem $\P_j^+(\x_{j,t})$ if there is no bankruptcy constraint)\;
  Update: for $t=t+1$
            $$\b{y}_{t+1}=\b{y}_t-\z,~~ \b{x}_{j,t+1}=\b{x}_{j,t}+\z.$$
 }
\caption{Trading Process}
\end{algorithm}
\vspace{10pt}

In Trading Process \ref{algtrading}, we use $\p_t=(p_{1,t},\ldots, p_{I,t})$, $t=1,2,\ldots$, to denote the instantaneous price of the securities at time $t$. Given $\y_t$, $\p_t$ can be computed by (\ref{def_inst_price}). As $t\rightarrow \infty$, if $\x_{j,t}$ and $p_{i,t}$ converge to some limits $\x_{j}^*$ and $p_{i}^*$, respectively,  then we call $\x_j^*$ and $p_{i}^*$ the limiting wealth allocation for trader $j$ and the limiting price of the $i$-th security. Some important questions arise naturally: Does the wealth allocation $\x_{j,t}$ generated from the Trading Process \ref{algtrading} converge to some limiting allocation? If so, what property does the limiting wealth allocation have? Similarly, does the price generated from the Trading Process 1 converge and what property does the limiting price have?

To answer these questions, we first define the feasible set of all possible market states generated by the Trading Process
\ref{algtrading} as follows,
\begin{align}
\A=&\Big\{ \big( \x_1,\ldots, \x_J \big)~\Big|~\y +\sum_{j=1}^J \x_j = {w}_0^{\all}\cdot \e,~~U(\y) \geq U( W_0 \cdot \e),\notag\\
   &~~~~~~V_j(\x_j)\ge V_j(w_{j,0}\cdot\e),~(\x_j\geq\0, \textrm{if bankruptcy is prohibitted})~j=1,\ldots, J~ \Big\}.\label{def_cA}
\end{align}
It is not hard to prove that the feasible set $\A$ is a compact set. We then introduce the definition of \textit{Pareto optimality} of the wealth allocation with respect to the MU function $V_j(\cdot)$ for $j\in \{1,2,\ldots, J\}$.
\begin{definition}\label{def_pareto_optimal}
A wealth allocation $(\x_1^*, \ldots, \x_J^*)$ is called a Pareto optimal allocation if the following two conditions hold:
(i) $U( {w}_0^{\all} \cdot \e - \sum_{j=1}^J \x_j^* )\ge U(W_0\cdot\e)$; (ii) there does not exist any $(\x_1, \ldots, \x_J) $ such that $U( {w}_0^{\all}\cdot \e - \sum_{j=1}^J \x_j )\ge U(W_0\cdot\e)$ and $V_j(\x_j) \geq V_j(\x_j^*)$ for all $j\in \{1,\ldots, J\}$ with strict inequality holds for at least one $j^{\dagger}\in\{1,\ldots, J\}$.
\end{definition}

The Pareto optimal allocation means that, for each trader, it is not possible to increase his/her own utility without decreasing the utility of any other traders. All the Pareto optimal wealth allocations form the \textit{Pareto efficient set}. Note that, the concavity of $V_j(\cdot)$ guarantees that any Pareto efficient wealth allocation can be achieved by some weighting parameters $\pmb{\nu}=(\nu_1,\ldots,\nu_J)\in \mR^J_+ \setminus \{\0\}$ (see, e.g., \citealt{mas1995microeconomic}) by solving the following optimization problem:
\begin{align}
        \P_{\po}(\pmb{\nu }):~&~\max_{\x_1,\ldots, \x_J} ~~\sum_{j=1}^J \nu_j V_j(\x_j) \notag \\
        \textrm{Subject to}:~&~\y +\sum_{j=1}^J \x_j = {w}_0^{\all}\cdot \e,\notag\\
        ~&~U(\y) \geq U( W_0 \cdot \e), \label{PO:util_preverve}\\
        ~&~\x_j\ge\0,\quad j=1,\ldots,J~~(\textrm{if bankruptcy is prohibitted}).\notag 
\end{align}

We then present our main result which states that the Trading Process \ref{algtrading} converges to some
Pareto optimal allocation.
\begin{theorem}\label{thm_Uconv}
Under Assumptions \ref{assumption:U_mild}, \ref{assumption:V_j_mild} and \ref{assumption:differentiabiliy}, the wealth allocation $\x_{j,t}$ and $\y_t$ generated from Trading Process \ref{algtrading} for any given
trading sequence $\mathcal{S}\in\Phi$ converge to some limit
$\x^*_j$ and $\y^*$, i.e., $\lim \limits_{t\rightarrow \infty} \x_{j,t}=
{\x}^*_j$ for all $j=1,\dots, J$ and $\lim \limits_{t\rightarrow \infty} \y_{t}=
{\y}^*$, if either one of the following
conditions holds: (i) $\dom(U)=\mR^I$; or (ii) $\dom(U)$ can be
expressed as $\dom(U)=\{\y~|~ \y\ge\b{L}\}$ with $\b{L} <
W_0\cdot\e$ and satisfies $\underset{y_i\to L_i}{\lim}\frac{\partial
U}{\partial y_i}=+\infty$ and $\frac{\partial U}{\partial y_i}$ is
finite if $\y_i\ne L_i$. Furthermore, the limiting allocation
$\big(\x^*_1,\ldots, \x^*_J \big)$ must be a Pareto optimal
allocation.
\end{theorem}

In the above result, two alternative conditions \tb{(condition (i) and (ii)) are imposed to ensure that the limit of wealth allocation is an interior point in the domain of the utility function. Condition (i) is easy to understand, while condition (ii) is a generalization of the \textit{Inada condition}, which is commonly used in conventional utility optimization.} Theorem \ref{thm_Uconv} states that, as the Trading Process \ref{algtrading} goes on, eventually, no trader would like to modify their wealth position anymore, i.e., the wealth process $(\x_{1,t}, \ldots,
\x_{J,t})$ converges to some limiting wealth allocation. Moreover, the limiting wealth allocation $(\x_1^*, \ldots, \x_J^*)$ must locate on the Pareto optimal set defined in Definition \ref{def_pareto_optimal}. 
\tb{ 
The Pareto optimality of the limiting allocation formally confirms the theoretical foundations of prediction markets: a prediction market can indeed incentivize people to participate, and the limiting price is a ``good price'' in the sense that it is supported by a wealth allocation where the well-being (measured by utility level) of any trader cannot be improved anymore without decreasing some other traders' utility.
}
\begin{remark}\label{rem_infinite_trading}
\tb{
We emphasize that the requirement $\mathcal{S}\in\Phi$ is necessary for achieving overall Pareto optimality. The trading process will still converge if one or more traders cease trading from time $t>0$ and the remaining traders will still reach their own Pareto optimality. However, if a trader only interacts with the market maker a finite number of times, then he/she may still be willing to trade after his/her last transaction, but is deprived of that opportunity. Consequently, this trader's final wealth along with other traders' wealth may not be located in the Pareto optimal set.}
\end{remark}

A direct result of Theorem \ref{thm_Uconv} is that the price of the security also converges to some limiting price. Such a limiting price can be computed as follows.
\begin{corollary}
When the assumptions in Theorem \ref{thm_Uconv} are met and the trading process converges, 
the limiting price is given by,
\begin{align}
        \b{p}^* = \frac{\nabla U(\y^*)}{\nabla U(\y^*)^{\top}\e}, \label{def_U_price}
\end{align}
where $\y^*$ is the market maker's the limiting wealth allocation.
\end{corollary}

The price formula (\ref{def_U_price}) is a natural generalization of the one derived for the scalar-valued utility function in \cite{chen2007utility}. In the upcoming sections, we will study how the market forms prices by aggregating traders' beliefs. To represent these beliefs, we use $\pmb{\theta}=(\theta_1,\ldots,\theta_I) \in \Theta_I$ and $\pmb{\pi}_j=(\pi_{1,j},\ldots, \pi_{I,j})\in \Theta_I$ to denote the market maker's and the $j$-th trader's subjective beliefs (probabilities) on the outcomes, respectively. Unless otherwise stated, we assume that these beliefs remain constant throughout the trading process. We will use $\p_t = \big( p_{1,t}, p_{2,t},\ldots, p_{I,t} \big)$ to denote the price vector generated by Trading Process \ref{algtrading}, where $t=0,1,2\ldots$.

\section{Exponential Utility and Risk Measure-Based Market}\label{sec:expU_Risk}

The previous section demonstrated market states converging to Pareto optimal allocations but provided limited insight into how the market forms equilibrium prices. In this section, we explore two special cases: exponential utility-based and risk measure-based MU functions. We will derive the price convergence results for these market models.


\subsection{Exponential Utility Based Market}\label{sse:exponential}
We first consider the \textit{exponential utility functions}-based market. The exponential utility has a constant risk aversion coefficient, which leads to trading decisions independent of the trader's wealth level \cite{makarov2010note}. This feature eliminates the wealth effect in the trading process and allows us to clearly illustrate the impacts of traders' beliefs and risk preferences upon the limiting prices. \tb{It is important to note that adopting the exponential utility to price the security is equivalent to using the LMSR pricing rule, which is a widely adopted mechanism (see Appendix \ref{apdx:ssec:equivalence_mechanism} or \cite{chen2007utility}).} Specifically, given market state vectors $\y=(y_1,\ldots,y_I)$ and $\x_j=(x_{1,j},\ldots, x_{I,j})$ for $j=1,\ldots,J$, \tb{the} market maker's MU function and the $j$-th trader's utility function are
\tb{
\begin{align}
U(\y)\triangleq \sum_{i=1}^I \theta_i \frac{1- e^{-\beta \cdot
y_i}}{\beta}~~~\textrm{and}~~~V_j(\x_j)\triangleq \sum_{i=1}^I
\pi_{i,j}\frac{1- e^{-\alpha_j\cdot x_{i,j}}}{\alpha_j},~~j=1,\ldots,J \label{eq:exp_U_Vj}
\end{align}
respectively, where $\beta>0$ and $\alpha_j>0$ are the risk aversion coefficients for the market maker and the $j$-th trader, respectively.}




Since the exponential utility-based MU function satisfies the conditions outlined in Theorem \ref{thm_Uconv}, both wealth allocation and instantaneous prices converge to their respective limiting values. The following results characterize the evolution of market prices generated by Trading Process \ref{algtrading}.

\begin{proposition}\label{prop_exp_unique}
\tb{
In a market equipped with the exponential utility functions defined in (\ref{eq:exp_U_Vj}), given the market state $\{\x_{j,t}, \y_{t}\}$ with the price $\p_t$ at time $t$, it has the following results: (a) If the $j$-th trader trades with the market maker, then the optimal trading amount is $\z^*=(z^*_1,\ldots,z^*_I)$ with
\begin{align}
z_i^*= \frac{1}{\beta}\ln\left( \frac{ \big( \hat{\pi}_{i,j,t}\big/p_{i,t} \big)^{\frac{\beta}{\alpha_j+\beta}}  }
{ \sum \limits_{k=1}^I  p_{k,t} \left( \hat{\pi}_{k,j,t} \big/ p_{k,t}  \right)^{ \frac{\beta}{\alpha_j+\beta} }  } \right),
~~i=1,\ldots, I, \label{prop_exp_z}
\end{align}
where $\hat{\pi}_{i,j,t} \triangleq \pi_{i,j} \cdot e^{-\alpha_j x_{i,j,t}} \big/(\sum \limits_{k=1}^I \pi_{k,j} \cdot e^{-\alpha_j x_{k,j,t}})$. (b) The price is updated to 
\begin{align*}
p_{i,t+1} = \frac{(p_{i,t})^{\frac{\alpha_j}{\alpha_j+\beta}} \cdot(
\h{\pi}_{i,j,t} )^{\frac{\beta}{\alpha_j+\beta}}} { \sum \limits_{k=1}^I
(p_{k,t})^{\frac{\alpha_j}{\alpha_j+\beta}}\cdot(\hat{\pi}_{k,j,t})^{\frac{\beta}{\alpha_j+\beta}} },~~i=1,\ldots,I.
\end{align*}
(c) As $t\rightarrow \infty$, for any ${\mathcal{S}}\in\Phi$, price $\p_t$
converges to the limiting price $\p^*=\big( p^*_{1},\ldots, p^*_I
\big)$, where 
\begin{align}
p^*_i=\frac{ \left(\theta_i^{\frac{1}{\beta}}\prod \limits_{j=1}^J \pi_{i,j}^{\frac{1}{\alpha_j}} \right)^{ \frac{1}{\sum_{j=1}^J \frac{1}{\alpha_j} +\frac{1}{\beta} }}}{ \sum \limits_{k=1}^I  \left(\theta_k^{\frac{1}{\beta}}\prod \limits_{j=1}^J \pi_{k,j}^{\frac{1}{\alpha_j}} \right)^{ \frac{1}{\sum_{j=1}^J \frac{1}{\alpha_j} +\frac{1}{\beta} }} },~~i=1,\ldots, I.\label{prop_expp_pi}
\end{align}
}
\end{proposition}

The above result demonstrates that the limiting price corresponds precisely to the geometric mean of the risk-adjusted beliefs held by all participants, including the traders and the market maker. It is worth noting that this limiting price, $p_i^*$, remains unaffected by the chosen trading sequence, represented by $\mathcal{S}$, as long as $\mathcal{S}\in\Phi$. Price formulation (\ref{prop_expp_pi}) is closely connected to the concept of the Logarithm Opinion Pool mentioned in \tb{\cite{chakraborty2015market}}\footnote{\tb{The Logarithm Opinion Pool \tb{refers} to the normalized weighted geometric mean of different opinions which is originated from the early works such as \cite{Morris:MS1974} and \cite{Bordley:MS1982} in area of the opinion pooling.}}. In \cite{chakraborty2015market}, as it only focuses on the static trading model, it suggests one potential research direction as extending these findings to a market involving multiple agents engaging in repeated trading until market states converge. By providing an explicit outcome for this scenario, Proposition \ref{prop_exp_unique} effectively addresses such a question raised by \tb{\cite{chakraborty2015market}}. 

\begin{remark}\label{rem_risk_neutral}
\tb{
It is also important to note that, although the risk-neutral utility is a special case of the exponential utility,\footnote{If we let $\beta \rightarrow 0$ or $\alpha_j\rightarrow 0$ in (\ref{eq:exp_U_Vj}), then the exponential utilities become risk-neutral utilities, i.e., $U(\y)=\sum_{i=1}^I \theta_i y_i$ and $V_j(\x_j)=\sum_{i=1}^I \pi_{i,j}x_{i,j}$ for all $j=1,\ldots, J$.} the result in Proposition \ref{prop_exp_unique} does not hold for the risk-neutral case as it needs the assumption $\beta>0$ and $\alpha_j>0$ for all $j=1,\ldots, J$. Even when the market maker adopts the exponential utility, but the trader's utility is risk neutral, the whole trading process may not converge due to violating the strictly convex condition in Assumption \ref{assumption:V_j_mild}. Indeed, under the risk-neutral setting, after the $j$-th trader interacts with the market, the price of the $i$-th security will be modified exactly to the $j$-th trader's belief, $\pi_{i,j}$, which leads to price oscillation (see the detail in Appendix \ref{apdx:risk-neutral}).
}
\end{remark}

\subsection{Risk Measure-Based Market}\label{sse:risk-measure}

\tb{
We have demonstrated that the exponential utility-based market can converge to a unique limiting price, regardless of the trading sequence. This subsection delves deeper into the risk-measure-based MU function, which exhibits a similar property under specific conditions. Although some literature (e.g., \cite{hu2014multi,frongillo2015elicitation}) has investigated the risk-measure-based prediction market and developed convergence results, the existing studies provide limited information about the general convergence condition and the method to compute the limiting price. This section aims to complete these missing pieces.}

\tb{We mainly focus on the convex risk measures, which possess some promising properties (\cite{follmer2011stochastic}) that have emerged as a valuable tool in decision-making. However, directly utilizing the convex risk measure to construct the MU functions in trading problem $\P_j(\x_{j,t})$ may not satisfy the conditions outlined in Assumptions \ref{assumption:U_mild} and \ref{assumption:V_j_mild}, which leads to a potential non-convergence in the trading process. To illustrate such a case, we provide a counter-example in Appendix \ref{apdx:ssec_exp_risk_measure} (Example \ref{example_RM_counterexample}). In order to address this issue, we need additional mild conditions on the risk-measure-based MU functions. Specifically, we assume the $j$-th trader's MU function and the market maker's MU function are respectively constructed by some risk measures as follows,
\begin{align}
V_j(\x_j)=-\rho_j(\x_j),~j=1,\ldots,J,~~~\textrm{and}~~~U(\y)=-\rho_{0}(\y), \label{def_U_V_RM}
\end{align}
where $\rho_{j}(\cdot):\mathbb{R}^I \rightarrow \mathbb{R}$ for $j=0,1,\ldots,J$ are \textit{differentiable} convex risk measures. Recall that a convex risk measure has the following dual representation (see e.g., \cite{follmer2011stochastic}):
\begin{align}
\rho_j(\x)= \sup_{ \bm{p} \in \Theta_I} \big\{ - \bm{p}^{\top}\x - \alpha_j(\bm{p}) \big\} 
          =-\inf_{ \bm{p}  \in\Theta_I} \big\{   \bm{p}^{\top}\x + \alpha_j(\bm{p})\big\}, \label{def_U_V_RM_dual}
\end{align}
where $\alpha_j(\cdot):\Theta_I \to \mR$ is called the \textit{penalty function} for $j=0, \ldots, J$. We assume $\alpha_j(\cdot)$ is differentiable for all $j=0,\ldots,J$ and define its partial derivative as follows. For some $ \bm{p}=( p_1,p_2,\ldots,p_I)\in \Theta_I$, the derivative of $\alpha_j( \bm{p})$ is denoted as: 
\begin{align}
f_{i,j}(  p_i) \triangleq \frac{\partial \alpha_j(\bm{p}) }{\partial  p_i},~~~i=1,\ldots,I,~j=0,\ldots,J.\label{def_fij}
\end{align}
If we impose some mild conditions on $f_{i,j}(\cdot)$, the market constructed by the risk measures will also converge to a unique limiting state.
} 
\tb{
\begin{proposition}\label{prop:rm_unique}
If the functions $\{f_{i,j}(\cdot)\}|_{i=1,j=1}^{I,J}$ defined in (\ref{def_fij}) satisfy the following two conditions: (i) $f_{i,j}(\cdot)$ is  continuously differentiable and strictly increasing; and (ii) $f_{i,j}(1)$ is finite and $\underset{p_i\to 0}{\lim} f_{i,j}(p_i)=-\infty$ for all $i= 1,\ldots, I$ and
$j=0,\ldots, J$, then the MU function $U(\cdot) = -\rho_0(\cdot)$
satisfies Assumption \ref{assumption:U_mild}, the MU functions
$V_j(\cdot)=-\rho_j(\cdot)$, $j=1,\ldots,J$, satisfy Assumption
\ref{assumption:V_j_mild}. In addition, both $V_j(\cdot)$ and $U(\cdot)$ satisfy Assumption
\ref{assumption:differentiabiliy}. Thus, the market constructed by (\ref{def_U_V_RM}) meets 
the conditions in Theorem \ref{thm_Uconv} and the market state converges to a Pareto optimal allocation.
\end{proposition}
}
Based on the above result, to check whether the market constructed by the risk measures (\ref{def_U_V_RM}) converges, we only need to check whether the penalty functions $\{\alpha_j(\cdot)\}|_{j=0}^J$ meet the conditions in Proposition \ref{prop:rm_unique}. 

\tb{
We then focus on characterizing the limiting wealth allocation and the correspondent limiting price. We have the following result. 
\begin{theorem}\label{thm:alloc-unique}
In risk measure-based market which satisfies the condition in Proposition \ref{prop:rm_unique}, the following results are true: (i) The limiting wealth allocation is the unique solution of problem $\P_{\po}(\pmb{\nu})$ by setting $\pmb{\nu}=\e$. (ii) For any trading sequence ${\mathcal{S}}\in\Phi$, the unique limit price generated by Trading Process \ref{algtrading} is $\p^*=-\nabla \rho_0(\y^*)$ where $\y^*={w}_0^{\all} \cdot \e- \sum \limits_{j=1}^J \x_j^*$ and $(\x^*_1,\ldots, \x^*_J)$ are the solution of problem $\P_{\po}(\e)$. Furthermore, the price vector $\p^*$ is given by $\p^*=\arg \underset{ \bm{p} \in \Theta_I}{\min}~\sum \limits_{j=0}^J~\alpha_j(\bm{p})$ 
where $\alpha_j(\cdot)$ is defined in (\ref{def_U_V_RM_dual}) for $j=0,1,\ldots,J$.
\end{theorem}}
\tb{One surprising result of this theorem is that the limiting price can be independent of the initial wealth allocation. The root cause is the translation invariance property of risk measures. Indeed, considering the $j$-th trader's decision problem, the objective function $ \min_{\z }\{\rho_j(w_{j,0}\cdot\b{e}+\z)\}$ is equivalent to $ \min_{\z } \{\rho_j(\z)\}$.
Therefore, the initial wealth $w_{j,0}$ can initially be eliminated from the individual problem. Hence, all subsequent trades become independent of the initial wealth. A similar observation holds for the market maker. As a result, the limiting $\y^*$, which determines the limiting price $\p^*$, is independent of the initial wealth allocation.
}

\tb{
With the help of Theorem \ref{thm:alloc-unique}, it becomes feasible to explicitly compute the limiting price $\b{p}^*$ when the formulation of the risk measures $\rho_j(\cdot)$ is provided. We now consider a particular type of risk measure constructed using a dual formulation. We maintain the same notations as in Section \ref{sse:exponential}, employing $\{\theta_i\}|_{i=1}^I$ and $\{\pi_{i,j}\}|_{i=1}^I$ to represent the beliefs of the market maker and the $j$-th trader, respectively. The following result establishes that the limiting price can be expressed as the power-weighted mean of all participants' beliefs.}
\tb{
\begin{corollary}\label{cor:mean_exact} Let $h_j \geq 0$, $j=0,\ldots,J$, be some parameters. If the penalty functions $\alpha_j(\cdot)$ take the following form, $\alpha_0(\b{p})=-\frac{1}{\gamma}\sum \limits_{i=1}^I p_i \big( p_i/\theta_i\big)^{\gamma-1} h_0$ and $\alpha_j(\b{p})$$=$$-\frac{1}{\gamma}\sum \limits_{i=1}^I p_i \big(p_i/\pi_{i,j}\big)^{\gamma-1} h_{j}$ for $j=1,\ldots,J$ with some $\gamma<1$, then the limiting price is
\begin{align}
    p_i^*=\frac{ \left(h_0 \cdot \theta_i^{ \frac{1}{1-\gamma}} + \sum \limits_{j=1}^J
            h_{j}\cdot \pi_{i,j}^{\frac{1}{1-\gamma}}\right)^{1-\gamma}}{\sum \limits_{k=1}^I \Big( h_0 \cdot \theta_k^{
\frac{1}{1-\gamma}} + \sum \limits_{j=1}^J h_{j} \cdot \pi_{k,j}^{\frac{1}{1-\gamma}}\Big)^{1-\gamma}} \quad \label{power_mean_exact}
\end{align} 
for $i=1,\ldots,I$. 
\end{corollary}
}
\tb{
The price formula (\ref{power_mean_exact}) shows that the limit price is a \textit{weighted power mean} of the traders' beliefs where the weight is a trader dependent parameter $h_j$. When $\gamma \rightarrow 0$, the penalty function becomes the \textit{cross entropy}, i.e., $\alpha_0(\b{p})= -\sum \limits_{i=1}^I \theta_{i} \log(p_i) h_0$ with $\alpha_j = -\sum \limits_{i=1}^I \pi_{i,j}\log(p_i)h_{j}$ for $j=1,\ldots,J$, and the corresponding limiting price becomes the arithmetic weighted mean of trader beliefs. Regarding the selection of parameter $h_j$, one particular choice is the initial wealth of each trader. In some application, such as the artificial prediction market \cite{Chakravorti:2023}, these parameters can be viewed as the hyper-parameters which control the overall model performance.     
}

\section{Limiting Price for CRRA Utility}\label{sec:CRRA}

\subsection{Price Variability}\label{sse:crra}
\tb{
The preceding section demonstrates that, under certain conditions, the price sequence produced by the exponential utility-based and risk-measure-based markets can converge to a unique limit price for any trading sequence that satisfies the IP property. This motivates us to inquire if a similiar outcome can be extended to more commonly used utility functions. In this section, we mainly explore a market model constructed by the utility with constant relative risk aversion (i.e., CRRA utility). The CRRA utility enjoys widespread usage in economic modeling and analysis. In the CRRA utility-based market, we assume the market maker's and the $j$-th trader's MU functions are
\begin{align} 
U(\y)=\frac{1}{\gamma_m} \sum_{i=1}^I \theta_i \cdot y_i^{\gamma_m}~~\textrm{and}~~ 
V_j(\x_j)=\frac{1}{\gamma} \sum_{i=1}^I \pi_{i,j} \cdot x_{i,j}^{\gamma},~~j=1,\ldots,J, \label{def:CRRA}
\end{align} 
respectively, where $\gamma<1$ and $\gamma_m<1$ are the traders' and market maker's risk aversion parameters, respectively; $\y \in \mR^I$ and $\x_j \in \mR^I$ are the wealth vectors; $(\theta_1,\ldots,\theta_I) \in \Theta_I$ and $(\pi_{1,j},\ldots, \pi_{I,j})\in \Theta_I$ are the associated subjective beliefs as we defined in previous sections. 
}
\tb{
Under the above setting, the trader's decision problem $\P_j(\x_j)$ generally lacks a closed-form solution. Nonetheless, when all participants have identical risk parameters, the trader's decision can be solved as follows. 
\begin{proposition}\label{prop:crra_trader_solution}
In the CRRA uitlity-based market definied by (\ref{def:CRRA}) with $\gamma=\gamma_m$, given $\{\x_{j,t}\}|_{j=1}^J$ and $\y_t$ at time $t$, the solution of the $j$-th trader's decision problem $\P_j{(\x_{j,t})}$ is $z^*_i = (y_{i,t}\cdot \kappa_{i,j} - x_{i,j,t} \cdot \lambda^*_j )/(\lambda^*_j +\kappa_{i,j})$ for $i=1,\ldots,I$, and the market states are updated to
\begin{align}
y_{i, t+1}  = \frac{ \lambda_{j}^* }{ \kappa_{i,j} + \lambda_{j}^*}\big( x_{i,j,t} + y_{i,t} \big),~~
x_{i,j,t+1} = \frac{ \kappa_{i,j} }{\kappa_{i,j} + \lambda_{j}^*} \big( x_{i,j,t} + y_{i,t} \big), \label{def_crra_ytxt}
\end{align} 
where $\kappa_{i,j} \triangleq  \big(\pi_{i,j} \big/\theta_i\big)^{\frac{1}{1-\gamma}}$, for $i=1,\ldots, I$ and $j=1,\ldots,J$, and $\lambda_j^*>0$ is the solution to the following equation,
\begin{align}
\sum_{i=1}^I \left(\frac{  \lambda_j^*}{\lambda_j^* + \kappa_{i,j} }(x_{i,j,t} + y_{i,t})\right)^{\gamma} \theta_i =  W_0^{\gamma}.\label{crra_equation}
\end{align}
\end{proposition}
}
\tb{This proposition shows that in a CRRA-utility-based market, the initial wealth of each trader is recursively adapted into the wealth vector $\b{y}_t$ of the market maker, hence the limiting $\y^*$ may be affected by the initial wealth allocation, as opposed to the risk measure-based markets. The following example demonstrates how the initial wealth, risk parameters, and trading sequence could affect limiting price.}
\begin{example}\label{exam_crra}
\tb{
We consider a CRRA utility-based market with three assets ($I=3$) and two traders ($J=2$) whose belief are $\pmb{\pi}_1=(0.2, 0.2, 0.6)$ and $\pmb{\pi}_2=(0.6, 0.1, 0.3)$, respectively. The market maker's parameters are $\theta_1=\theta_2=\theta_3=1/3$, $W_0=5$ and $\gamma_m=\gamma$. The two traders' initial wealth are $w_{1,0}=w^1$ and $w_{2,0}=w^2$ where we vary the parameters $(w^1, w^2)$ to examine their impact on the limiting prices. In Trading Process \ref{algtrading}, we consider two trading sequences: $\mathcal{S}_1=\{1, 2, 1, 2,\ldots\}$ and $\mathcal{S}_2=\{2,1,2,1,\ldots\}$.\footnote{Note that if the same trader arrives twice in a row, the second trade is vacuous because the first one is already utility maximizing.} 
Figure \ref{fig:exam_crra_util} depicts the utility value trajectories generated by these trading processes for different $(w^1,w^2)$ when $\gamma=0.5$. The Pareto efficient frontier (represented by the solid line) is obtained by solving problem $\P_{\po}(\pmb{\nu})$ for different $\pmb{\nu}$. This figure illustrates that, for a given wealth distribution and fixed trading sequence (either $\mathcal{S}_1$ or $\mathcal{S}_2$), the utility value trajectories indeed converge to points on the Pareto efficient frontier, confirming the result in Theorem \ref{thm_Uconv}. However, even for identical initial wealth, different trading sequences may generate different utility paths (and hence different limiting prices). 
}
\tb{
Figure \ref{fig:exam_crra_price} demonstrates the impact of the risk parameter. It shows the price trajectories of $p_{1,t}$ and $p_{3,t}$ when $w^1=w^2=10$ for the trading sequence $S_1$. This illustration makes it clear that a change in the risk aversion parameter, e.g., from $\gamma=0.5$ to $\gamma=0.7$, has an effect on the limiting prices. Table \ref{table_exam_crra} provides details of the limiting prices for different $(w^1, w^2)$ and $\gamma$. A closer examination reveals that even with fixed initial wealth, the limiting price may vary under different trading sequences, although both $\mathcal{S}_1$ and $\mathcal{S}_2$ satisfy the IP property.\footnote{It is worth noting that when there are only two traders, $\mathcal{S}_1$ and $\mathcal{S}_2$ are effectively the only possible sequences that satisfy the IP property. The impact of the trading sequence is also discussed in \cite{sethi2016belief}.} However, among these factors, the trading sequence has less impact compared to the initial wealth and the risk parameter.
}

\begin{figure}
\caption{The utility and price trajectories in Example
\ref{exam_crra}}
\centering
\begin{subfigure}{0.48\textwidth}
    \includegraphics[width = \linewidth]{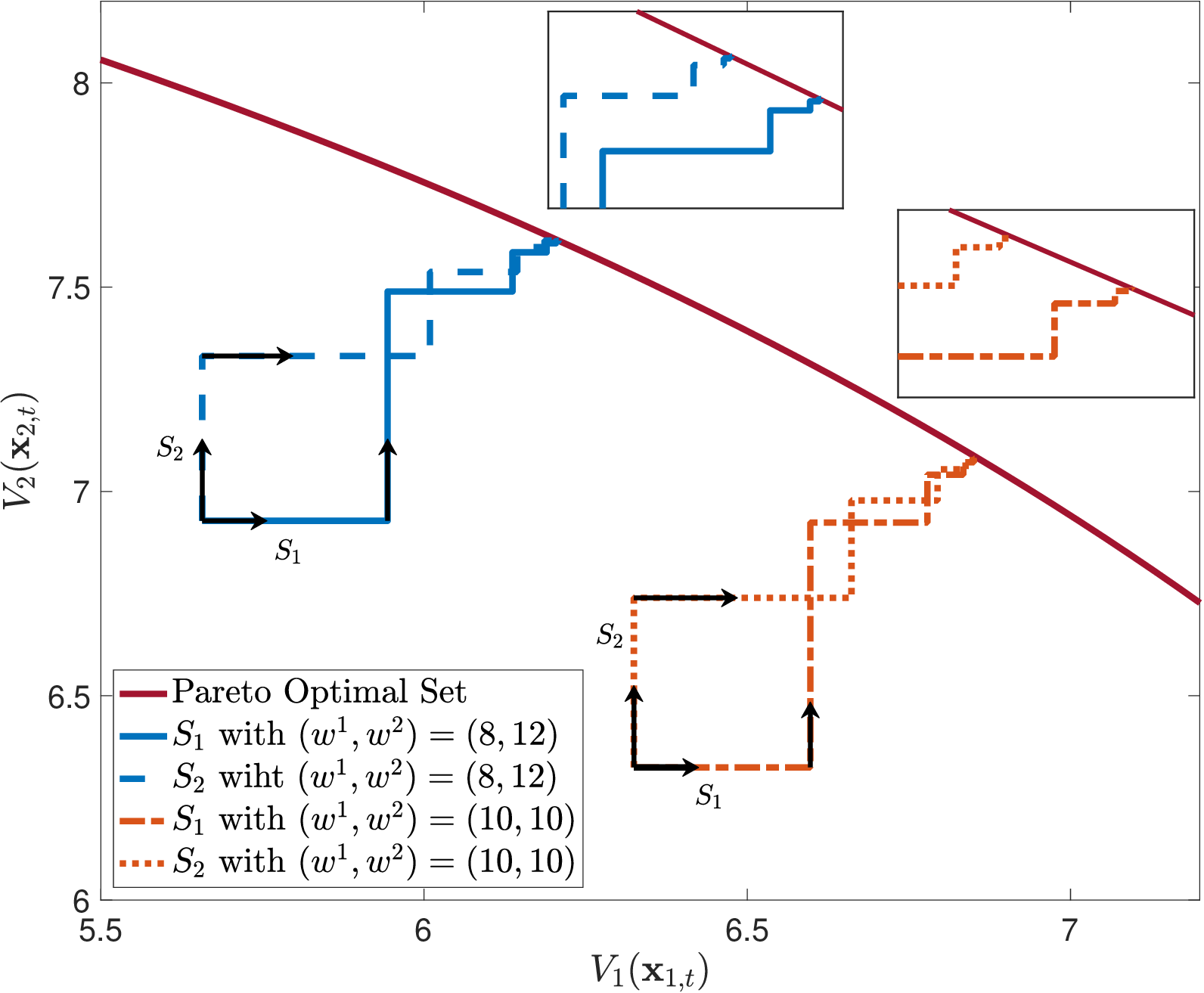}
    \caption{\small Trajectory of the utility}
    \label{fig:exam_crra_util}
\end{subfigure}
\begin{subfigure}{0.485\textwidth}
    \includegraphics[width = \linewidth]{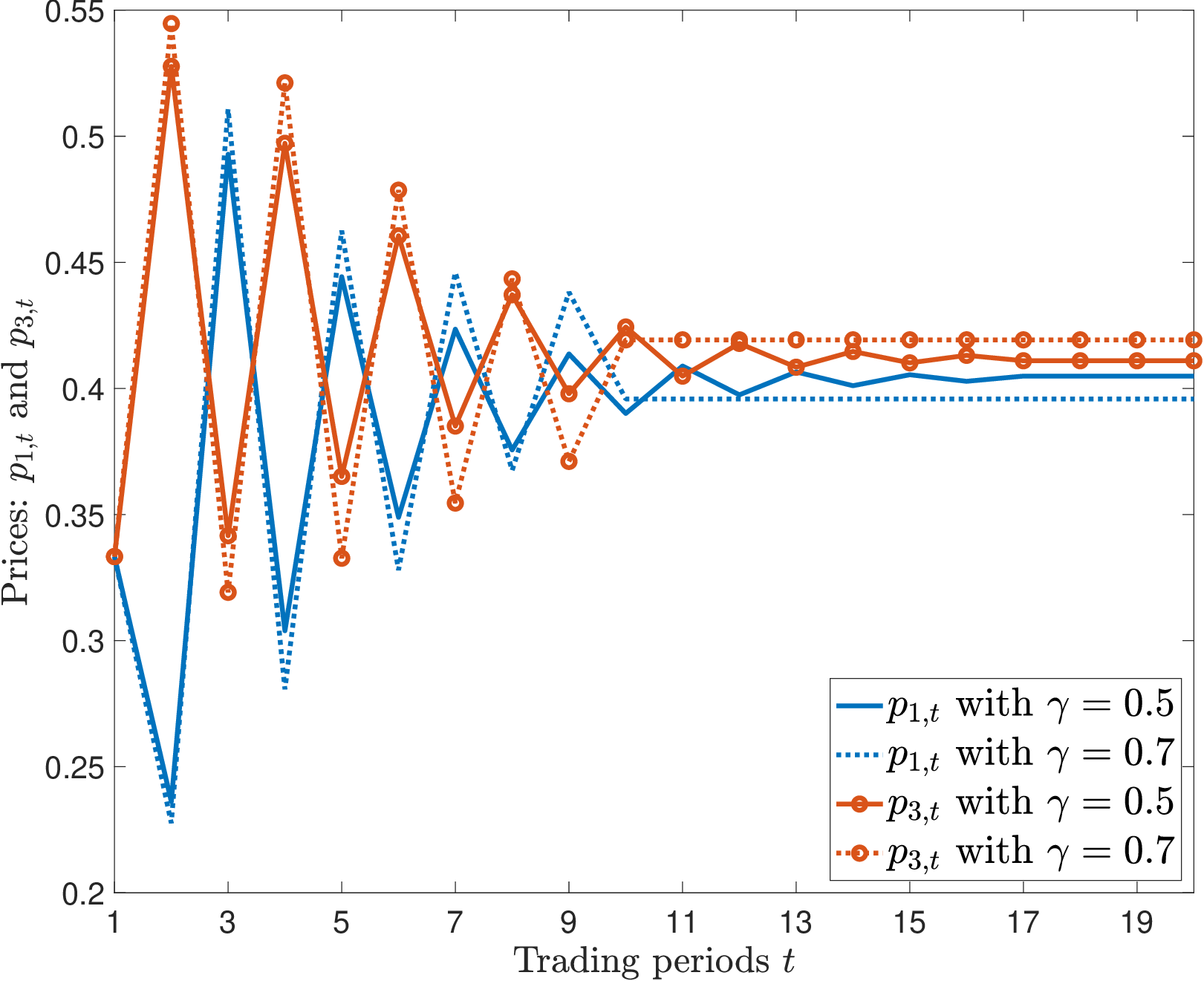}
    \caption{\small Trajectory of the price}
    \label{fig:exam_crra_price}
\end{subfigure}
\label{fig_exam_crra}
\end{figure}

\begin{table}
\small
\caption{\tb{Limiting wealth allocation and price in Example
\ref{exam_crra}} }\label{table_exam_crra} 
\centering
\begin{tabular}{cc c c ccc}
\toprule
\multicolumn{2}{c}{Wealth} &  Risk     &     &\multicolumn{3}{c}{Limiting Price}  \\
\cmidrule(r){1-2}  \cmidrule(r){3-3}  \cmidrule(r){5-7}  
 $w^1$ &  $w^2$          &  $\gamma$ &       & $p_1^*$&  $p_2^*$  &  $p_3^*$ \\
\toprule
\multirow{2}*{10}&\multirow{2}*{10} & \multirow{2}*{0.5~(0.7)} 
& $\mathcal{S}_1:$  & 0.4044~{(0.3959)}   & 0.1839~(0.1849)   & 0.4117~(0.4192)   \\
                 &                  &
& $\mathcal{S}_2:$  & 0.4045~(0.4004)  & 0.1838~(0.1841)    & 0.4117~(0.4155)    \\
\midrule
\multirow{2}*{8} &\multirow{2}*{12} & \multirow{2}*{0.5~(0.7)} 
& $\mathcal{S}_1:$  & 0.4436~(0.4449)  & 0.1745~(0.1726)    & 0.3819~(0.3825)  \\
                 &                  &                    
& $\mathcal{S}_2:$  & 0.4440~(0.4800)  & 0.1743~(0.1741)    & 0.3817~(0.3459)   \\
\midrule
\multirow{2}*{12} &\multirow{2}*{8} & \multirow{2}*{0.5~(0.7)} 
& $\mathcal{S}_1:$  & 0.3624~(0.3463)   & 0.1933~(0.1959)    & 0.4443~(0.4578)   \\
                 &                  &                    
& $\mathcal{S}_2:$  & 0.3621~(0.3776)   & 0.1933~(0.2044)    & 0.4446~(0.4180)   \\
\bottomrule
\end{tabular}
\end{table}

\end{example}

\tb{As demonstrated in Example \ref{exam_crra}, the trading sequence can impact the limiting price. A key observation for understanding this phenomenon is that $\mathcal{S}_1$ can be equivalently viewed as $\mathcal{S}_2$ preceded by an extra Trader 1 action. Such an action results in an initial market state different from that of the original $\mathcal{S}_2$. As Proposition \ref{prop:crra_trader_solution} suggests, different initial wealth results in a different limiting price; thus the limiting prices for the two sequences $\mathcal{S}_1$ and $\mathcal{S}_2$ are different. To illustrate, we consider Example \ref{exam_crra} with $w^1=w^2=10$. The extra action of $\mathcal{S}_1$ (the first trade) yields market states $\x_1 = (7.49,7.49.13.50)$, $\x_2=(10,10,10)$, and $\y=(7.51,7.51,1.50)$. After the first trade, sequence $\mathcal{S}_1$ becomes identical to $\mathcal{S}_2$. However, for $\mathcal{S}_2$, the initial condition is $\x_1 = (10,10,10)$, $\x_2=(10,10,10)$, and $\y=(5,5,5)$, which is different from the states yield from $\mathcal{S}_1$ after the first trading, thus the two sequences result in different limiting prices.
}

\tb{
Practically, traders interact with the market maker randomly, leading to a random trading path. Consequently, the limiting price becomes a trading-sequence-dependent random variable. However, our next result characterizes the limiting price for a given trading sequence and demonstrates that price invariance can be achieved when the trader population is large.
\begin{theorem}\label{thm_crra_limit}
For the CRRA-utility-based market, we have the following results:\\ 
\noindent(i) Given a trading sequence $\mathcal{S}\in \Phi$, the Trading Process \ref{algtrading} converges to the limiting wealth allocation $\{\x_j^*\}$ characterized by problem $\P_{\po}(\pmb{\nu}^*)$ with some weights $\pmb{\nu}^* = (\nu_1^*,\ldots, \nu_J^*)\in \mathbb{R}^J_+\setminus \{\0\}$; and the limiting price is given by
\begin{align}
p_i^*  = \xi^*_i \Big/ \sum_{k=1}^I \xi_k^*, ~~~i=1,\ldots,I,\label{crra_lim_price}
\end{align}
where $\pmb{\xi}^*=(\xi_1^*,\ldots, \xi_I^*)$ is the unique solution of the following equations:
\begin{align}
\xi_i^* &=  \Bigg( \frac{ \sum \limits^J_{j=1} (\nu_j^*)^{\frac{1}{1-\gamma}} \cdot \pi_{i,j}^{\frac{1}{1-\gamma}} 
+  (\xi_i^*)^{\frac{\gamma-\gamma_m}{(1-\gamma)(1-\gamma_m)} }\cdot\mathcal{L}(\pmb{\xi}^*,\pmb{\theta})\cdot W_0 \cdot \theta_i^{\frac{1}{1-\gamma_m} } }{w_0^{\all}} \Bigg)^{1-\gamma},~~ i=1,\ldots, I \label{crra_xi}
\end{align} 
where $\mathcal{L}(\pmb{\xi},\pmb{\theta}) \triangleq \left(\sum \limits_{k=1}^I \theta_k(\theta_k/\xi_k)^{\frac{\gamma_m}{1-\gamma_m}}\right)^{-\frac{1}{\gamma_m}}$.\\
\noindent(ii) In Trading Process \ref{algtrading}, if traders' beliefs $\pmb{\pi}_j$ and initial wealth $w_{j,0}$ are independently and identically distributed random variables for $j=1,\ldots, J$; with each trader having an equal probability of trading with the market maker at each trading instance, then the solution $\{\xi_i^*\}|_{i=1}^I$ defined in Eq. (\ref{crra_xi}) and the associated limiting price $\{p_i^*\}|_{i=1}^I$ generated by each trading sample path converge almost surely to a pair of constant vectors as the trader population $J \rightarrow \infty$.
\end{theorem}
}
\tb{
The first result in Theorem \ref{thm_crra_limit} provides an explicit method for computing the limiting price $\p^*$ for a sepecified trading sequence with the associated weighting parameters $\pmb{\nu}^*$. It is worth noting that the functional form of the limiting price (\ref{crra_lim_price}) and (\ref{crra_xi}) remain consistent, as the influence of the trading sequence is solely reflected through the weights $\pmb{\nu}^*$. The second result in Theorem \ref{thm_crra_limit} is an interesting product of the price formula. One can consider it a prediction market analogue of the \textit{``Law of Large Numbers''}. It implies that in a randomly trading market with a large trader population, the variation of limiting price induced by different trading sequences becomes negligible. This underscores the significance of trader participation in prediction markets, as it not only enhances liquidity but also increases the reliability of limiting prices. Imagine a prediction market that yields predictions (limiting price) that are heavily dependent on the trading sequence. In such a case, one cannot know whether its prediction truly represents the overall beliefs or only happens to be a biased result of a specific trading sequence. Such uncertainty will undermine the interpretability of prices and jeopardize the prediction accuracy.
}

\tb{
Theorem \ref{thm_crra_limit} also provides several potentially useful insights for market maker designing. Combining Eqs. (\ref{crra_lim_price}) and (\ref{crra_xi}), the limiting price has the following decomposition:
\begin{align}
p_i^* = \Bigg(\sum_{j=1}^J \frac{\left( \frac{\nu_j}{\sum_{k=1}^I\xi_k^*} \right)^{\frac{1}{(1-\gamma)}}}{w_0^{\all}}\left(\pi_{i,j}\right)^{\frac{1}{1-\gamma}}+ (p_i^*)^{\frac{\gamma-\gamma_m}{ (1-\gamma)(1-\gamma_m)}} \cdot\mathcal{L}(\bm{p}^*,\boldsymbol{\theta})\frac{W_0}{w_0^{\all} } \theta_i^{\frac{1}{1-\gamma_m}}\Bigg)^{1-\gamma},\quad i=1,\ldots,I.\label{crra_p_decompose}
\end{align}
This decomposition can be explained as follows. First, a smaller initial wealth $W_0$ can reduce the market maker's impact on the limiting price. Second, a market maker with higher risk aversion (smaller $\gamma_m$) also diminishes its influence on the limiting price because $(p_i^*)^{\frac{\gamma-\gamma_m}{(1-\gamma)(1-\gamma_m)}}$ 
is a decreasing function of $\gamma_m$ for $p_i^*\in(0,1)$. Third, the function $\mathcal{L}(\p^*,\pmb{\theta})$ exhibits an intriguing suppressive effect designed to mitigate the price distortion caused by the market maker' externality. To explain further, $\mathcal{L}(\p^*,\pmb{\theta})$ reaches its maximum value of $1$ when $\p^*=\pmb{\theta}$. When the market consensus $\p^*$ significantly deviates from the market maker's belief $\pmb{\theta}$, the term $\mathcal{L}(\p^*,\pmb{\theta})$ lessens its impact, effectively reducing the market maker's influence on the limiting price. It is worth noting that such a suppressive effect stems from the utility preserving condition (\ref{PO:util_preverve}), underscoring the additional benefits of the MU-based mechanism.
}

\tb{With the above results, in the next section, we investigate how to approximate the limiting price under a finite trader scenario through choosing an appropriate weight $\pmb{\nu}^*$.}



\subsection{Heuristic Weighting Parameters and Price Formula}\label{sse:price_crra}
\tb{
The previous section has emphasized the vital role of determining the weighting parameters, $\pmb{\nu}^*$, in solving problem $\P_{\po}(\pmb{\nu}^*)$ to establish the limiting price. Given the practical difficulty of obtaining the ``true'' weighting parameters $\pmb{\nu}^*$ before trading converges, we propose an approximation using deterministic weights. Drawing inspiration from previous analysis, which highlights Pareto optimality of the limiting wealth allocation and underscores the significance of both the initial wealth distribution $w_{j,0}$ and the risk parameter $\gamma$ in shaping the limiting price (e.g., see Table \ref{table_exam_crra}), we introduce the concept of \textit{Pareto Optimal Induced} (POI) weights, denoted as $\hat{\pmb{\nu}} = (\hat{\nu}_1,\ldots, \hat{\nu}_J)$ defined as:
\begin{align}
\hat{\nu}_j = \big(w_{j,0}\big)^{ 1-\gamma },~~j=1,\ldots,J. \label{def_crra_weight}
\end{align}
Applying Theorem \ref{thm_crra_limit}, we replace $\pmb{\nu}^*$ with the POI weights $\hat{\pmb{\nu}}$ in Eq. (\ref{crra_xi}) to calculate the associated POI limiting price, denoted as $\hat{\p}=(\hat{p}_1,\ldots, \hat{p}_I)$, with its components given as follows:
\begin{align}
\hat{p}_i = \hat{\xi}_i \Big/ \sum_{k=1}^I \hat{\xi}_k, ~~~i=1,\ldots,I, \label{crra_approx_price}\
\end{align}
where $\hat{\pmb{\xi}}=(\hat{\xi}_1,\ldots, \hat{\xi}_I)$ is the unique solution of the following equation, 
\begin{align}
\hat{\xi}_i =  \Bigg(\frac{ \sum \limits^J_{j=1} w_{j,0} \cdot \pi_{i,j}^{\frac{1}{1-\gamma}} 
+ \hat{\xi}_i^{\frac{\gamma-\gamma_m}{(1-\gamma)(1-\gamma_m)} }\cdot\mathcal{L}(\hat{\pmb{\xi}}, \pmb{\theta} )\cdot W_0 \cdot \theta_i^{\frac{1}{1-\gamma_m}}}{  w_0^{\all} } \Bigg)^{1-\gamma}, ~~i=1,\ldots,I.\label{crra_approx_xi}
\end{align} 
Now we provide some interpretation for the POI limiting price. Eq (\ref{crra_approx_xi}) implies that the price $\hat{p}_i$ is directly linked to a weighted average that considers all participants' wealth and risk-adjusted beliefs regarding the $i$-th asset. More specifically, the term $\sum \limits_{j=1}^J w_{j,0} \cdot \pi_{i,j}^{\frac{1}{1-\gamma}}$ represents the avarage risk-adjusted belief contributed from all the traders weighted by their own wealth. In contrast, the term $\hat{\xi}_i^{\frac{\gamma-\gamma_m}{(1-\gamma)(1-\gamma_m)} }\cdot\mathcal{L}(\hat{\pmb{\xi}}, \pmb{\theta} )\cdot W_0 \cdot \theta_i^{\frac{1}{1-\gamma_m}} $ which combines the risk difference and belief divergence, controls the contribution of market maker's risk-adjusted belief in such a price.} 


\tb{
The POI price formula (\ref{crra_approx_price}) and (\ref{crra_approx_xi}) encompass specific market scenarios as special cases. When both market maker and traders adopt the logarithmic utilities, i.e., $\gamma=\gamma_m=0$ in (\ref{def:CRRA}), Eq. (\ref{crra_approx_xi}) will simplify to:
\begin{align}
\hat{\xi}_i= \frac{ \sum \limits^J_{j=1} w_{j,0} \cdot \pi_{i,j} 
+ \mathcal{L}_0(\hat{\pmb{\xi}},\pmb{\theta}) \cdot W_0 \cdot \theta_i  }{ w_0^{\all} },~~~i=1,\ldots,I, \label{crra_xi_log}
\end{align}
where $\mathcal{L}_0(\hat{\pmb{\xi}}, \pmb{\theta} )= \exp(-\sum \limits_{k=1}^I \theta_k\log(\theta_k/\hat{\xi}_k))$ representing the exponential \textit{Kullback-Leibler Divergence} (KLD) between $\pmb{\theta}$ and $\hat{\pmb{\xi}}$. If we further set $\mathcal{L}_0(\hat{\pmb{\xi}}, \pmb{\theta})\equiv 1$ then the price formula (\ref{crra_approx_price}) becomes the wealth-weighted average of market beilefs discussed in \cite{sethi2016belief}.}

\tb{We would like to highlight that the solution methodology for deriving the CRRA-utility based limiting price, e.g., Eq. (\ref{crra_xi}) and the POI weights (\ref{def_crra_weight}) remain applicable even when market participant's MU functions extend beyond the CRRA family of utilities. One example is the hybrid market model where the market maker adopts exponential utility, as given by Eq. (\ref{eq:exp_U_Vj}), while the traders adopt logarithmic utility. In this scenario, the POI limiting price remains the same as (\ref{crra_approx_price}), but the equations for $\hat{\pmb{\xi}}$ become:
\begin{align}
\h{\xi}_i = \frac{\sum \limits_{j=1}^J w_{j,0} \cdot \pi_{i,j} + \frac{1}{\beta} \h{\xi}_i \log \left( \theta_i\Big/\frac{\h{\xi}_i}{\sum_{k=1}^I \h{\xi}_k} \right)}{\sum \limits_{j=1}^J w_{j,0} },\quad i=1,2,...,I.\label{log_exp_xi}
\end{align}
We provide more details of Eq. (\ref{log_exp_xi}) in Appendix \ref{apdx:sse:log-exp-utility}. Our numerical experiments (given in Appendix \ref{apdx:sse:log-exp-utility}) demonstrate that the price formula (\ref{log_exp_xi}) outperforms the conventional wealth-weighted average price reported in \cite{sethi2016belief}. The CRRA-utility-based pricing formula can also be extended to a more comprehensive market model where all participants adopt utilities with \textit{hyperbolic absolute risk aversion} (HARA utility), which finds wide applications in economic modeling. A brief discussion of this model is presented in Appendix \ref{apdx:HARA}.
}

\subsection{Evaluation of Approximation Scheme}\label{ssec:numerical_evaluation}
\tb{In this section, we first employ simulation to validate the accuracy of the POI weights (\ref{def_crra_weight}) and the price formula (\ref{crra_approx_price}) in estimating the true weighting coefficients and the true limiting price, respectively. Following this validation, we proceed to examine how the risk parameters and initial wealth impact the accuracy of our approximation.
}

\tb{
We consider a CRRA-utility-based market with $I=3$, $W_0=10$, and $\theta_1=\theta_2=\theta_3=1/3$. Traders are generated randomly as follows. The $j$-th trader's belief is sampled from $\pmb{\pi}_j = \alpha \cdot \bar{\pmb{\pi}}+ (1-\alpha) \cdot \pmb{\epsilon}/|\pmb{\epsilon}|$, where $\bar{\pmb{\pi}}$ is the baseline belief evenly sampled from $(0.6, 0.2, 0.2)$ and $(0.2, 0.2, 0.6)$; $\pmb{\epsilon}$ is uniformly distributed random noise on $[0,1]^3$, and $\alpha=0.5$ controls the market belief structure. We conduct simulations according to Trading Process \ref{algtrading}, with each trader randomly selected with equal probability to interact with the market maker at each time period. For each set of parameters, we perform $N = 5\times J$ rounds of simulation, and we denote the $k$-th sample of the limiting wealth allocation as $\{\x_1^{\sim}(k), \ldots, \x_J^{\sim}(k)\}|_{k=1}^N$. 
\begin{itemize}
\item \textit{Evaluation by limiting wealth:} To verify the accuracy of the POI weights (\ref{def_crra_weight}), we first solve the problem $\P_{\po}(\hat{\pmb{\nu}})$ whose solution is denoted as $\{\hat{\x}_1,\ldots,\hat{\x}_J \}$. If the heuristic weight $\hat{\pmb{\nu}}$ provides an accurate approximation for the ``true'' weighting parameter, then the wealth allocation $\{ \hat{\x}_1, \ldots, \hat{\x}_J \}$ should be close to the sample limiting wealth allocation $\{\x_1^{\sim}(k),\ldots,\x_J^{\sim}(k)\}|_{k=1}^N$ generated by the simulation. We introduce the following quantity to measure the difference for each sample,
\begin{align*}
\delta_{\x}(k) \triangleq  \sum \limits_{j=1}^J \|\x_j^{\sim}(k)
-\hat{\x}_j\|\big/\sum\limits_{j=1}^J \|\x_j^{\sim}(k)\|,
\end{align*}
for $k=1,\ldots,N$. The sample mean and variance of $\delta_{\x}(k)$ is denoted by $\textrm{E}[\delta_{\x}]$ and $\textrm{Var}[\delta_{\x}]$, respectively. 
\item \textit{Evaluation by limiting price:} We denote the price associated with the $k$-th sample of limting wealth as $\p^{\sim}(k)$ and we compute the POI price $\hat{\p}$ by using (\ref{crra_approx_price}). To measure the difference between these prices, we employ the Kullback-Leibler Divergence (KLD), i.e., given two prices $\p^{a},\p^{b} \in \Delta_I$, the KLD of these prices is defined as $\D(\p^{a},\p^{b}) \triangleq \sum_{i=1}^I p^{a}_i \ln \left(p^{a}_i/p^{b}_i \right)$. We then denote the KLD between the $k$-th sample price $\p^{\sim}(k)$ and the POI price $\hat{\p}$ as $\delta_{\p}(k) \triangleq \D(\p^{\sim}(k), \hat{\p} )$ for $k=1,\ldots,N$. We then compute the sample mean and sample variance of $\delta_{\p}(k)$ as $\text{E}[\delta_{\p}]$ and $\textrm{Var}[\delta_{\p}]$, respectively.
\end{itemize}
}
\tb{
Table \ref{table_verify_approx} reports the comparative results for different setups ($\gamma_m,\gamma \in\{-2, 0.5\}$ and  $J\in\{50,100,200, 400\}$). For each set of risk parameters, denoted as ($\gamma_m$, $\gamma$), we list the results for different values of $J$. The results clearly show that, regardless of the different settings, the POI weights (\ref{def_crra_weight}) and the POI price (\ref{crra_approx_price}) perform exceptionally well, exhibiting minimal disparities in both wealth and price. A closer examination reveals several noteworthy patterns. Firstly, for a fixed set of risk parameters $\gamma_m$ and $\gamma$, increasing the trader population $J$ enhances the accuracy of our approximation. This enhancement is evident as all error and variation indicators decrease with the increase in $J$. Of particular significance is the observation that Table \ref{table_verify_approx} also highlights a notable decrease in approximation variances, $\textrm{Var}[\delta_{\x}]$ and $\textrm{Var}[\delta_{\p}]$, with a larger population. This reduction in variance can be attributed to the decrease in variation stemming from the trading sequence. Consequently, this pattern confirms the second result outlined in Theorem \ref{thm_crra_limit}, which posits that a larger population leads to a more stable limit price. Secondly, for a given trader population, the approximation error is smaller in scenarios with $\gamma_m=-2$ compared to those with $\gamma_m=0.5$, suggesting that our approximation is more favorable for higher levels of risk aversion. 
}

\begin{table}
\small
\caption{Verification of approximation scheme}\label{table_ccsm}
\centering
\sisetup{table-number-alignment=center, exponent-product=\times}
\begin{tabular}{ccc l l l l }
\toprule
\multicolumn{3}{c}{Market Par.} & \multicolumn{2}{c}{Price Diff.} & \multicolumn{2}{c}{Wealth Diff.}\\
\cmidrule{1-3}   \cmidrule(lr){4-5} \cmidrule(lr){6-7}
$\gamma_m$ & $\gamma$ & $J$ & $\textrm{E}[\delta_{\p}]$ & $\textrm{Var}[\delta_{\p}]$ & 
       $\textrm{E}[\delta_{\x}]$ & $\textrm{Var}[\delta_{\x}]$ \\
\cmidrule[1pt]{1-3}   \cmidrule[1pt](lr){4-5} \cmidrule[1pt](lr){6-7}
-2        & -2        & 50        & \num{1.71E-05} & \num{5.75E-10} & \num{5.75E-04} & \num{9.30E-08} \\
-2        & -2         & 100       & \num{8.84E-06} & \num{2.61E-10} & \num{2.48E-04} & \num{1.76E-08} \\
-2        & -2         & 200       & \num{2.54E-06} & \num{1.35E-11} & \num{1.84E-04} & \num{1.31E-08} \\
-2        & -2       & 400       & \num{1.51E-06} & \num{4.52E-12} & \num{1.26E-04} & \num{6.06E-09} \\
\cmidrule[1pt]{1-3}   \cmidrule[1pt](lr){4-5} \cmidrule[1pt](lr){6-7}
-2        & 0.5       & 50        & \num{1.04E-04} & \num{1.35E-08} & \num{9.69E-04} & \num{2.31E-07} \\
-2        & 0.5       & 100       & \num{4.33E-05} & \num{1.75E-09} & \num{4.29E-04} & \num{4.10E-08} \\
-2        & 0.5       & 200       & \num{3.20E-05} & \num{1.17E-09} & \num{2.50E-04} & \num{1.79E-08} \\
-2        & 0.5       & 400       & \num{1.91E-05} & \num{5.48E-10} & \num{1.35E-04} & \num{6.37E-09} \\
\cmidrule[1pt]{1-3}   \cmidrule[1pt](lr){4-5} \cmidrule[1pt](lr){6-7}
0.5       & -2        & 50        & \num{5.08E-05} & \num{3.60E-08} & \num{2.77E-03} & \num{6.64E-06} \\
0.5       & -2        & 100       & \num{1.84e-05} & \num{1.78e-08} & \num{9.70E-04} & \num{1.42E-06} \\
0.5       & -2        & 200       & \num{5.34e-06} & \num{2.00E-10} & \num{5.09E-04} & \num{2.45E-07} \\
0.5       & -2        & 400       & \num{2.01E-06} & \num{7.94E-11} & \num{3.66E-04} & \num{8.11E-08} \\
\cmidrule[1pt]{1-3}   \cmidrule[1pt](lr){4-5} \cmidrule[1pt](lr){6-7}
0.5       & 0.5       & 50        & \num{3.83E-05} & \num{1.20E-09} & \num{2.89E-03} & \num{2.03E-06} \\
0.5       & 0.5       & 100       & \num{1.95E-05} & \num{2.23E-10} & \num{1.34E-03} & \num{3.24E-07} \\
0.5       & 0.5       & 200       & \num{8.61E-06} & \num{3.38E-11} & \num{6.70E-04} & \num{6.12E-08} \\
0.5       & 0.5       & 400       & \num{2.15E-06} & \num{3.47E-12} & \num{3.18E-04} & \num{1.77E-08} \\
\bottomrule
\end{tabular}\label{table_verify_approx}
\end{table}

\tb{
We proceed to examine how the risk parameters and initial wealth affect the quality of our approximation. To better illustrate the impact of these parameters, we introduce the conventional wealth-weighted price formula from \cite{sethi2016belief} as a benchmark, denoted as $\breve{\mathbf{p}} = (\breve{p}_1,\ldots, \breve{p}_I)$, where
\begin{align}
\breve{p}_i = \frac{ \theta_i\cdot W_0+ \sum \limits_{j=1}^J w_{j,0} \pi_{i,j} }{W_0 + \sum\limits_{j=1}^J w_{j,0}},~~i=1,\ldots, I.\label{price_wealth_weight}
\end{align}
We follow a simulation procedure similar to that in Table \ref{table_verify_approx}, using a fixed $J=50$ and $\gamma_m=-0.5$. To highlight the impact of traders' risk parameter $\gamma$, we vary its values and compare the resulting limiting prices obtained from the simulation with two heuristic price formulas: (\ref{crra_approx_price}) and (\ref{price_wealth_weight}).
}
\begin{figure}[h!]
\centering
\caption{Comparison of two heuristics for different risk parameters}
\begin{subfigure}{0.49\textwidth}
    \includegraphics[width = 0.89\linewidth]{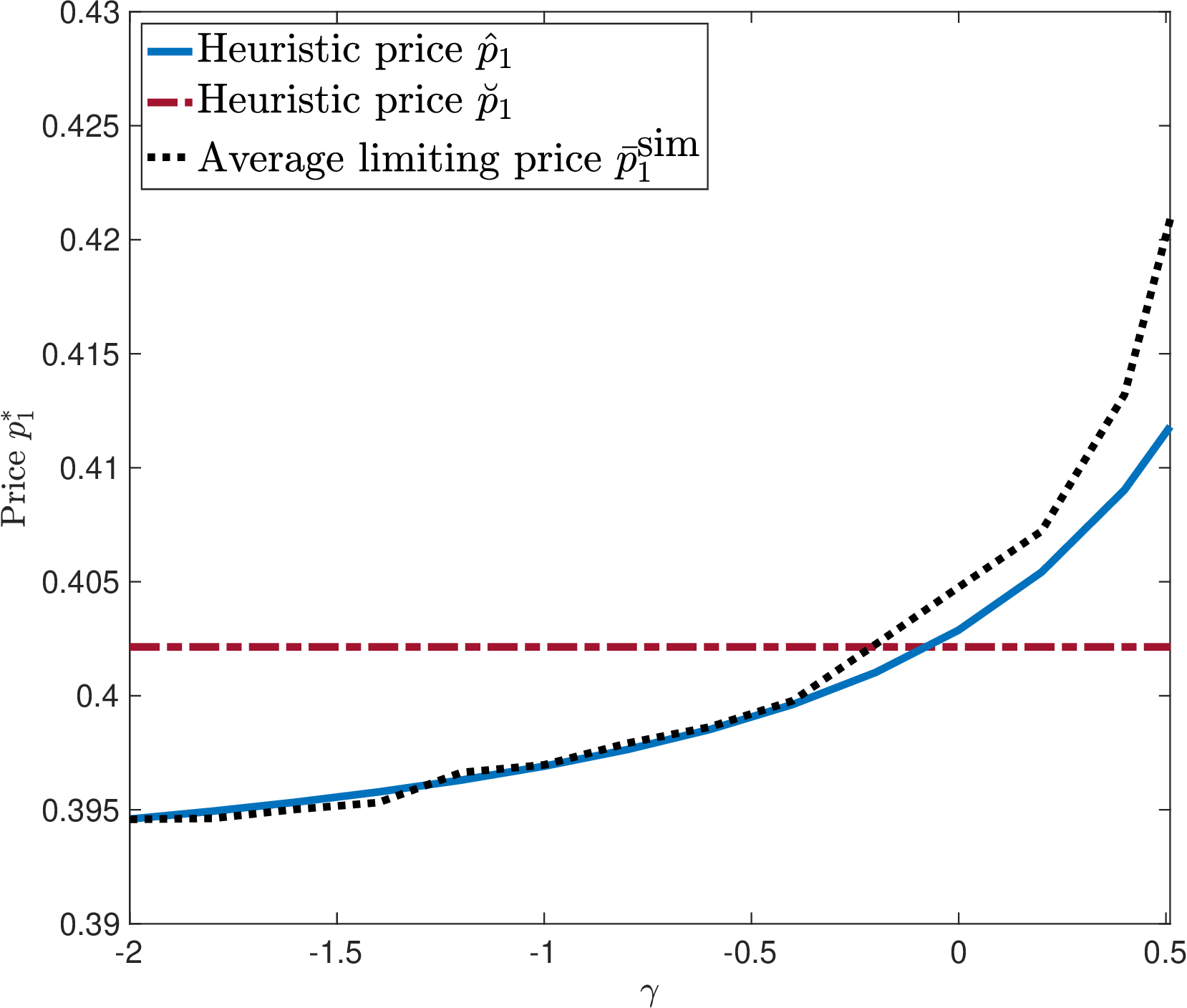}
    \caption{\small Price $p_1^*$ and $\gamma$ }
    \label{fig:price_risk}
\end{subfigure}
\begin{subfigure}{0.49\textwidth}
    \includegraphics[width = 0.92\linewidth]{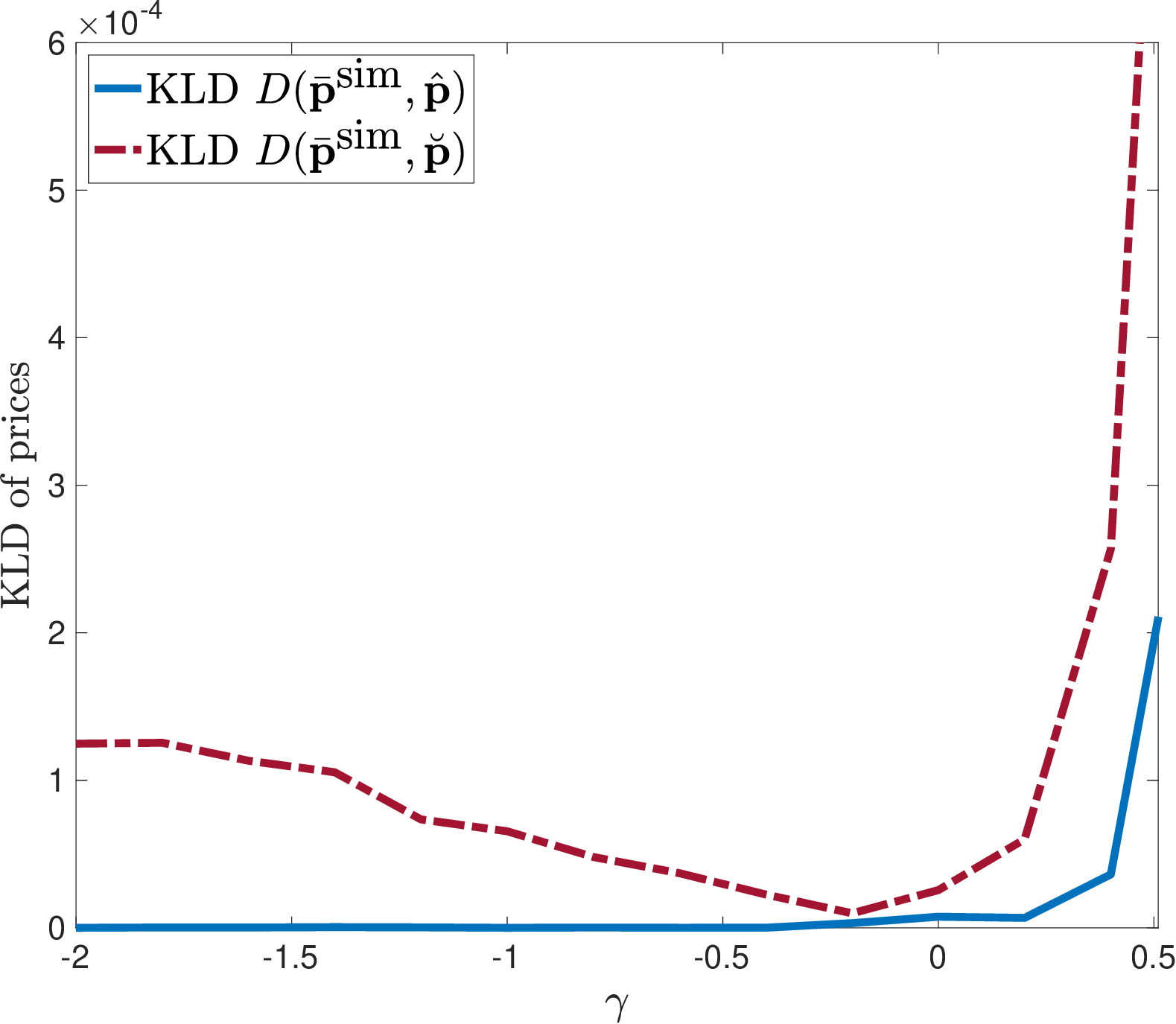}
    \caption{\small KLD and $\gamma$ }
    \label{fig:kld_crra}
\end{subfigure}
\label{fig_tr_risk_price}
\end{figure}

\tb{
In Figure \ref{fig:price_risk}, the red curve represents the average price of the first security resulting from $N=250$ rounds of simulations, which exhibits an increasing trend as $\gamma$ increases. Importantly, our POI price formula (\ref{crra_approx_price}) effectively captures this trend, as indicated by the solid blue curve. In contrast, the conventional wealth-weighted formula (\ref{price_wealth_weight}) fails to exhibit such a distinct pattern as it does not account for the risk parameter. Figure \ref{fig:kld_crra} illustrates the approximation errors of the two heuristic prices where $\D(\bar{\p}^{\sim},\hat{\p})$ and $\D(\bar{\p}^{\sim},\breve{\p})$ measure the KLD between the simulated average price and the two heuristic prices, respectively. Notably, when $\gamma$ deviates from $0$, especially when it is below zero, our price formula (\ref{crra_approx_price}) significantly outperforms the conventional wealth-weighted formula (\ref{price_wealth_weight}).
}

\begin{figure}[h!]
\centering
\caption{Comparison of two heuristics for different initial wealth}
\begin{subfigure}{0.49\textwidth}
    \includegraphics[width = 0.9\linewidth]{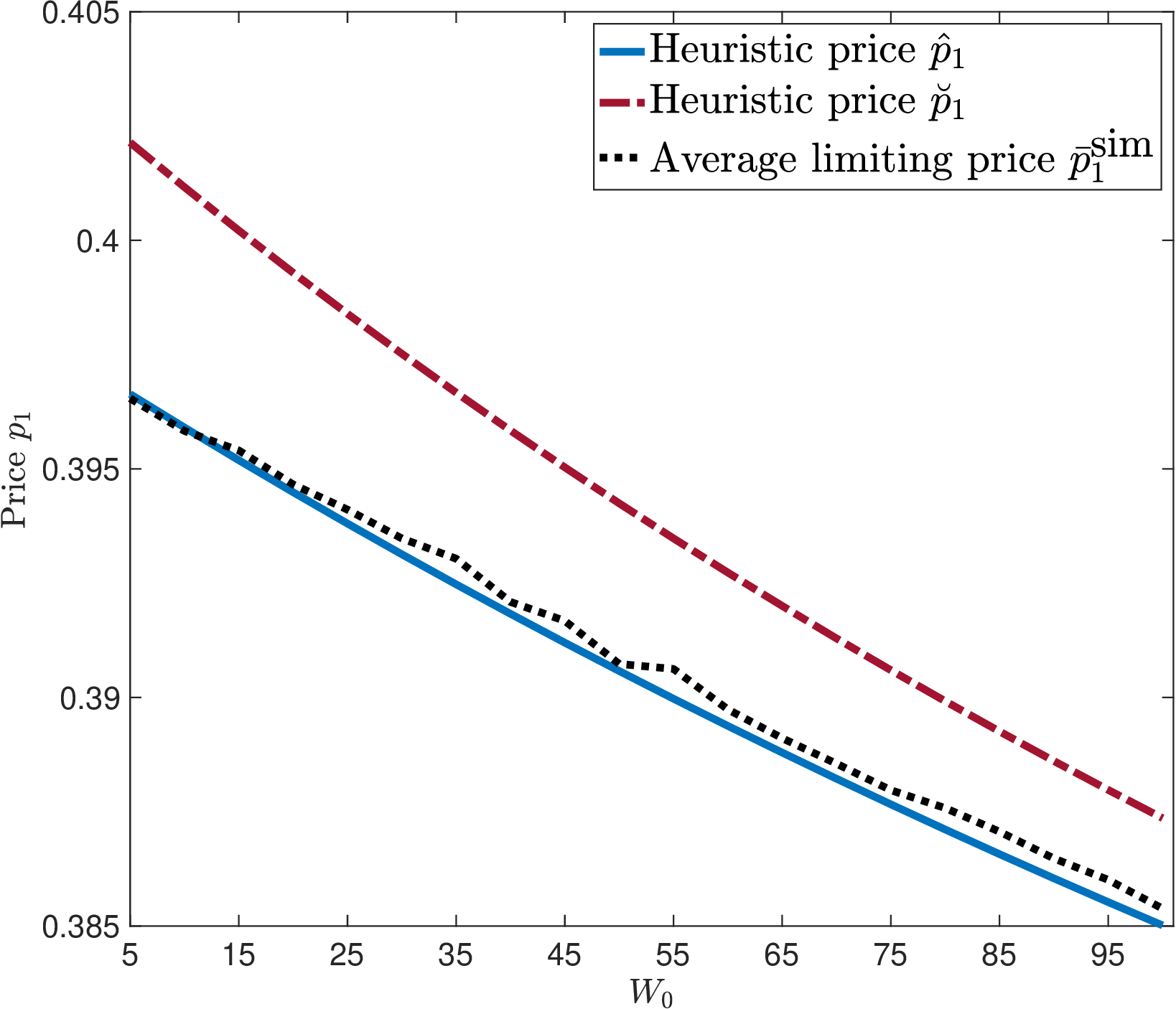}
    \caption{\small Price and market maker's wealth}
    \label{fig:price_wealth}
\end{subfigure}
\begin{subfigure}{0.49\textwidth}
    \includegraphics[width = 0.94\linewidth]{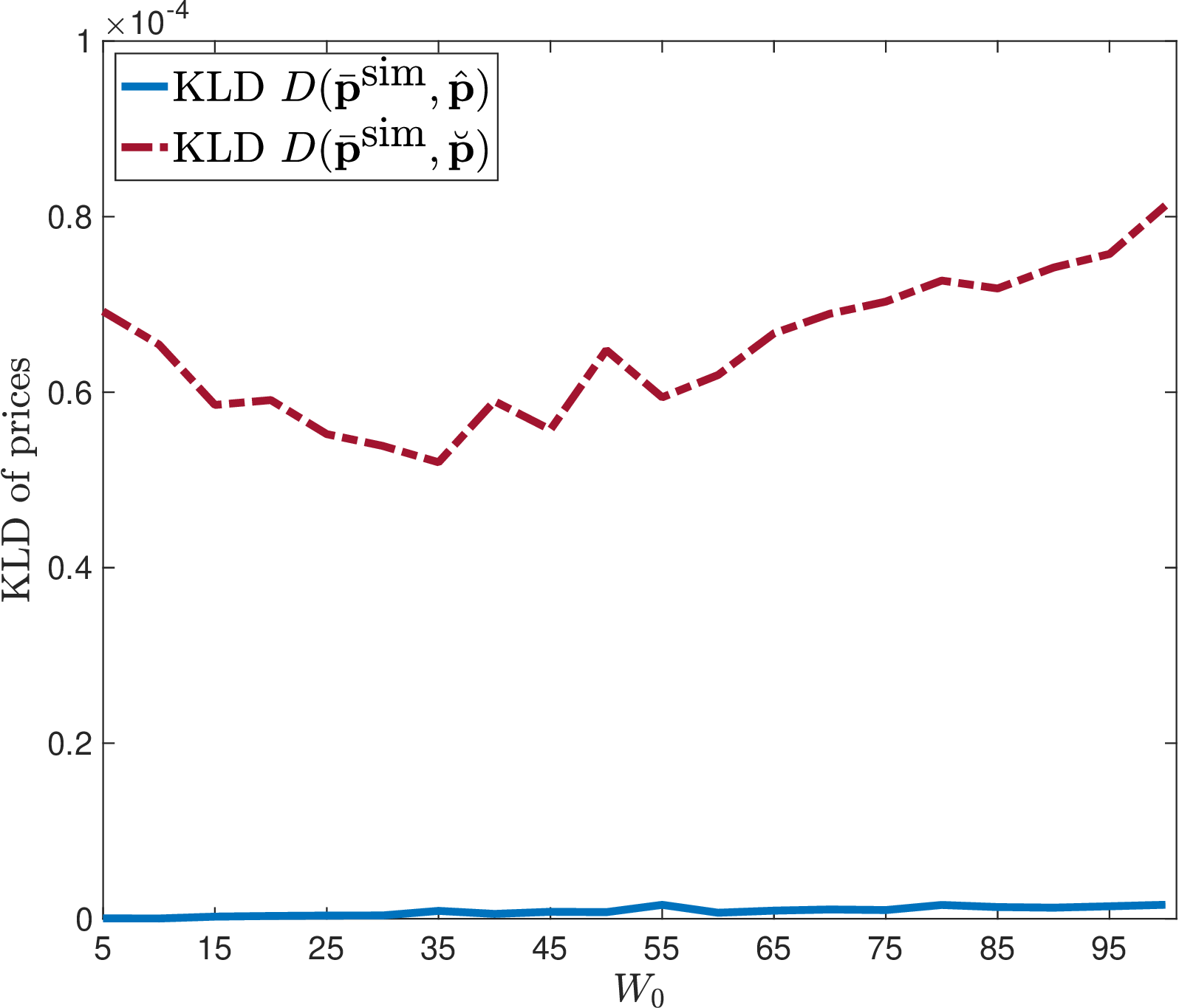}
    \caption{\small KLD  and market maker's wealth}
    \label{fig:price_wealth_kld}
\end{subfigure}
\end{figure}
\tb{
We then investigate the impact of the market maker's initial wealth $W_0$, which is a key design variable for providing liquidity. Holding $\gamma_m=\gamma=-2$, we incrementally raise $W_0$ from $5$ to $100$. In Figure \ref{fig:price_wealth}, the limiting price decreases as $W_0$ increases. This is because the price becomes increasingly biased towards the market maker's own belief (i.e., $\theta_i=1/3$) as its wealth grows. Both formulas (\ref{crra_approx_price}) and (\ref{price_wealth_weight}) display a similar trend. However, our price formula (\ref{crra_approx_price}) closely tracks the limiting price, while the conventional one (\ref{price_wealth_weight}) tends to overestimate it significantly. Figure \ref{fig:price_wealth_kld} provides a more detailed comparison of these approximation errors by evaluating the KLD. It becomes evident that our POI price formula (\ref{crra_approx_price}) not only demonstrates greater accuracy but also exhibits much higher stability than formula (\ref{price_wealth_weight}) when varying the market maker's wealth $W_0$. This characteristic highlights that the quality of our heuristic price formula (\ref{crra_approx_price}) is not sensitive to fluctuations in the market maker's wealth level. Such a property may benefit the market maker in customizing the market-making mechanism. 
}

\tb{
In summary, the combination of the POI weights (\ref{def_crra_weight}) with Eq. (\ref{crra_xi}) offers an accurate estimation of the limiting price across various market scenarios. Although such a pricing formula is not entirely analytical, the parameters $\xi_i^*$, given in Eqs. (\ref{crra_xi}) provide comprehensive information on how the price is aggregated based on participants' beliefs, risk attitudes, and wealth levels. }

\section{Price-Volume Relationship and Nonstationary Market}\label{sec:price-volume-nonstationary}

In this section, we provide analysis of the price-volume relationship resulted from our model and discuss the market behavior in a non-stationary market.

\tb{
\noindent\textbf{The price-volume relationship:}~Like microstructure analysis in the financial market, studying the relationship between the price and trading volume reveals how prices evolve during trading. To examine this, we consider the $t$-th round of trading where the $j$-th trader interacts with the market maker in Trading Process \ref{algtrading}. Although it might seem like only one trader is trading at a time, we can think of this as a group of traders. For instance, the $j$-th trader represents a group of traders holding the similar beliefs with the wealth being the aggregate wealth. Therefore, the order size in each round of trading represents an aggregate order for a group of traders. For simplicity, we consider a market with binary securities ($I=2$). The $j$-th trader solves problem $\P_j(\x_{j,t})$ to determine the optimal decision $\z^*=(z_1^*, z_2^*)$. Since the two securities are mutually exclusive, the trading volume is only related to the net demand $\textrm{D}_{t}\triangleq \Delta q_1^* -\Delta q_2^*=z^*_1- z^*_2$ where $q_i^*$ is the amount of the order.\footnote{The last equality is from the definition of $z^*$.} The mutual exclusiveness also implies the prices $\p_t=(p_{1,t}, p_{2,t})$ satisfies $p_{1,t}=1-p_{2,t}$. We then examine the relationship between the net demand $\text{D}_t$ and the security's price $p_{1,t}$ for different market settings.}

\tb{As the exponential MU-function-based market admits a closed-form solution for trading problem $\P_j(\x_j)$, it is an ideal vehicle to study the price-volume relationship. Using the expression (\ref{prop_exp_z}) in Proposition \ref{prop_exp_unique}, we can compute the net demand as,
\begin{align}
\textrm{D}_{t} =z^*_1-z^*_2&= \frac{1}{\alpha_j +\beta} 
\ln \left( \frac{ \hat{\pi}_{1,j,t}(1- p_{1,t})}{(1 -\hat{\pi}_{1,j,t}) p_{1,t} } \right) 
=\frac{1}{\alpha_j +\beta} \left( \ln \left( \frac{\hat{\pi}_{1,j,t}}{1 - \hat{\pi}_{1,j,t}}\right) 
-\ln \left( \frac{p_{1,t}}{1-p_{1,t}} \right) \right),\label{eq:exp_vol}
\end{align}
where $\hat{\pi}_{i,j,t}$ is defined in Proposition \ref{prop_exp_unique}. We observe that the volume $\textrm{D}_t$ is proportional to the difference between the log odds of the $j$-th trader's belief and the security price. The demand $\text{D}_t$ decreases as the price $p_{1,t}$ increases, indicating a higher demand for the security at a lower price. 
}
\tb{
Conversely, for a fixed price $p_{1,t}$, the demand $\text{D}_t$ depends on the trader's utility-adjusted belief $\hat{\pi}_{1,j,t}$. A trader trades more security if $\hat{\pi}_{1,j,t}$ deviates further from the price. It is also evident that the net trade volume is discounted by a factor $\frac{1}{\alpha_j+\beta}$. That is to say, even if two traders have identical subjective beliefs, the trader who is more risk-averse trades less security than the one with less risk aversion. This finding further demonstrates the importance of including the risk parameter in our approximated pricing formula (\ref{crra_approx_xi}).
}

\begin{figure}
\centering
\caption{Demand as function of price for CRRA utility}
\includegraphics[width = 220pt]{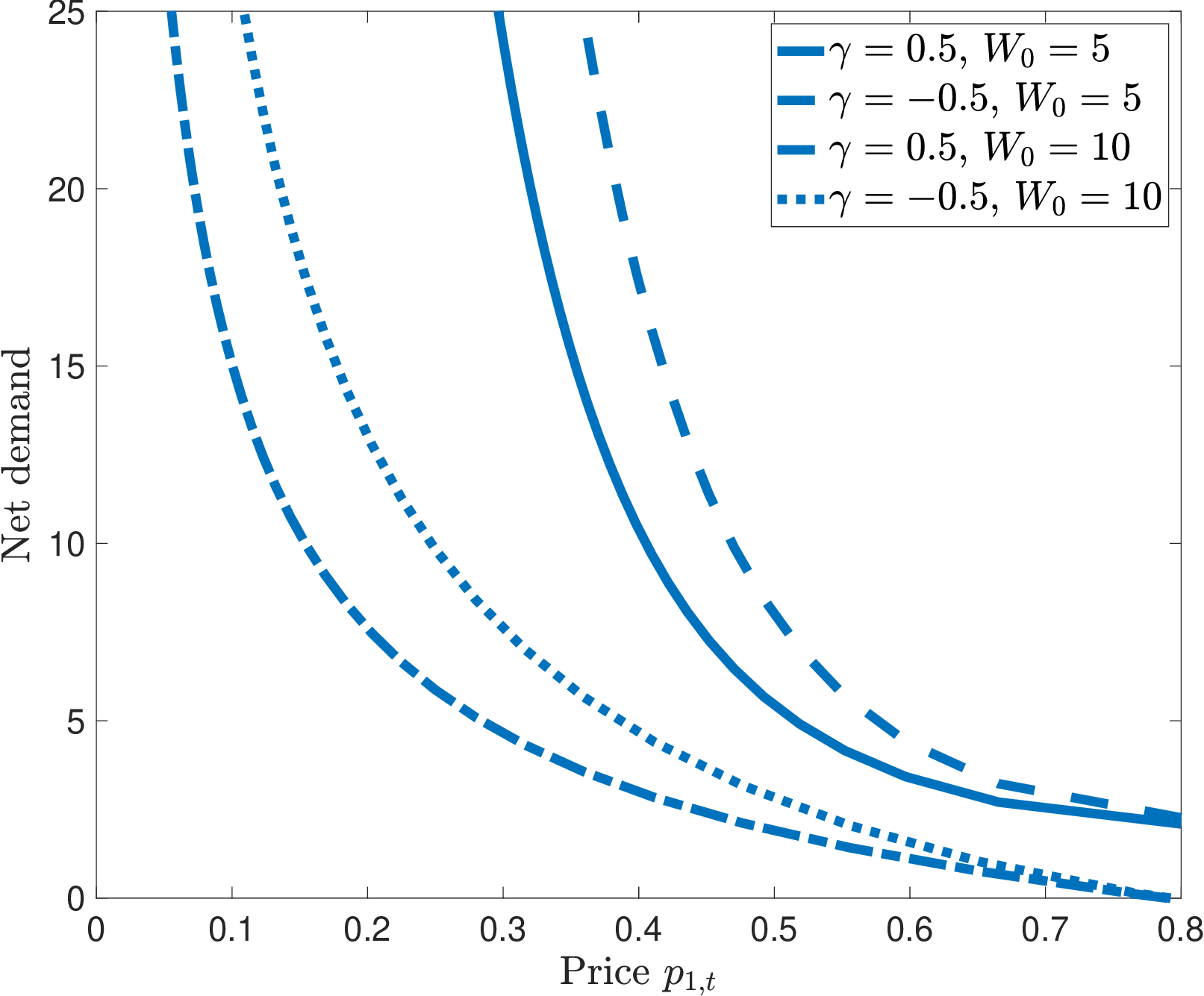}
\label{fig:crra_demand_price}
\end{figure}

\tb{
The above price-volume relationship also holds for the CRRA utility-based market. We slightly modify the market in Example \ref{exam_crra} by aggregating the first two securities as a single security, which gives a binary market. That is, the two traders' beliefs become $\pi_{1}=(0.4,0.6)$ and $\pi_{2}=(0.7,0.3)$. Figure \ref{fig:crra_demand_price} plots trader 2's net demand as a function of $p_{1,t}$ for diffferent paramters, i.e., $\gamma_m, \gamma\in\{0.5, -0.5\}$ and $W_0\in\{5,10\}$. Clearly, we can observe the pattern that the net demand is negatively correlated with the price and the risk aversion parameter. In addition, this figure also indicates that when the market maker has larger wealth, it may provide higher trading volume (higher liquidity).}

\begin{figure}[!h]
\caption{Utility and price trajectories after shock }
\begin{subfigure}{0.48\textwidth}
    \includegraphics[width = \linewidth]{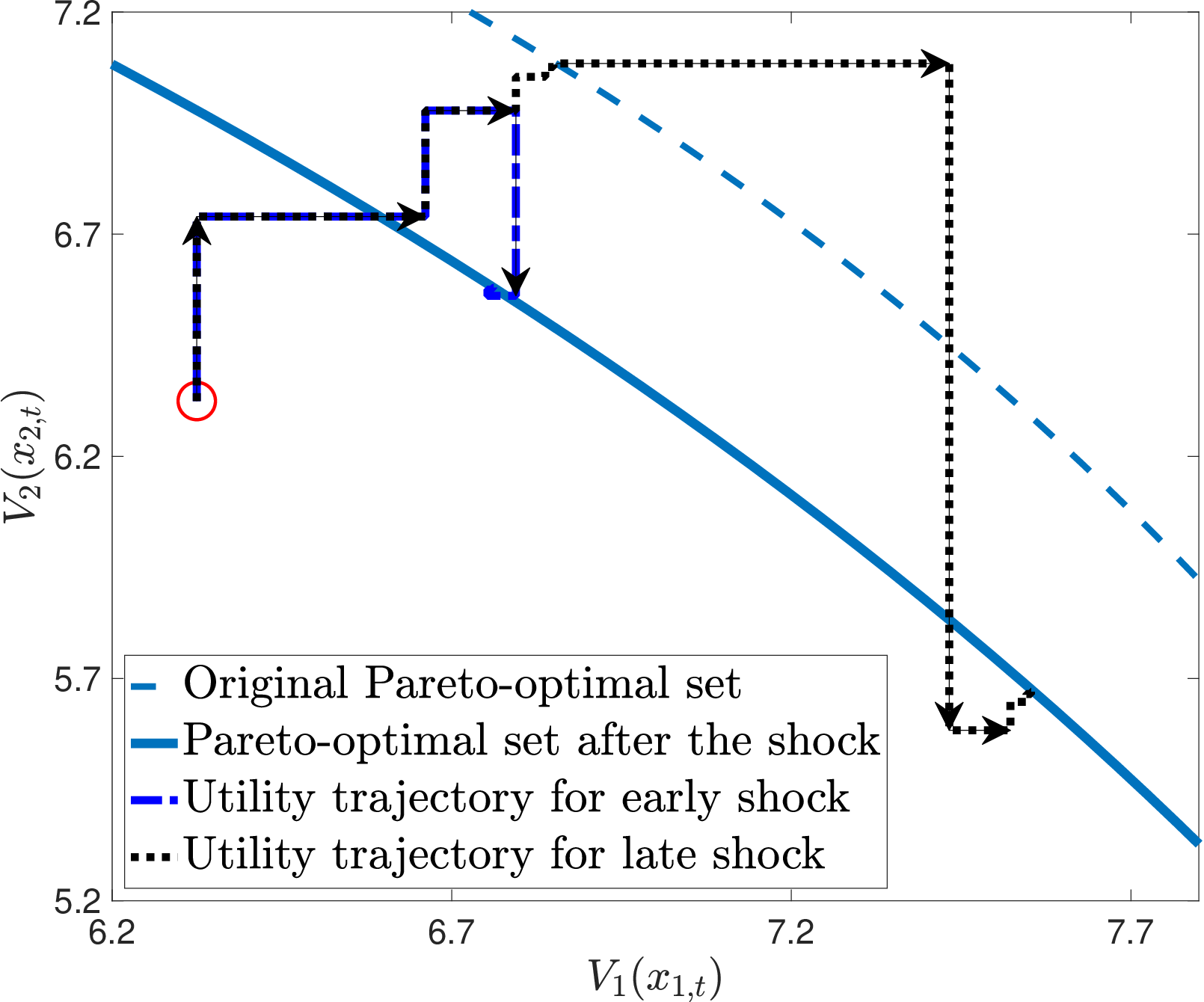}
    \caption{Utility trajectories}
    \label{fig:exam_crra_utility_shock}
\end{subfigure}
\begin{subfigure}{0.485\textwidth}
    \includegraphics[width = \linewidth]{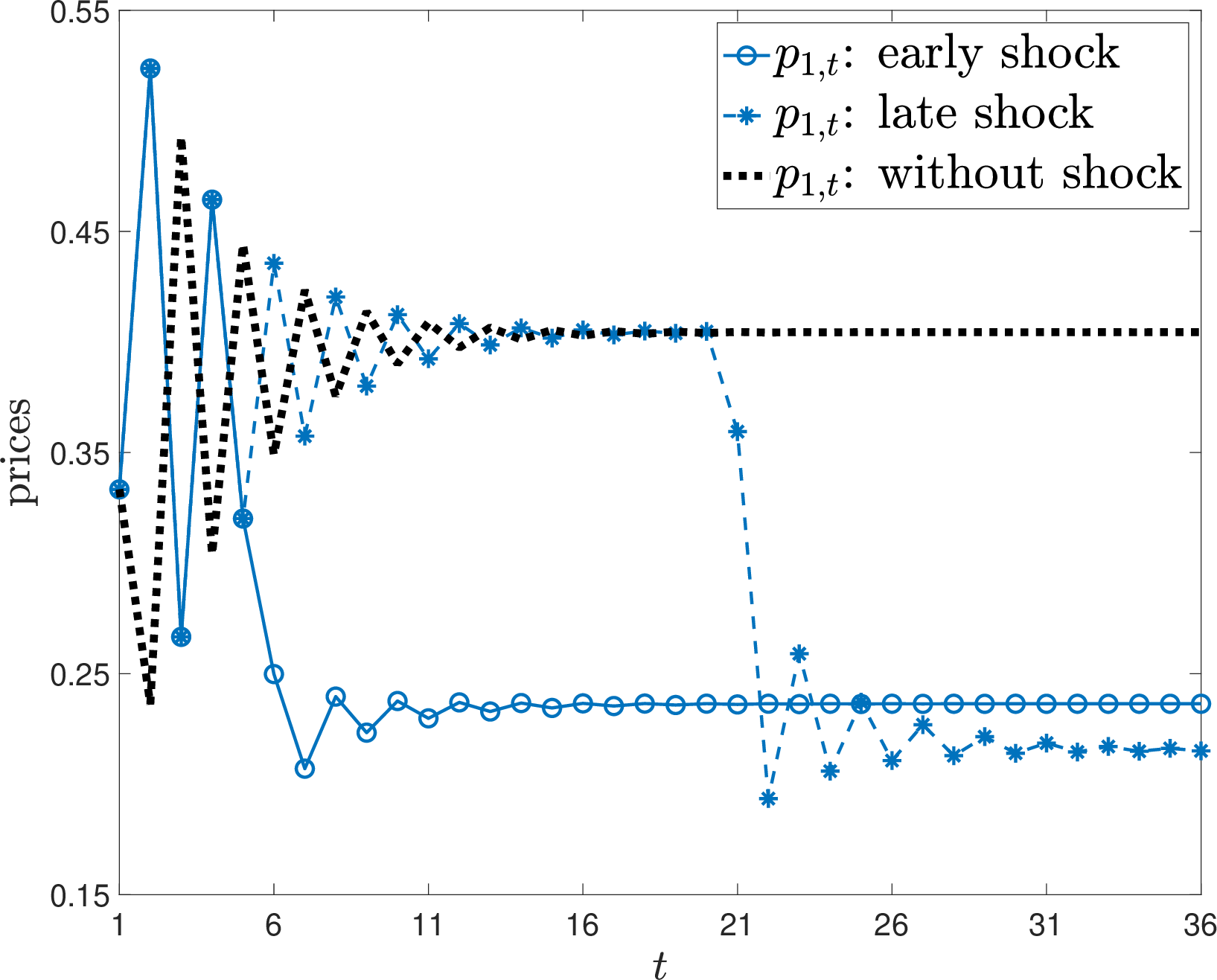}
    \caption{Price trajectories}
    \label{fig:exam_crra_price_shock}
\end{subfigure}
\label{fig_exam_crra_shock}\label{fig:exam_crra_shock}
\end{figure}

\tb{\noindent\textbf{Nonstationary market:} All the previous analysis is based on the stationary assumption, i.e., we assume that no new information comes to the market during trading. However, in reality, most prediction markets run for several months, and when an event shocks the market, traders' beliefs are significantly changed. To show how new information affects security prices, we consider two situations: the shock happens at an early stage (before price convergence), and the shock happens at a late stage. We adopt Example \ref{exam_crra} for illustration. The shock changes the traders' beliefs on the second security, i.e., the original beliefs, $\pi_1=(0.2,0.2,06)$ and $\pi_2=(0.6, 0.1, 0.3)$ are modified to $\hat{\pi}_1=\text{normalize}\{(0.2, 0.2, 0.6)+ (0,1,0)\}= (0.1, 0.6, 0.3)$ and $\hat{\pi}_2=\text{normalize}\{(0.6, 0.1, 0.3)+(0, 1, 0)\}=(0.3, 0.55, 0.15)$, respectively. 
Figure \ref{fig:exam_crra_utility_shock} displays the trajectories of two traders' utilities. The solid line and dash-dot line indicate the Pareto optimal frontiers of utility values generated by the original and modified beliefs, respectively. We can observe that, regardless of when the shock occurs, the utilities ultimately converge to the new Pareto-optimal set (i.e., the solid curve and the dashed curve indicate the Pareto-optimal sets for the market with and without the shock, respectively). However, the position of the limiting points varies. Figure \ref{fig:exam_crra_price_shock} illustrates the trajectory of the first security's price, $p_{1,t}$. Similar to the utility values, when new information enters the market, the price $p_{1,t}$ responds rapidly to this new information. It is important to note that the limiting prices resulting from early and late shocks are different. This is a straightforward consequence of Theorem \ref{thm_crra_limit}, which indicates that all traders' initial conditions and beliefs determine the limiting price. If we consider the time at which the shock occurs as the new starting point, early and late shocks produce different initial conditions, thus potentially leading to different limiting prices.
}

\section{Conclusion}\label{se:conclusion}
\tb{ 
This paper studies the prediction market convergence properties through the utilization of the multivariate utility (MU)-based market-making mechanism. This mechanism not only consolidates various existing market-making methods under specific conditions but also establishes a framework for examining the dynamic trading process within a market featuring a single market maker and a finite number of traders. Within this framework, we address one fundamental question arising in the prediction market: the formation of the limiting price (final price) by all market participants. We establish, under mild conditions, that traders' wealth processes converge to a limiting wealth distribution, resulting in Pareto optimal utility levels for all participants. This outcome enables us to explore the limiting price across different market types. In exponential utility markets, we present an explicit formula for the limiting price. For convex risk-measure-based markets, we introduce a method to design MU functions via penalty functions and demonstrate that the limiting price is uniquely determined as a weighted power mean of trader beliefs. Regarding CRRA utility-based markets, we characterize the limiting price through a system of equations. Additionally, we present a theoretical result indicating that the impact of trading sequence variation becomes negligible as the trader population increases. This novel finding enables us to formulate an approximate price formula, suggesting that the limiting price can be construed as a wealth-weighted risk-adjusted average of all participants' beliefs. Numerical experiments validate the high accuracy and robustness of this approximation.
}

\tb{
There are several promising directions for future research. Firstly, although this work has derived price formulas under various theoretical settings, a critical next step is the empirical validation of these results using real market data.
}

\tb{
Secondly, our model is built on the stationary assumption that participants' beliefs remain constant. In real-world trading, traders and market maker typically adaptively revise their beliefs based on new information. While our preliminary analysis in Section \ref{sec:price-volume-nonstationary} indicates that changes in traders' beliefs can result in significant shifts in the limiting price, it remains unknown how traders and market maker should incorporate this new information into their decision-making process as part of a learning process. A recent study by \cite{Birge:OR2021} analyzes the profit maximization problem for market maker in the spread betting market, where market maker may learn the distribution of event outcomes during trading. To address the potential profit loss that Bayesian policies may suffer in the presence of strategic bettors, they propose a new policy that balances learning with bluff-proofing. Building on this line of research, our model has the potential to incorporate learning features for both traders and market maker.
}

\tb{Thirdly, in our model, we assume that traders' decision problems are myopic, meaning that they focus on single-period decision problems. However, in the actual market, some traders exhibit forward-looking behavior. For example, some professional traders may possess significant insights into future market movements. Therefore, it is worthwhile to explore a model that incorporates traders who tackle forward-looking decision problems involving multi-period considerations. In such a scenario, the decision horizon spans multiple periods, which introduces more price uncertainty resulting from the trading activities of other market participants. Establishing a tractable multi-period decision problem for traders is a crucial step in exploring this forward-looking aspect in the market.
}

\tb{
Lastly, it is well-known that classical risk-averse preferences cannot fully explain certain decision-making behaviors observed in empirical studies and psychological experiments \cite{tversky1992advances}. Prospect theory (PT) is an influential alternative theory that accounts for psychological factors such as loss aversion, reference points, and probability distortion in individual decision-making (\cite{kahneman2013prospect,barberis2013thirty}). Recently, \cite{YuGao:Betting+PT} adopt a PT-based utility and construct an equilibrium model to explain pricing anomalies in betting markets. In the context of pricing models, \cite{Boer:MS2022} incorporate the PT-based demand function into a dynamic pricing model with demand learning and propose an asymptotically optimal dynamic pricing policy. Incorporating PT-based traders into the current analysis could provide a more realistic representation of real-world trading. Furthermore, it would be intriguing to explore how such traders influence the limiting price. However, establishing market convergence could be challenging due to the non-convex nature of the PT-based model. To address this challenge, several key assumptions made in this work may require refinement.
}

\bibliographystyle{plainnat}
\bibliography{prediction_market2023}





\appendix
\section{Proof of Main Results for Market Convergence Analysis}\label{apdx:sec:main_proof}

\subsection{Proofs of Main Results in Section \ref{sec:market_model}}\label{apdx:proof:sec:market_model}
The proof of Proposition \ref{prop:equivalence_diff_mechanism} is provided in \ref{apdx:ssec:equivalence_mechanism}. We begin with Proposition \ref{prop:equivalent_form}.

\noindent\textbf{$\bullet$ Proof of Proposition \ref{prop:equivalent_form}:}
\proof Given states $w_{j,t}$, $\q_{j,t}$, $W_t$ and $\Delta \q$, we introduce functions $h_U(\cdot):\mathbb{R}\rightarrow \mathbb{R}$ as $h_U(w)= U( (W_t+w)\cdot \e-(Q_t+ \Delta \q))$ and $h_V(\cdot):\mathbb{R}\rightarrow \mathbb{R}$ as
$h_V(w)=V_j((w_{j,t}-w)\cdot \e+(\q_{j,t}+\Delta \q))$ for some $w\in \mR$. Assumption \ref{assumption:U_mild} implies that $h_U(\cdot)$ is a strictly increasing and concave function. Now we consider the two level optimization problem $\bar{\P}_{j}(w_{j,t}, \q_{j,t})$. In lower-level problem (\ref{eq:barP_max_constraint}), the constraint $h_U(w)\geq U(W_0\cdot \e)$ actually defines a feasible interval of $w$, e.g., $w\in [h_U^{-1}(U(W_0\e)),\infty)$ where $h_U^{-1}(\cdot)$ is the inverse
function of $h_U(\cdot)$. On the other hand, Assumption \ref{assumption:V_j_mild} implies that $h_V(w)$ is a strictly
decreasing function of $w$. Based on these properties, when $\Delta
\q$ is fixed, replacing the objective function ``$\min_w~w$'' by
``$\max_w~h_V(w)$'' in the lower-level problem defined in constraint
(\ref{eq:barP_max_constraint}) does not change its optimal solution
for $\Delta w$. After the replacement, the subproblem can be viewed
as a partial optimization of the master problem for a given $\Delta
\q$, therefore, it can be viewed as the following problem which
optimizes $\Delta w$ and $\Delta\q$ simultaneously,
\begin{align*}
\max_{\Delta w\in\mR, \Delta \q\in\mR^I }~&~ V_j( (w_{j,t} - \Delta w)\cdot \e + ( \q_{j,t}+ \Delta \q))\notag\\
\textrm{subject to}~&~U\big((W_{t} + \Delta w)\cdot \e -(\Q_{t}+ \Delta \q)\big)\ge U(W_0\e),\\
            ~&~(W_{t} + \Delta w)\cdot \e -(\Q_{t}+ \Delta \q)\in\dom(U),\\
            ~&~ (w_{j,t} - \Delta w)\cdot \e + ( \q_{j,t}+ \Delta \q)\in \dom(V_j).
\end{align*}
Letting $\x_{j,t}= w_{j,t}\cdot \e + \q_{j,t}$, $\y_t= W_{t}\cdot\e - \Q_{t}$
and $\z= \Delta \q - \Delta w \cdot\e$ in above problem gives problem
$\P_j(\x_{j,t})$. The uniqueness of the solution follows from the strict
concavity of $V_j(\cdot)$. \hfill\qed
\vspace{10pt}

\noindent \textbf{$\bullet$ Proof of Proposition \ref{prop:inst_price}:} 
\proof
We first prove the proposition under condition (i). Given the market maker's wealth level $W$ and asset position (where we ignore the subscript $t$), we denote the neighborhood of $\Q$ as $\mathbb{B}(\Q;r)=\{\b{a}\in \mR^I~|~\|\b{a} - \Q\|\leq r \}$ for some $r>0$. First, we show that, if $\y= W\cdot\e-\Q$ is an interior point of $\dom(U)$ and $U(\y) = U(W_0 \cdot\e)$, then there exists some $r>0$ such that, any order $\Delta \q$ satisfying $\Q + \Delta \q \in \mathbb{B}(\Q;r)$ can be priced by
solving the following equation,
\begin{align}
U( (W + \Delta w)\cdot \e -(\Q+\Delta\q)) = U(\y + \Delta w \cdot \e -\Delta
\q)=U(W_0\cdot\e), \label{eq:implicit_eq_price}
\end{align}
where $\Delta w$ is the price of the order $\Delta \q$. Indeed, let
$2r>0$ be any number that is less than the minimal distance between
$\y$ and boundary of $\dom(U)$. Then $\y -\Delta\q$ will be an
interior point of $\dom(U)$ for any order $\|\Delta \q\| \le r$. If
$U(\y - \Delta\q))$$<$$U(W_0\e)$, then the monotonicity and
continuity of $U(\cdot)$ imply that there exists a $\Delta w$
satisfies Eq.(\ref{eq:implicit_eq_price}). Thus, the quantity
$\Delta w$ solves the pricing problem $\P_{\cmin}(\Delta \q)$ and it
is indeed the price of order $\Delta \q$. On the other hand, if
$U(\y - \Delta\q)> U(W_0\cdot \e)$, since $\|\Delta \q\| \le r$, it has
$-r\cdot \e - \Delta\q \le \0$. The monotonicity of $U(\cdot)$ implies
$U(\y - r\e -\Delta\q) $$\le$$ U(\y)= U(W_0\cdot\e)$. Thus, there exists
a $\Delta w \in[-r,0)$ such that Eq.
(\ref{eq:implicit_eq_price}) holds due to continuity and
monotonicity of $U(\cdot)$. The above result means that for any
small order $\Delta \tilde{\q}$ satisfying $ \Q +\Delta \tilde{\q}
\in\mathbb{B}(\Q;r)$, its price $\Delta \tilde{w}$ can be determined
by Eq. (\ref{eq:implicit_eq_price}). Now, we consider small
order $\epsilon \cdot \e_i$ satisfying $\Q +\epsilon \cdot \e_i \in
\mathbb{B}(\Q;r)$. The price $\Delta w^i$ of the order $\epsilon
\e_i$ is the solution of the equation
\begin{align}
U(\y +\Delta w_i \cdot \e - \epsilon \cdot \e_i ) ~=~U(W_0\cdot \e).
\label{prop:inst_price_equ1}
\end{align}
Recall that the instantaneous price of $i$-th security is the unit
cost of purchasing an infinitesimal $\epsilon \e_i$, i.e., it is
$p_i=\frac{d \Delta w^i}{d \epsilon}=\lim_{\epsilon \rightarrow
0}\frac{\Delta w^i}{\epsilon}$. We apply implicit differentiation
rule with respect to $\epsilon$ on equation
(\ref{prop:inst_price_equ1}), which gives,
\begin{align}
&-\frac{\partial U}{\partial y^i} + \frac{d \Delta w^i}{d \epsilon
}\sum_{k=1}^I \frac{\partial U}{\partial y^k}\frac{\partial
y^k}{\partial \Delta w^i}=0 ~\implies~~p^i = \lim_{\epsilon
\rightarrow 0}\frac{d \Delta w^i}{d \epsilon}=\frac{\frac{\partial
U}{\partial y^i}}{\sum \limits_{k=1}^I\frac{\partial U}{\partial y^k}}.
\label{def_Uprice}
\end{align}
Note that the conditions (ii) in Proposition \ref{prop:inst_price}
just ensure that $\y$ to be an interior in $\dom(U)$. Thus, the
proof for this condition is similar to condition (i).\hfill \qed

\vspace{10pt}
\subsection{Proof of Theorem \ref{thm_Uconv}}\label{apdx:proof_thm_Uconv}
Before we prove Theorem \ref{thm_Uconv}, we need the following result.
\begin{lemma}\label{lem_A_compact}
Under Assumptions \ref{assumption:U_mild} and
\ref{assumption:V_j_mild}, the feasible market states set $\A$ is a compact set.
\end{lemma}
\proof The set $\A$ is closed as it results from the intersection of closed sets, and it's convex due to the convex nature of $U(\cdot)$ and $V_j(\cdot)$. To prove the compactness, we only need to demonstrate that $\A$ is bounded. Suppose $\A$ is unbounded, which implies the existence of a recession direction $(\b{d}_1, \b{d}_2, \ldots, \b{d}_J)$ such that $\b{d}_{j'}\neq 0$ for some $j'$. This means that for any $(\h{\x}_1, \ldots, \h{\x}_J)\in \A$ and any large number $\beta>0$, it follows that $(\h{\x}_1 + \beta\cdot \b{d}_1, \ldots, \h{\x}_J + \beta \cdot\b{d}_J)\in \A$. However, due to Assumption \ref{assumption:V_j_mild}, it has $\mathrm{recc}(V_j)\subseteq\mR^I_+$, for all $j$, which implies $\b{d}_j\geq \0$ for $j=1,\ldots, J$. Let $\h{\x}_j = w_{j,0}\e$. Assumption \ref{assumption:U_mild} for $U(\cdot)$ implies that, for some $\beta>0$, it has $U(W_0 \cdot \e - \beta \cdot \sum\limits_{j=1}^J \b{d}_j) < U( W_0 \cdot \e)$, which contradicts with the assumption $(\h{\x}_1 + \beta \cdot\b{d}_1,  \ldots, \h{\x}_J + \beta \cdot \b{d}_J )\in \A$ for all $\beta > 0$. Thus, the set $\A$ is bounded. Note that, if the no-bankruptcy constraint $\x_j\geq \0$ is involved, the compactness of the set $\mathcal{A}$ is straightforward result.

We then proceed to prove Theorem \ref{thm_Uconv} through three steps: (I) Firstly, we prove the existence of a sub-sequence within the utility values generated by the trading sequence that converges to a specific limiting utility level. (II) We then establish that the limiting allocation corresponding to such a convergent subsequence achieves Pareto optimality. (III) Finally, we show that any convergent subsequence derived from the original bounded sequence converges to the same limit.

To ensure this proof fit more general situation, we consider the no-bankruptcy constraints, $\x_j\geq 0$ for all $j=1,\ldots,J$, in the Trading Process \ref{algtrading}. In other words, we solve problem $\P^+j(\x_{j,t})$ in each round of trading. It is worth noting that this proof can be readily adapted for cases without the no-bankruptcy constraint. The detailed proof unfolds as follows.

\noindent \textbf{(I)} Given a trading sequence $\mathcal{S}\in \Phi$, the wealth position $(\x_{1,t},\ldots, \x_{J,t})$ for
$t=1,2,\ldots$ generated from the Trading Process \ref{algtrading} is confined in the compact feasible set $\A$ (see Lemma
\ref{lem_A_compact}). From Bolzano-Weierstrass theorem, we know
there exists a subsequence $(\x_{1, t_n}$, $\ldots$, $\x_{J,t_n})$
for some $\{t_1,\ldots, t_n, \ldots\}$ $\subseteq$ $\{1,2,3...\}$
that converges to a limiting wealth distribution $(\h{\x}_{1}$,
$\ldots$, $\h{\x}_J)$, i.e., it has $\lim \limits_{n\rightarrow
\infty} \x_{j,t_n}=\h{\x}_j$ for $j=1,\ldots,J$.

\noindent \textbf{(II)} We then show that the limiting allocation
$(\h{\x}_{1}, \ldots,\h{\x}_J)$ is a Pareto optimal allocation. As there are two conditions to guarantee the convergence, we treat them seqerately. 

We first consider the case (i) (i.e., the case $\dom(U)=\mR^I$). The argument for case (ii) is identical to case (i), once we show that $\y_t$ will be bounded away from the domain boundary for all $t$. Given limiting allocation $(\h{\x}_{1}$, $\ldots$, $\h{\x}_J)$, the market maker's limiting wealth position is $\h{\y}={w}_0^{\all} \cdot \e -\sum_{j=1}^J \h{\x}_j $. We first prove the following claim. For any $j\in\{1,\ldots,J\}$, given $j$-th trader's wealth $\h{\x}_j=(\h{x}_{1,j},\ldots, \h{x}_{I,j})$, there does not exist vector $\h{\d}=(\h{d}_1,\ldots, \h{d}_I) \in \mR^I$ which satisfies the following system of
inequalities,
\begin{align}
\begin{dcases}
~\nabla V_j(\h{\x}_j)^{\top} \h{\d} > 0, \\
~\nabla U(\h{\y})^{\top} \h{\d} \leq 0, \\
~\h{\b{H}}_j \h{\d} \ge \0,\\
\end{dcases}\label{thm_converge_inequ}
\end{align}
where $\nabla V_j(\h{\x}_j)$ $=$ $\left(\frac{\partial V_j(\x)}{\partial x_1}, \ldots, \frac{\partial V_j(\x)}{\partial x_I} \right)\big|_{\x=\h{\x}_j}$,  $\nabla U(\h{\y})$ $=$ $\left(\frac{\partial U(\y)}{\partial y_1}, \ldots,  \frac{\partial U(\y)}{\partial y_I} \right)\big|_{\y=\h{\y}}$, and 
\begin{align*}
\h{\b{H}}_j =\textrm{diag}\big(\h{H}_{1,j}, \ldots, \h{H}_{i,j},
\ldots, \h{H}_{I,j} \big)\in \mR^{I\times I}
\end{align*}
is a diagonal matrix with diagonal elements being $\h{H}_{i,j} =0$
if $\h{x}_{i,j}>0$ and $\h{H}_{i,j}=1$ if $\h{x}_{i,j}=0$.

We then prove the following claim. If there exists $\d\in \mR^I$ which satisfies
(\ref{thm_converge_inequ}) for some $j\in\{1,\ldots, J\}$, then $\d$
must satisfy the following condition,
\begin{align}
\begin{dcases}
~\nabla V_j(\h{\x}_j)^{\top} \d > 0, \\
~\nabla U(\h{\y})^{\top} \d <0, \\
~\h{\b{H}}_j \d \ge \0,\\
~d_i > 0,\quad\mbox{for all }i\mbox{ such that } \h{H}_{i,j}=1. \\
\end{dcases}\label{thm_converge_inequ_strict}
\end{align}
First, we require the following result: for any $\y\in\dom(U)$, it holds that $\frac{\partial U}{\partial y_i}>0$ for all $i$. To establish this fact, consider an arbitrary $\y\in\dom(U)$. We define a function $f(\cdot):\mathbb{R}\rightarrow \mathbb{R}$ as $f(t;\y,\e_i) \triangleq U(\y+t\e_i)$. Since $U(\cdot)$ is concave and monotonic, $f(t;\y,\e_i)$ is also concave and strictly increasing with respect to $t$. Therefore, we have $f'(t;\y,\e_i)>0$ for all $t\in\mathbb{R}$, and thus $\frac{\partial U(\y)}{\partial y_i} = f'(0;\y,\e_i)>0$. This establishes the monotonicity result, $\frac{\partial U}{\partial y_i}>0$ for all $i$, as the choice of $\y \in \dom(U)$ is arbitrary.

We proceed to establish the claim presented in (\ref{thm_converge_inequ_strict}), which leads to three cases: (a) $\nabla U(\hat{\y})^{\top} \hat{\d}=0$ and $\hat{H}_{i,j}d_i>0$ for some $i$, (b) $\nabla U(\hat{\y})^{\top} \hat{\d}>0$ and $\hat{\b{H}}_j \hat{\d}=0$, and (c) $\nabla U(\hat{\y})^{\top} \hat{\d}=0$ and $\hat{\b{H}}_j \hat{\d}=0$. The first two cases can be easily addressed using the fact that $\frac{\partial U}{\partial y_i}>0$ for $i$. We will focus on the case where $\nabla U(\h{\b{y}})^{\top}\h{\d}=0$ and $\h{\b{H}}_j\h{\d}=\0$. In this scenario, we can construct a vector $\d=\h{\d}+\pmb{\epsilon}$, where $\pmb{\epsilon}=l\cdot \big[(\h{\b{H}}_j-\b{I})+k\cdot \h{\b{H}}_j\big]\cdot\e$ for some small positive numbers $k$ and $l$. Clearly, $\d$ satisfies the condition (\ref{thm_converge_inequ_strict}). Given that $\h{\y}$ lies within the interior of $\dom(U)$, we can utilize the vector $\d$ to enhance the expected utility for trader $j$ in Trading Process \ref{algtrading} while maintaining feasibility. Specifically, for some $\delta>0$ and small number $\alpha$, it has
\begin{align*}
\begin{dcases}
    ~V_j(\h{\x}_j+\alpha\d)-V_j(\h{\x}_j)=\alpha \nabla V_j(\h{\x}_j)^{\top}\d+o(\alpha)>2\delta,\\
    ~U(\h{\y}-\alpha\d)- U(\h{\y})=-\alpha \nabla U(\h{\y})^{\top}\d+o(\alpha)>0, \\
    ~\h{\x}_j+ \alpha\d > \0,~\h{\x}_j+ \alpha\d \in \dom(V_j),\\
    ~\h{\y}-\alpha\d\in\dom(U),
\end{dcases}
\end{align*}
where the above inequalities can be derived from (\ref{thm_converge_inequ_strict}) and by noting that the linear term dominates the higher-order term in the Taylor expansion when $\alpha$ is sufficiently small. 

Next, we demonstrate that $\alpha\d$ is a feasible direction that can drive the utility of trader $j$ to surpass their limit $V_j(\h{\b{x}}j)$ at some time $t_N$. Indeed, there exists some positive integer $N$, such that the wealth vector $\b{x}{j,t_N}$ is sufficiently close to $\h{\b{x}}_j$. Then, we have:
\begin{align*}
\begin{dcases}
~V_j(\x_{j,t_N}+\alpha\d)-V_j(\x_{j,t_N})>\delta,\\
~U(\y_{t_N}-\alpha\d)-U(\y_{t_N})>0, \\
~\x_{j,t_N}+ \alpha\d > \0, ~{\x}_{j,t_N}+ \alpha\d \in \dom(V_j),\\
~\y_{t_N} - \alpha\d \in \dom(U),\\
~V_j(\h{\x}_j)-V_j(\x_{j,t_N})<\delta/2,
\end{dcases}
\end{align*}
where the first two inequalities are a result of the continuity of $V_j(\cdot)$ and $U(\cdot)$, and the last inequality arises from the convergence property: $\lim_{n\rightarrow }V_j(\x_{j,t_n})\rightarrow V_j(\h{\x}_j)$. 
TTherefore, $\alpha\d$ represents a feasible trade for trader $j$ at time $t_N$, causing the post-trade utility to exceed their limit, as demonstrated below:
\begin{align*}
V_j(\x_{j,t_N}+\alpha\d)-
V_j(\h{\x}_j)=V_j(\x_{j,t_N}+\alpha\d)-V_j(\x_{j,t_N})+V_j(\x_{j,t_N})-V_j(\h{\x}_j)>\delta-\delta/2=\delta/2>0.
\end{align*}
This inequality contradicts the fact that $V_j(\x_{j,t})$ should be upper-bounded by the limit $V_j(\h{\x}_j)$ for all $t$.

We then show that if the condition (ii) in Theorem \ref{thm_Uconv}
is met, then the subsequential limit of market maker wealth $\h{\y}$
is bounded away from the domain boundary. Indeed, assume $\h{\y}$ lies
on the boundary of $\dom(U)$, then $\frac{\partial U}{\partial
y_i}\big|_{\h{\y}}=\infty$ for some $i$ while $\frac{\partial
U}{\partial y_{i'}}\big|_{\h{\y}}<+\infty$, $y_{i'}>0$ for some
$i'\ne i$. Otherwise, $\h{\y}=\b{L}\implies U(\h{\y})<U(W_0\cdot \e)$,
which is a contradiction. Since $\h{y}_i = L_i<W_0$ and $\h{\y} +
\sum_{j=1}^J \h{\x}_j = \big(W_0+\sum_{j=1}^J w_{j,0}\big)\cdot \e$ it has
$\h{x}_{i,j} > w_{j,0} > 0$ for at least one $j$. In other words,
$\h{x}_{i,j}$ is bounded away from trader's utility boundary. Let
$\d = M\cdot \e_{i'} -\e_i$, then for trader $j$ and some $M$ sufficiently
large, it has
\begin{align*}
        \begin{dcases}
            ~\nabla V_j(\h{\x}_j)^{\top} \d > 0, \\
            ~\nabla U(\h{\y})^{\top} \d < 0. \\
        \end{dcases}
\end{align*}
Then, for some $\delta>0$ and $\alpha$ which are sufficiently small,
it has
\begin{align*}
        \begin{dcases}
            ~V_j(\h{\x}_j+\alpha\d)-V_j(\h{\x}_j)=\alpha \nabla V_j(\h{\x}_j)^{\top}\d+o(\alpha)>2\delta,\\
            ~U(\h{\y}-\alpha\d)- U(\h{\y})=-\alpha \nabla U(\h{\y})^{\top}\d+o(\alpha)>0, \\
            ~\h{x}_{i,j}+\alpha d_i>0,~\h{x}_{i,j}+\alpha d_i \in \dom(V_j)\\
            ~\h{\y}-\alpha\d\in\dom(U).
        \end{dcases}
\end{align*}
Since $\x_{j,t_n}\to \h{\x}_j$ for all $j$, we can find a $N$
sufficiently large such that,
\begin{align*}
        \begin{dcases}
            ~V_j(\x_{j,t_N}+\alpha\d)-V_j(\x_{j,t_N})>\delta,\\
            ~U(\y_{t_N}-\alpha\d)-U(\y_{t_N})>0, \\
            ~x_{i,j,t_N}-\alpha d_i>0,x_{i,j,t_N}-\alpha d_i \in \dom(V_j),\\
            ~\y_{t_N} - \alpha\d \in \dom(U),\\
            ~V_j(\h{\x}_j)-V_j(\x_{j,t_N})<\delta/2.
        \end{dcases}
\end{align*}
Again, we achieve a non-trivial utility gain for trader $j$ using
feasible trade  $\alpha\d$. 

As a summary at this stage, we've demonstrated that the existence of a vector $\d$ satisfying condition (\ref{thm_converge_inequ}) leads to condition (\ref{thm_converge_inequ_strict}). This, in turn, results in an improvement in the utility value for $V_j(\h{\x}_j),$ contradicting the fact that ${\h{\x}_j}$ is the limiting wealth allocation. Therefore, at this stage, we conclude that condition (\ref{thm_converge_inequ}) must be infeasible.

Now we reformulate the infeasibility condition given by (\ref{thm_converge_inequ}) as the alternative of Farkas' Lemma \cite{bertsekas2003convex}. Let $\mathbf{A}_j=\big(-\nabla U(\h{ \y }),  ~\h{\b{H}}_j\big)^{\top}$.
Recall our previous claim which says that there does not exist $\d$
satisfying the following conditions
\begin{align}
    \mathbf{A}_j \d \geq \0,~~-\nabla V_j(\h{\x}_j)^{\top} \d<0.
\end{align}
Applying the Farkas' Lemma gives the alternative of the claim, i.e.,
there exists a vector $ \left(
\begin{array}{c}
    a_j \\
    \b{g}_j \\
    \end{array}
    \right)\in \mR^{1+I}_+$ which satisfies
\begin{align}
\b{A}_j^{\top} \left(
\begin{array}{c}
a_j \\
\b{g}_j \\
\end{array}
\right) = -\nabla V_j(\h{\x}) ~\Rightarrow ~ \nabla V_j(\h{\x}_j)+
\h{\b{H}}_j \b{g}_j - a_j \nabla U(\y^*) =\0
\label{thm_converge_alt1}
\end{align}
for all $j=1,\ldots, J$. We then consider the aggregate optimization
problem,
\begin{align*}
  \P_{\po}(\pmb{\nu }):~&~\max_{\x_1,\ldots, \x_J} ~~\sum_{j=1}^J \nu_j V_j(\x_j)\\
  \textrm{Subject to:}~&~\y +\sum_{j=1}^J \x_j = {w}_0^{\all} \cdot \e,\notag\\
  ~&~U(\y) \geq U( W_0 \cdot \e),\notag\\
  ~&~\x_j\ge\0,\quad j=1,\ldots,J~~(\textrm{if bankruptcy is prohibitted}),
\end{align*}
where $\pmb{\nu}=(\nu_1,\ldots,\nu_J)^{\top}\in \mR_+^I \setminus
\{\0\}$ is some weighting parameter. Because $V_j(\cdot)$'s are
strictly concave, it is well known that a wealth allocation
$(\x^*_1,\x^*_2,\ldots, \x^*_J)\in\mathcal{A}$ is Pareto optimal if
and only if it is a solution to $\P_\po$ for certain
$\pmb{\nu}=(\nu_1,\ldots,\nu_J)^{\top}>0$. We now show that
allocation $(\h{\x}_1,\cdots, \h{\x}_J)$ is a solution of problem
$\P_{\po}(\pmb{\nu})$. Since problem $\P_{\po}(\pmb{\nu})$ is a convex optimization problem,
its optimal solution $(\x_1^{*},\ldots, \x_J^*)$ is completely
characterized by the Karush-Kuhn-Tucker (KKT) conditions as follow,
\begin{align}
    \begin{dcases}
        \nu_j\nabla V_j(\x^*_j) + \pmb{\eta}_j - \mu\nabla U(\y^*)=\0, ~~~j=1,\cdots, J, \\
        \textrm{diag}(\pmb{\eta}_j) \x^*_j =\0,~~ j=1,2,\ldots, J,\\
        \mu \big(U(\y^*)-U(W_0 \cdot \e)\big)=0,\\
        \sum_{j=1}^J \x^*_j + \y^{*} =  \b{w}_0,
    \end{dcases}\label{thm_converge_KKT}
\end{align}
for some $\mu>0$ and $\pmb{\eta}_j\in \mR_+^I$, $j=1,\ldots, J$.
Noticing the relationship (\ref{thm_converge_alt1}) holds for all
$j=1,\ldots,J$. Multiplying (\ref{thm_converge_alt1}) by $\nu_j$,
it is not hard to verify the limit allocation $(\h{\x}_1,\ldots,
\h{\x}_J)$ together with the following Lagrange multipliers, $\mu=
1$, $\nu_j =\frac{1}{a_j}$ and  $\pmb{ \eta }_j
=\frac{1}{a_j}\h{\b{H}}_j \b{g}_j$ for $j=1,\ldots, J$,
satisfy the KKT condition (\ref{thm_converge_KKT}). Clearly, it must
have ${a}_j\ne0$ for all $j$, otherwise (\ref{thm_converge_alt1})
cannot hold for $\nabla V_j(\x^*_j)>\b{0}$ and  $\b{g}_j\ge\b{0}$.
Indeed, if $a_j=0$ for some $j$, then $\nabla
V_j(\h{\x}_j)=\h{\b{H}}_j\b{g}_j\le\b{0}$, which contradicts with
the monotonicity of $V_j$. That is to say,
$(\h{\x}_1,\ldots,\h{\x}_J)$ is a Pareto optimal allocation.\\

\noindent \textbf{(III)} At last, we then show that every convergent
subsequence of $(\x_{1,t}, \ldots, \x_{J,t})$ shares an identical
limit. Suppose there is another wealth allocation
$(\bar{\x}_1,\ldots, \bar{\x}_J)\in \A$ satisfying
$(\bar{\x}_1,\ldots, \bar{\x}_J) \ne(\h{\x}_1,\ldots, \h{\x}_J)$ and
achieving the same utility level as $V_j(\bar{\x}_j) =
V_j(\h{\x}_j)=V_j^*$ for all $j=1,\ldots, J$. Then we can construct
another allocation $(\check{\x}_1,\ldots, \check{\x}_J)$ as
$\check{\x}_j = (\h{\x}_j + \bar{\x}_j)/2$ for $j=1,\ldots, J$.  Due
to the convexity of $\A$,  it has $(\check{\x}_1,\ldots,
\check{\x}_J)\in \A$. Moreover, for any $j=1,\cdots, J$, it has
\begin{align*}
V_j(\check{\x}_j )  \ge \frac{1}{2}(V_j(\h{\x}_j)+ V_j(\bar{\x}_j))=
V_j^*
\end{align*}
with strict inequality holds for at least one $j\in\{1,\ldots, J\}$,
due to the strict concavity of $V_j(\cdot)$. Clearly this
contradicts with the Pareto optimality of $(\h{\x}_1,\ldots,
\h{\x}_J)$. The uniqueness of $(\h{\x}_1,\ldots, \h{\x}_J)$ implies
that every convergent subsequence of $(\b{x}_{1,t},\ldots,
\b{x}_{J,t})$ for $t=1,2,\ldots$ converges to the identical limiting
allocation $(\h{\b{ x }}_1,\ldots, \h{\b{ x }}_J )$, which further
implies the original bounded sequence $(\x_{1,t}, \ldots, \x_{J,t})$
also converges to the limiting allocation $(\h{\x}_1,\ldots,
\h{\x}_J)$ (e.g., see page 58 in
\citealt{abbott2001understanding}). \hfill \qed

\vspace{10pt}

\subsection{Proofs of Main Results in Section \ref{sec:expU_Risk}}\label{apdx:ssec_exp_risk_measure} 
\noindent\textbf{Proof of Proposition \ref{prop_exp_unique}}. 
\proof{Proof.} Given the utility function (\ref{eq:exp_U_Vj}) with the wealth vector $\y_t=\big( y_{1,t}, \ldots, y_{I,t} \big)$, we can explicitly compute the instantaneous price vector $\p_t=\big(p_{1,t}, \ldots, p_{I,t}\big)$ using Eq. (\ref{def_U_price}) as follows:
\begin{align}
p_{i,t} =\frac{ \theta_i e^{-\beta y_{i,t} } }{\sum \limits_{k=1}^I \theta_k e^{-\beta y_{k,t}}}. \label{eq:price_exp}
\end{align}
for $i=1,\ldots, I$.

We then prove the result (a). The $j$-th trader interacts with the market maker at trading time $t$ and solves the problem
$\P_j(\x_{j,t})$ with the market states $\x_{j,t}$ and $\y_t =
{w}^{\all}_0 \cdot \e - \sum_{j=1}^J \x_{j,t}$. Under the assumption of
exponential utility in (\ref{eq:exp_U_Vj}), after eliminating the negative sign in the objective function, 
problem $\P_j(\x_{j,t})$ becomes,
\begin{align}
\P_j(\x_{j,t}):~~\min_{\z}~&~ \sum_{i=1}^I \hat{\pi}_{i,j} \cdot e^{-\alpha_j z_i}\notag\\
 \textrm{Subjec to}&~\sum_{i=1}^I \hat{\theta}_i \cdot e^{\beta z_i}  \leq e ^{-\beta W_0},\label{proof_exp_contr1}
\end{align}
where $\hat{\pi}_{i,j} \triangleq  \pi_{i,j} e^{-\alpha_j \cdot x_{i,j,t}}$
and $\hat{\theta}_i \triangleq  \theta_i e^{-\beta \cdot y_{i,t}}$. Due to the convexity, the following Karush-Kuhn-Tucker (KKT) conditions fully characterize the solution as,
\begin{align}
&-\alpha_j \hat{\pi}_{i,j} e^{-\alpha_j z^*_i} + \zeta^*  \beta \cdot \hat{\theta}_i e^{\beta z^*_i}=0 ~\Rightarrow~~z_i^* = 
\frac{1}{\alpha_j +\beta } \ln \left( \frac{\alpha_j\hat{\pi}_{i,j} }{\zeta^*  \beta  \hat{\theta}_{i} } \right),~~i=1,\ldots,I,\label{KKT1}\\
&\sum_{i=1}^I \hat{\theta}_{i} e^{\beta z^*_i} = e^{-\beta W_0 }, \label{KKT2}
\end{align}
where $\zeta^*>0$ is the langrange multiplier. The condition (\ref{KKT2}) is from the fact that $\zeta^*\not =0$. Otherwise, the problem admits no solution. It implies that the constraint (\ref{proof_exp_contr1}) always holds equality in each trading round. Thus, before the $j$-th trader interacts with market maker, the condition (\ref{KKT2}) also holds, i.e., it has
\begin{align}
\sum_{i=1}^I \hat{\theta}_i e^{\beta z^*_i} = \sum_{i=1}^I \hat{\theta}_i e^{\beta z_i}|_{z_i=0,i=1,\ldots,I} = \sum_{k=1}^I \hat{\theta}_k. \label{proof_exp_contr2}
\end{align}
In addition, the price formula (\ref{eq:price_exp}) implies that $p_{i,t}=\frac{\hat{\theta}_i }{ \sum_{k=1}^I \hat{\theta}_k}$
for $i=1,\ldots, I$. Using this pricing formula and subtituting (\ref{KKT1}) into condition (\ref{proof_exp_contr2}) give the lagrange multiplier
\begin{align*}
\zeta^* = \frac{\alpha_j}{\beta}\left( \sum_{i=1}^I p_{i,t} \left( \frac{\hat{\pi}_{i,j}}{\hat{\theta}_{i}}\right)^{\frac{\beta}{\alpha_j+\beta}}  \right)^{\frac{\alpha_j+\beta}{\beta}}.
\end{align*}
Combining the above expression with (\ref{KKT1}) gives the solution (\ref{prop_exp_z}). 

We then prove the result (b). After the trading, the market state becomes
$\x_{j,t+1}=\x_{j,t}+\z^*$ and $\y_{t+1}=\y_t - \z^*$. We then
compute the updated price $p_{i,t+1}$ by Proposition
\ref{prop:inst_price} which gives
\begin{align}
p_{i,t+1}= \frac{ \frac{ \partial U(\y_{t+1}) }{ \partial y_{i,t+1}
} }{ \sum_{k=1}^I  \frac{ \partial U(\y_{t+1}) }{ \partial y_{k,t+1}
}   } = \frac{ \hat{\theta}_i e^{\beta z_i^*}}{\sum_{k=1}^I
\hat{\theta}_k e^{\beta z_k^*} }
=\frac{\hat{\theta}_i^{\frac{\alpha_j}{\beta
+\alpha_j}}(\hat{\pi}_{i,j})^{\frac{\beta}{\beta +\alpha_j}}}
{\sum_{k=1}^I (\hat{\theta}_k)^{\frac{\alpha_j}{\beta +\alpha_j}}(
\hat{\pi}_{k,j})^{\frac{\beta}{\beta
+\alpha_j}}}.\label{def_price_t+1}
\end{align}
Note that, at time $t$, the instantaneous price is
$p_{i,t}=\frac{\hat{\theta}_i}{\sum_{k=1}^I \hat{\theta}_k}$. Then,
replacing $\hat{\theta}_i$ by $p_{i,t}(\sum_{k=1}^I \hat{\theta}_k)$
in (\ref{def_price_t+1}) gives the updated price.

To prove result (c), we compute the limiting price $p_i^*$ when $t \rightarrow
\infty$. The exponential utility naturally meet the convergence
conditions in Theorem \ref{thm_Uconv}. Thus, the wealth process
$\{\x_{j,t}\}_{t=1,2,\ldots}$ indeed converges to some Pareto
optimal allocation as $t\rightarrow \infty$. Let
$\pmb{\nu}=(\nu_1,\ldots,\nu_J)\in \mR^I_+\setminus \{\0\}$.
We then solve problem $\P_{\po}(\pmb{\nu})$ for the Pareto
optimal wealth allocation,
\begin{align}
 \max_{\x_1,\ldots,\x_{J}, \y}~&~\sum_{j=1}^J \nu_j V_j(\x_j)= -\sum_{j=1}^J\frac{ \nu_j }{\alpha_j}\sum_{i=1}^I\pi_{i,j} e^{-\alpha_j x_{i,j}}\notag\\
\textrm{Subject to}  &~ -\frac{1}{\beta}\sum_{i=1}^I \theta_i e^{-\beta y_i} \ge -\frac{1}{\beta}e^{-\beta W_0} \label{prop_exp_const_U}\\
        &~ \sum_{j=1}^J \x_j + \y = {w}^{\all}_0 \cdot \e, \notag
\end{align}
where $\x_j=(x_{1,j}, x_{2,j},\ldots, x_{I,j})$ and
$\y=(y_1,\ldots,y_I)$. The last equality constraint helps to
eliminates $\y$ in this problem. The KKT condition characterizes the
optimal solution as follows,
\begin{align}
 \nu_j \pi_{i,j} e^{-\alpha_j x_{i,j}^*}- \lambda \theta_i B e^{\beta\sum_{j=1}^J x_{i,j}^*}=0, ~~i=1,2,\ldots, I,~\textrm{and}~~j=1,\ldots, J,
 \label{prop_exp_kkt}
\end{align}
where $\lambda \geq 0$ is the Lagrange multiplier for constraint
(\ref{prop_exp_const_U}) and $B \triangleq  e^{-\beta W_0}$. Note
that it must have $\lambda>0$, as the constraint
(\ref{prop_exp_const_U}) is binding at optimal solution. We then let
$M_i =\lambda \theta_i e^{\beta \sum_{j=1}^J x_{i,j}^*}$ in equation
(\ref{prop_exp_kkt}), which leads to $x_{i,j}^* = \frac{1}{\alpha_j}
\ln \left(\frac{\nu_j \pi_{i,j} }{ B \cdot M_i } \right)$.
Substituting $x_{i,j}^*$ back to the definition of $M_i$ gives,
\begin{align}
\frac{1}{\beta}\ln \left(  \frac{M_i}{ \lambda \theta_i} \right)=
\sum_{j=1}^J x_{i,j}^* = \sum_{j=1}^J \frac{1}{\alpha_j}\ln\left(
\frac{\nu_j \pi_{i,j}}{ B\cdot M_i}  \right) \Rightarrow~M_i =
(\lambda \theta_i)^{\eta} \prod_{j=1}^J \left(\frac{\nu_j
\pi_{i,j}}{B}\right)^{\eta_j},\label{prop_exp_Mi}
\end{align}
where $\eta_j\triangleq \frac{ 1/\alpha_j }{\sum_{j=1}^J
\frac{1}{\alpha_j}+\frac{1}{\beta}}$ and $\eta \triangleq 
\frac{1/\beta }{\sum_{j=1}^J\frac{1}{\alpha_j}+\frac{1}{\beta}}$. We
then compute the instantaneous price associated with the Pareto
optimal wealth $x_{i,j}^*$ which gives
\begin{align*}
p^*_i & =  \frac{ \theta_i e^{-\beta( w^{\all}_0 - \sum_{j=1}^J
x_{i,j}^* ) } }{ \sum_{k=1}^I \theta_k e^{-\beta(w^{\all}_0 -
\sum_{j=1}^J x_{k,j}^* ) } }
 =\frac{M_i}{\sum_{k=1}^I M_k},~i=1,\ldots, I,
\end{align*}
where $w_0^{\all} = W_0 +\sum_{j=1}^J w_{j,0}$. Substituting equation
(\ref{prop_exp_Mi}) into the above formulation gives the result
(\ref{prop_expp_pi}). 
\hfill\qed

\vspace{10pt}
\noindent \textbf{Example of Risk Measure Based-MU Function}

\begin{example}\label{example_RM_counterexample}
We present an example to illustrate that, without additional conditions, the convex risk measure-based MU function may fail to satisfy the convergence condition in Theorem \ref{thm_Uconv}.

In this example, we consider the widely used risk measure called \textit{average value at risk} (AVaR). Given $\mathbf{x} \in \mathbb{R}^I$ and a probability vector $\pmb{\pi} \in \Theta_I$, the AVaR of $\x $ is defined as 
$\text{AVaR}_\beta(\x) = \max{ \p \in \Theta_I} \left\{ -\p^\top \x ~|~\0 \leq \p \leq \frac{1}{\beta} \pmb{\pi} \right\}$, where $\beta \in (0,1)$ represents the confidence level. In a binary setting with $\pmb{\pi} = (0.5,0.5)$ and $\beta=0.95$, the probability $\p=(p_1,p_2)$ has a feasible set $\mathcal{A}_p \triangleq \{ (p_1,p_2)~|~ p_1+p_2=1,~p_1\in [0.4737,0.5263]\}$.

We then consider the direction $\d = (-2,3)$. For any $\alpha>0$, we have $-\p^\top(\mathbf{x}+\alpha \mathbf{d}) = -\p^\top \x - \alpha \cdot (-2p_1+3p_2)$. Note that, for any $\p\in \mathcal{A}_p $, it has $(-2p_1+3p_2)>0$. Thus,as $\alpha\rightarrow \infty$, $\text{AVaR}_\beta(\x + \alpha \d)\rightarrow \infty$, proving that $\d$ is a direction of recession for the function $\text{AVaR}_\beta(\x)$. Consequently, the function $\text{AVaR}\beta(\mathbf{x})$ violates condition (ii) in Assumption \ref{assumption:V_j_mild}. Furthermore, since AVaR is a coherent risk measure, the translation-invariant property implies $\text{AVaR}_\beta(\alpha\cdot\x) = \alpha \cdot \text{AVaR}_{\beta}(\x )$ for any $\alpha \in \mathbb{R}$. This property contradicts the strict concavity condition for non-equivalent vectors, i.e., condition (i) in Assumption \ref{assumption:V_j_mild}.

\end{example}

\vspace{10pt}
\noindent \textbf{Proof of Proposition \ref{prop:rm_unique}}.
\proof{Proof.}
As this proof applies for all $\rho_j(\cdot)$, $j=0,1\ldots,J$, we
simply drop the index $j$ in the following proof. First, we show
that $\rho$ is well-defined for any $\x \in \mR^I$. Indeed, for
$\x=(x_1,\ldots, x_I) \in\mR^I$, KKT conditions characterize the
optimizing probability $\b{q}^*$ for the dual representation
(\ref{def_U_V_RM_dual}) as,
\begin{align*}
&\x + \nabla \alpha(\b{q}^*)- \lambda^* \cdot \e +\pmb{\mu}^* =
\0,~{\b{q}^*}^{\top} \e=1,~{\b{q}^*}^{\top} \pmb{\mu}^*=0,~\pmb{\mu}^*\ge\0,
\end{align*}
where $\pmb{\mu}$ is the Lagrange multiplier. One can see that the
KKT system is still complicated. However, by temporarily assuming
$\pmb{\mu}^*=\0$ and noting that
$f_i(q_i)=\frac{\partial\alpha(\b{q})}{\partial q_i}$ is an
invertible function due to its strict monotonicity, we see that the
solution $\b{q}^*$ and $\lambda^*$ should satisfy:
\begin{align}\label{optimality:risk_measure}
        \sum_{i=1}^I f^{-1}_i(\lambda-x_i) = 1,
\end{align}
where $f^{-1}_i(\cdot)$ is the inverse function of $f_i(\cdot)$.
Clearly, if $\overline{\lambda}$ large enough, it has
$\overline{\lambda} - x_{i'}= f_{i'}(1)$ $\Rightarrow$
$f^{-1}_{i'}(\overline{\lambda}-x_{i'})=1$ for some $i'$, which
further implies that $\sum_{i=1}^I f^{-1}_i(\overline{\lambda} -
x_i)>1$. On the other hand, it has $\underset{\lambda
\to-\infty}{\lim}\sum_{i=1}^I f^{-1}_i(\lambda - x_i)=0<1$. Thus, we
can find a $\lambda^*$ that uniquely solves the equation
(\ref{optimality:risk_measure}). Furthermore, since $\nabla^2
\alpha(\b{q})= \mathrm{diag}\big(f'_1(p_1),\ldots, f'_I(p_I)\big)$
which is positive definite for any $\b{q} \in\Theta_I$, we know that
$\alpha(\cdot)$ is strictly convex, therefore the optimal solution
is unique. As a result, it indeed confirms our previous assumption
$\pmb{\mu}^*=\0$. Moreover, if $x_i=-\infty$ for some $i$, by setting
the corresponding $p^*_i=0$, we may still solve
(\ref{optimality:risk_measure}) for some  $\lambda^*\in\mR$. Next,
we show that the recession cone of $\rho$ is contained in the
non-negative orthant. Assume $\d\in\mathrm{recc}(\rho)$ with $d_i<0$
for some $i$. Without loss of generality, we  assume $d_1<0$. Then,
for some $\tau$, it has
\begin{align*}
\rho(\x+ \tau \cdot \d) &= -\inf_{\b{q} \in \Theta_1}\{\b{q}^{\top} \x +\tau\cdot \b{q}^{\top} (\d - d_1\e_1)+ \tau d_1\b{q}^{\top} \e_1+\alpha(\b{q})\}\\
                     &\ge - x_1 - \tau d_1 - \alpha(\e_1)\\
                     &= - x_1 - \tau d_1 - \int_0^1 f_1(s) ds -  \sum_{i=2}^I \underset{p \to 0}{\lim}\int_0^{p} f_i(s) ds.
\end{align*}
Due to the strict monotonicity of $f_i(\cdot)$, it has
    \begin{align*}
        \int_0^1 f_1(t)dt< f(1) ~\mbox{ and }~\underset{p\to 0}{\lim}\int_0^{p} f_i(t)dt \le \underset{p\to
        0}{\lim}~p f_i(p)<0,
    \end{align*}
which implies $ \rho(\x + \tau \cdot\d)>-x_1 - \tau d_1 - f(1)$. As
$f(1)$ is finite, $d_1<0$ and the fact that $\tau$ can be
arbitrarily large, we see that such $\d$ cannot be a direction of
recession of  $\rho$.

As for the strict convexity of  $\rho$. First, we show that, if
$\x_1$ and $\x_2$ are not equivalent wealth, the correspondent
optimal probabilities $\b{q}_1^*$ and $\b{q}_2^*$ generated from the
dual problem (\ref{def_U_V_RM_dual}) do not coincide, i.e., it has
$\p_1^*\ne\p_2^*$. Indeed, from the above discussion, we know that
\begin{align}\label{gradient_non_equal}
\nabla\alpha(\b{q}_1^*) =
\lambda_1^* \cdot \e-\x_1,~~\nabla\alpha(\b{q}_2^*) = \lambda_2^*\cdot \e-\x_2,
\end{align}
for some $\lambda_1, \lambda_2 \in \mR$. Since $\x_1$ and $\x_2$ are
not in equivalent class, (\ref{gradient_non_equal}) implies
\begin{align*}
        \nabla \alpha(\b{q}_1^*) \ne \nabla\alpha(\b{q}_2^*) \implies \b{q}_1^*\ne\b{q}_2^*,
\end{align*}
due to strict monotonicity of $f_i(\cdot)$,~ $i=1,\ldots,I$. If
$\x_1$ and $\x_2$ are not in equivalent class, then $\tau
\x_1+(1-\tau)\x_2$ will not be in the equivalent class with $\x_1$
and $\x_2$ for all $t\in(0,1)$. Then, let $\b{q}^*$, $\b{q}_1^*$ and
$\b{q}_2^*$ be the solutions of problem (\ref{def_U_V_RM_dual}) for
$\tau\x_1+(1-\tau)\x_2$, $\x_1$ and $\x_2$, respectively. Then, it
has $\b{q}^* \ne \b{q}_1^*$ and $\b{q}^*\ne \b{q}_2^*$, which
further implies,
\begin{align*}
\rho(\tau \x_1+(1-\tau) \x_2) &= -{\b{q}^*}^{\top} (\tau \x_1+(1 - \tau)\x_2) - \alpha(\q^*)\\
                              &= \tau \big(-{\q^*}^{\top}\x_1-\alpha(\q^*)\big) + (1-\tau)\big(-{\q^*}^{\top} \x_2 - \alpha(\q^*)\big)\\
                              &<\tau \big(-{\q_1^*}^{\top}\x_1-\alpha(\q_1^*)\big) + (1-\tau)\big(-{\q_2^*}^{\top} \x_2 - \alpha(\q_2^*)\big)\\
                              &=\tau\rho(\x_1)+(1-\tau)\rho(\x_2).
\end{align*}
The above inequality implies the strict convexity of $\rho(\cdot)$.\hfill\qed
\vspace{10pt}

\noindent\textbf{Proof of Theorem \ref{thm:alloc-unique}}
\proof Using Theorem \ref{thm_Uconv}, the limiting allocation can be identified
by solving the following problem for some weights
$(\nu_1,\nu_2,\ldots,\nu_J) \in \mR^J_+$,
\begin{align}
\underset{\x_1,\ldots,\x_J}{\min}~&~\sum_{j=1}^{J}~\nu_j \rho_j(\x_j)\notag\\
        \textrm{Subjec to}~&~ \rho_{0}(\y) \le \rho_{0}(W_{0}\cdot\e), \label{equ_thm:alloc-unique}\\
                          ~&~ \y+\sum_{j=1}^{J} \x_j = {w}_0^{\all}\cdot \e.\notag
\end{align}
Then optimality condition implies that, the optimal allocation
$(\x_1^*,\ldots,\x_J^*)$ should satisfy,
\begin{align*}
\nu_j\nabla\rho_j(\x_j^*) = \mu \nabla \rho_{0}({w}_0^{\all}\cdot \e
-\sum_{j=1}^J \x_j^*), \quad\forall j=1,\ldots,J,
\end{align*}
where $\mu\ge 0$ is the Lagrange multiplier for constraint
(\ref{equ_thm:alloc-unique}). On the other hand, since
$\rho_j(\cdot)$ is a convex risk measure for all $j=0,\ldots,J$, it
has $\nabla_{\x_j} \rho_j(\cdot)\ge\0$ and $\nabla_{\x_j}
\rho_j(\cdot)^{\top} \e=1$ for all $j=0,\ldots,J$, which implies that
$\nu_1=\cdots=\nu_J=\mu$. (Otherwise problem
(\ref{equ_thm:alloc-unique}) is unbounded, which will not give any
Pareto wealth allocation.) Clearly, this is exactly the optimality
conditions that characterizes the solution of
$\P_{\po}^{\mathrm{RM}}$.

As previous analysis shows that $\nu_1=\cdots=\nu_J=\mu$, we consider
(\ref{equ_thm:alloc-unique}) with all weights in the objective being
$1$. To simplify the notation, we let $\x_0=\y$. Then, using dual
representation of convex risk measure and considering the
Lagrangian, problem $\P_{\po}^{\mathrm{RM}}$ becomes
\begin{align}
\underset{\x_0,\x_1,\ldots,\x_{J}}{\max} \Big\{ \sum_{j=0}^{J}
\underset{\q_j \in \Theta_I}{\inf}~\{\q_j^{\top}\x_j+\alpha_j(\q_j)\}~|~
\sum_{j=0}^{J} \x_j = w_0^{\all}\cdot\e \Big\}.\label{dual_minmax}
\end{align}
Denote by $\q_j(\x_j)$ the optimizing probability generated from
dual representation (\ref{def_U_V_RM_dual}) for a given $\x_j$ for all
$j=0,\ldots,J$. Obviously, the gradient of  $\rho_j(\cdot)$ at
$\x_j$ is exactly $\nabla_{\x_j} \rho_j(\x_j)=\q_j(\x_j)$. Thus, if
$\x_0$ and $(\x_1^*,\ldots,\x_J^*)$ are the optimal solution of
problem $\P_{\po}^{\mathrm{RM}}$, it has,
\begin{align*}
\begin{dcases}
\nabla \rho_0(\x_0^*) = \nabla \rho_1(\x_1^*) =\cdots= \nabla \rho_{J}(\x_{J}^*)\\
\sum_{j=0}^{J} \x_j^* = w_0^{\all}\cdot\e.
\end{dcases}
\end{align*}
The above property means all the dual variables $\q_j(\x_j)$ should be identical at optimal solution. Thus, we may rewrite problem (\ref{dual_minmax}) in the following equivalent form,
\begin{align*}
\underset{\x_0,\ldots,\x_{J}}{\max}~\underset{ \q \in
\Theta_I}{\inf}~\Big\{ \q^{\top}
\sum_{j=0}^{J}\x_j+\sum_{j=0}^{J}\alpha_j(\q) ~\Big|~~\sum_{j=0}^{J}
\x_j = w_0^{\all}\cdot\e\},
\end{align*}
which further implies that the optimal objective value of problem (\ref{dual_minmax}) is independent to $(\x_0^*, \ldots, \x_J^*))$. Thus, we can find the limiting price by $\p^*= \arg\min_{\q \in\Theta_I}~\{\sum_{j=0}^{J} \alpha_j(\q)\}$. 
\hfill\qed
\vspace{10pt}

%
\noindent \textbf{Proof of Corollary \ref{cor:mean_exact}}. 
\proof
We focus on the case of $\gamma\ne 0$, the case of $\gamma=0$ can be proved
in the similar fashion. It can be verified that $\alpha_j(\cdot)$
meets the conditions in Theorem \ref{thm:alloc-unique} for all $j$.
Thus, the limiting price can be identified by solving
\begin{align}
\max_{\q \in \Theta_I}~&~~\sum_{i=1}^I
~q_i\Big(\frac{q_i}{\theta_{i}}\Big)^{\gamma-1} h_0
+\sum_{j=1}^J\sum_{i=1}^I~ q_i
\Big(\frac{q_i}{\pi_{i,j}}\Big)^{\gamma-1} h_j.
\end{align}
Let $q^*=(q_1,\ldots,q_I)$ be the optimal solution of above problem.
The optimality condition implies that
\begin{align*}
{q_i^*}^{\gamma-1}\cdot \Big(\theta_i^{1-\gamma}h_0 + \sum_{j=1}^J
\pi_{i,j}^{1-\gamma}h_j\Big)=\lambda^*~~ ~\Rightarrow~~{q_i^*} &=
\Big( \frac{  \theta_{i}^{1-\gamma}h_0 +  \sum_{j=0}^J
\pi_{i,j}^{1-\gamma}h_j}{\lambda^*} \Big)^{\frac{1}{1-\gamma}}
\end{align*}
for $i=1,\ldots,I$. From proof of Theorem \ref{thm:alloc-unique}, we
know that the gradient of all $\rho_j(\cdot)$, $j=0,\ldots,J$, share
the same vector $\q^*$. Thus, the limiting price is
\begin{align*}
p_i^*= \frac{q_i^*}{\sum_{k=1}^I q_k^*} = \frac{
\Big(\theta_{i}^{1-\gamma}h_0+ \sum_{j=1}^J h_j
\pi_{i,j}^{1-\gamma}\Big)^{\frac{1}{1-\gamma}}}{\sum_{k=1}^I\Big(\theta_{k}^{1-\gamma}h_0+
\sum_{j=1}^J h_j \pi_{k,j}^{1-\gamma}\Big)^{\frac{1}{1-\gamma}}},
\end{align*}
for all $i=1,\ldots,I$, which completes the proof.\hfill\qed
\vspace{10pt}

\subsection{Proofs and Additional Results in Section \ref{sec:CRRA}}\label{apdx:sec:crra_result}

\noindent \textbf{Proof of Proposition \ref{prop:crra_trader_solution}}
\proof Given $U(y)= \frac{1}{\gamma}\sum_{i=1}^I \theta_i y_i^{\gamma}$ and
$V_j(\x_j)=\frac{1}{\gamma}\sum_{i=1}^I \pi_{i,j}x_{i,j}^{\gamma}$, we first prove the result (i) by solving problem $\mathcal{P}_j(\x_{j,t})$ for the $j$-th trader,  
\begin{align*}
\max_{z_1,\ldots,z_I}~&~\frac{1}{\gamma} \sum_{i=1}^I \pi_{i,j}(x_{i,j,t}+z_i)^{\gamma}\\
s.t.~&~ \sum_{i=1}^I \theta_i \big( y_{i,t} - z_i\big)^{\gamma} \geq W_0^{\gamma}.
\end{align*} 
Introducing Lagrange multiplier $\lambda$ and checking the first order optimality condition for $z^*_i$ give
\begin{align*}
&\pi_{i,j}( x_{i,j,t} + z^*_i )^{\gamma-1} -\lambda \theta_i ( y_{i,t} -z^*_i )^{\gamma-1}=0.
\end{align*} 
for $i=1,\ldots,I$. Clearly, the above condition yields solution only when $\lambda>0$, which further implies
\begin{align*}
\frac{x_{i,j,t} + z^*_i }{y_{i,t} - z^*_i} =\frac{\kappa_{i,j}}{\hat{\lambda}}
~~\Rightarrow~~z^*_i= \frac{ y_{i,t} \cdot \kappa_{i,j}- \hat{\lambda} \cdot x_{i,j,t}}{\hat{\lambda} + \kappa_{i,j} }
\end{align*}
for $i=1,\ldots, I$, where $\lambda^*=\lambda^{\frac{1}{1-\gamma}}$. Then we have the result (\ref{def_crra_ytxt}). The Lagrange multiplier $\lambda^*$ can be found by solving from the constraint when it is binding.  \hfill\qed

\noindent \textbf{Proof of Theorem \ref{thm_crra_limit}}
\begin{proof} The CRRA utility-based MU functions defined in (\ref{def:CRRA}) satisfy the conditions in Theorem \ref{thm_Uconv}, ensuring the convergence of the market state to a Pareto optimal point. This Pareto optimal wealth allocation can be described by problem $\P_{\po}(\pmb{\nu})$ with a set of weighting parameters $\pmb{\nu}^*=(\nu_1^*, \ldots, \nu_J^*)$:
\begin{align}
\P_{\po}(\pmb{\nu}^* ):~&~ \max_{\y,\x_1,\ldots, \x_J}~~~\frac{1}{\gamma}\sum_{j=1}^J (\nu_j^* \cdot \sum_{i=1}^I \pi_{i,j}x_{i,j}^{\gamma})\notag\\
\textrm{Subject to:}~&~ \y+ \sum_{j=1}^J \x_j ={w}_0^{\all}\cdot \e, \label{crra_c1}\\
                   ~&~  \sum_{i=1}^I \theta_i y_i^{\gamma} \geq W_0^\gamma. \label{crra_c2}
\end{align}
This problem is convex, and we can verify its optimality conditions as follows: 
\begin{align}
&\nu_j^*  \pi_{i,j} (x_{i,j}^*)^{\gamma-1} =\xi_i~~
\Rightarrow~~
x_{i,j}^*=\left( \frac{\nu_j^* \pi_{i,j} }{\xi_i}\right)^{ \frac{1}{1-\gamma}},~i=1,\ldots,I;~j=1,\ldots, J, \label{thm_crra_limit_x}\\
&\eta \theta_i (y^*_i)^{\gamma_m -1} = \xi_i~~ \Rightarrow~~
y^*_i= \left( \frac{\eta \theta_{i} }{\xi_i}\right)^{
\frac{1}{1-\gamma_m} },~~i=1,\ldots,I, \label{thm_crra_limit_y}
\end{align}
where $\pmb{\xi}=(\xi_1,\xi_2,\ldots,\xi_I)$ and $\eta \geq 0$ are the Lagrange multipliers for constraints (\ref{crra_c1}) and (\ref{crra_c2}), respectively. From (\ref{thm_crra_limit_x}) and (\ref{thm_crra_limit_y}), if $\eta=0$ or $\xi=0$, then then $x^*_{i,j}=0$ and $y_i^*$ for all $i$ and $j$. Under this case, constraint (\ref{crra_c1}) can not be satisfied. Thus, we have $\xi_i>0$ for all $i$ and $\eta>0$. Note that, $\eta>0$ means that the cosntraint (\ref{crra_c2}) is binding. We substitute (\ref{thm_crra_limit_y}) in constraint (\ref{crra_c2}) which gives 
\begin{align}
\eta = \frac{ W_0^{1- \gamma_m}}{\left(\sum_{i=1}^I \theta_i (\theta_i/\xi_i)^{\frac{\gamma_m}{1-\gamma_m}} \right)^{\frac{1-\gamma_m}{\gamma_m}}}.\label{thm_crra_limit_eta}
\end{align}
Substituting (\ref{thm_crra_limit_x}) and (\ref{thm_crra_limit_y}) to Eq.(\ref{crra_c1}) yields
\begin{align}
\xi_i = \Bigg( \frac{ \sum \limits_{j=1}^J ( \nu_j^* \pi_{i,j})^{\frac{1}{1-\gamma}} 
+ \xi_i^{\frac{\gamma - \gamma_m}{(1- \gamma_m)(1-\gamma)}}\cdot( \eta \theta_i)^{\frac{1}{1-\gamma_m}} }{ w_0^{\all}} \Bigg)^{1-\gamma},\label{thm_crra_limit_xi}
\end{align} 
for $i=1,\ldots,I$. Combining (\ref{thm_crra_limit_eta}) into Eq. (\ref{thm_crra_limit_xi}) gives the equations for $\xi_i$: 
\begin{align}
\xi_i &=  \Bigg( \frac{ \sum \limits^J_{j=1} (\nu_j^*)^{\frac{1}{1-\gamma}} \cdot \pi_{i,j}^{\frac{1}{1-\gamma}} 
+  (\xi_i)^{\frac{\gamma -\gamma_m}{(1-\gamma)(1-\gamma_m)} }\cdot\mathcal{L}(\pmb{\xi},\pmb{\theta})\cdot \theta_i^{\frac{1}{1-\gamma_m} }\cdot W_0 }{w_0^{\all}} \Bigg)^{1-\gamma},~~i=1,\ldots,I, \label{thm_crra_limit_equ_xi}
\end{align}
where $\mathcal{L}(\pmb{\xi},\pmb{\theta}) \triangleq \left(\sum_{i=1}^I \theta_i(\theta_i/\xi_i)^{\frac{\gamma_m}{1-\gamma_m}}\right)^{-\frac{1}{\gamma_m}}$. We denote the solution of system of equations (\ref{thm_crra_limit_equ_xi}) as $\pmb{\xi}^*=(\xi^*_1,\ldots,\xi^*_I)$. As the constraint (\ref{crra_c2}) is binding, this optimization problem can be viewed as a convex optimization problem with equality constraints. From Proposition 3.1.1 in \cite{Bertsakas:book:nonlinear}, we know that this problem always has unique Lagrange multipliers, which further implies Eq. (\ref{thm_crra_limit_equ_xi}) always admits the solution $\pmb{\eta}^*$ together with $\eta^*$. To compute the limiting price, we use Proposition \ref{prop:inst_price}, which gives
\begin{align}
p_i^* = \frac{\theta_i \cdot (y_i^*)^{\gamma_m-1} }{ \sum \limits_{k=1}^{I} \theta_k \cdot (y_k^*)^{\gamma_m-1} }=\frac{\xi^*_i,
}{ \sum \limits_{k=1}^{I} \xi_k^*} \label{thm_crra_limit_pi}
\end{align}
where the last equality is from (\ref{thm_crra_limit_y}). This completes the proof for the result (i).

Then we focus on result (ii). For any $i\in\{1,\ldots, I\}$, we define $s_{i,j}\triangleq (\nu_j^*\cdot\pi_{i,j})^{1/(1-\gamma)}$ for $j=1,\ldots,J$. The assumption of this result says that $\pi_{i,j}$ and $w_{0,j}$ for $j=1,\ldots,J$ are independently sampled from a given distribution. Furthermore, given these traders' parameters, the distribution of the weight parameters $\{\nu_j^*\}|_{j=1}^J$ is purely determined by different trading sequences. Note that changing the indices of any two traders does not affect the distribution of $\{\nu_j^*\}|_{j=1}^J$. That is to say, the homogeneity of all traders implies $\{s_{i,1}, s_{i,2},\ldots, s_{i,J}\}$ is \textit{sequence of exchangeable random variable} (see \cite{klenke2013probability}). It is known that an infinite exchangeable sequence is almost surely Ces\`{a}ro convergent to a constant (see, e.g., \cite{klenke2013probability} or Proposition 1.1 in \cite{Tappe:JPS2021}). That is to say, the Ces\`{a}ro summation of $s_{i,j}$ satisfies, $\lim\limits_{J\to\infty}\frac{1}{J}\sum\limits_{j=1}^J s_{i,j} \xrightarrow{a.s.} C_i$ almost surely where $C_i$ is a finite number. Let $\pmb{\xi^*}|_{J}=(\xi^*_1|_J, \ldots, \xi^*_I|_J)$ be solution to the Eq. (\ref{thm_crra_limit_equ_xi}) for any fixed $J$. Then Eq. (\ref{thm_crra_limit_equ_xi}) can be written as,
\begin{align}
\xi^*_i|_J&=\Bigg( \frac{ \sum \limits^J_{j=1} s_{i,j} + M(\pmb{\xi^*}|_{J})}{w_0^{\all}} \Bigg)^{1-\gamma}~~
\Rightarrow ~~\xi^*_i|_{J}=\Bigg( \frac{ \frac{1}{J}\sum \limits^J_{j=1} s_{i,j} + \frac{1}{J}M(\pmb{\xi}^*|_J)}
     {\frac{1}{J}\sum\limits_{j=1}^J w_{0,j} + \frac{W_0}{J} } \Bigg)^{1-\gamma}, \label{thm_crra_xi_J}
\end{align}
where $M(\pmb{\xi}^*|_J)=(\xi_i^*|_J)^{\frac{\gamma -\gamma_m}{(1-\gamma)(1-\gamma_m)} }\cdot\mathcal{L}(\pmb{\xi}^*|_J,\pmb{\theta})\cdot \theta_i^{\frac{1}{1-\gamma_m} }\cdot W_0 $. Recall that the $j$-th trader's initial wealth is sampled from a given distribution where we use $\bar{w} \triangleq \E[\bar{w}_{j,0}]$ to denote its mean value. Note that market maker's parameters (i.e., $\{\theta_i\}|_{i=1}^I$ and $W_0$) are constant and $\pmb{\xi}^*|_J$ is the finite solution of equations (\ref{thm_crra_limit_equ_xi}) for any $J$. Clearly, it has $\lim \limits_{J\rightarrow \infty}~\frac{1}{J} M(\pmb{\xi}^*|_J) \rightarrow 0$ and  $\lim \limits_{J\rightarrow \infty}~\frac{W_0}{J} \rightarrow 0 $. Morever, due to the Law of large number, it has $ \lim \limits_{J\rightarrow \infty}~\frac{1}{J}\sum\limits_{j=1}^J w_{0,j} \xrightarrow{a.s.} \bar{w}$ almost surely. Thus, as $J\rightarrow \infty$, Eq. (\ref{thm_crra_xi_J}) implies $\lim \limits_{J\rightarrow \infty} \xi^*_i|_J \xrightarrow{a.s.} (C_i/\bar{w})^{1-\gamma}$ almost surely. We consider the price formula (\ref{thm_crra_limit_pi}) and use the above result, it has 
\begin{align*}
\lim_{J\rightarrow \infty} p_i^* = \lim_{J\rightarrow \infty} \frac{\xi^*_i|_J}{\sum \limits_{k=1}^J \xi^*_k|_J } = \frac{C_i^{1-\gamma}}{\sum \limits_{k=1}^I C_k^{1-\gamma} },~~i=1,\ldots, I,
\end{align*} 
which completes the proof of the second result.
\end{proof}
%

\subsection{Price Formula for Market with Logarithm-Exponential Utilities}\label{apdx:sse:log-exp-utility}
In this section, we explore a market model featuring mixed utility functions. Specifically, we consider a setup where the MU functions for the market maker (market maker) and the $j$-th trader are defined as follows: 
\begin{align}
U(\y)= \sum_{i=1}^I \theta_i \frac{1- e^{-\beta \cdot y_i}}{\beta} ~~\textrm{and}~~
V_j(\x)=\sum_{i=1}^I \pi_{i,j} \ln(x_{i,j}), ~~j=1,\ldots, J, \label{def_U_log_exp}
\end{align}
respectively. Here, $\y=(y_1,\ldots, y_I)$ represents the wealth vectors for the market maker, and $\x_j=(x_{1,j},\ldots, x_{I,j})$ represents the wealth vector for the $j$-th trader. The followin results characterizes the limiting wealth allocation and limiting price of such a market.

\begin{proposition}\label{prop:exp-log}
For the Logarithm-Exponential market model defined by (\ref{def_U_log_exp}), the limiting wealth allocation is characterized by problem $\P_{\po}(\pmb{\nu}^*)$ for some weighting parameter $\pmb{\nu}^*=(\nu_1^*,\ldots, \nu_J^*)\in \mathbb{R}_+^J\setminus \{\0\}$ and the limiting price $\p^*=(p_1^*,\ldots,p_I^*)$ is given by:
\begin{align}
p^*_i = \xi_i^*/\sum \limits_{k=1}^I \xi_k^*,~~i=1,\ldots,I, \label{log-exp-price}
\end{align}
where $\pmb{\xi}^*=(\xi_1^*, \ldots, \xi_I^*)$ is the solution of the following equations,
\begin{align}
\xi_i^* = \frac{\sum \limits_{j=1}^J \nu_j^* \cdot \pi_{i,j} + \frac{1}{\beta} {\xi}_i^* \log \left( \theta_i\Big/\frac{{\xi}_i^*}{\sum_{k=1}^I {\xi}_k^*} \right)}{\sum \limits_{j=1}^J w_{j,0} },~~i=1,\ldots,I. \label{log-exp-xi}
\end{align}
\end{proposition}

\proof The MU function defined in (\ref{def_U_log_exp}) satisfies the convergence condition stated in Theorem \ref{thm_Uconv}. Therefore, in this market model, the market state converges to a Pareto-optimal wealth allocation, which can be determined by solving problem $\P_{\po}(\pmb{\nu})$ for some weighting parameter $\pmb{\nu}^*=(\nu_1^*,\ldots, \nu_J^*)$. In this context, problem $\P_{\po}(\pmb{\nu}^*)$ can be formulated as follows:
\begin{align*}
\P_{\po}(\pmb{\nu}^*):~&~\max_{\y,\x_1,\ldots,\x_J}~~~\sum_{j=1}^J (\nu_j^* \cdot \sum_{i=1}^I \pi_{i,j} \ln (x_{i,j})) \notag\\
\textrm{Subject to:}~~&~ \y+ \sum_{j=1}^J \x_j ={w}_0^{\all}\cdot \e, \\
                    ~~&~  -\frac{1}{\beta} \sum_{i=1}^I \theta_i \exp(-\beta y_i )
                   \geq -\frac{1}{\beta} \exp(-\beta W_0),
\end{align*} 
where the decision varibles are $\y=(y_1,\ldots, y_I)$ and $\x_j=(x_{1,j},\ldots, x_{I,j})$ for $j=1,\ldots, J$. We may eliminate $W_0$ from this problem by introducing a new decision variable $\z=(z_1,\ldots, z_I)$, where $z_i = y_i-W_0$ for $i=1,\ldots,I$. This transforms problem $\P_{\po}(\pmb{\nu}^* )$ into the following form:
\begin{align}
\P_{\po}(\pmb{\nu}^* ):~&~\max_{\z,\x_1,\ldots,\x_J}~~~\sum_{j=1}^J (\nu_j^* \cdot \sum_{i=1}^I \pi_{i,j} \ln (x_{i,j})) \notag\\
\textrm{Subject to:}~~&~ \z + \sum_{j=1}^J \x_j =(\sum_{j=1}^J w_{j,0})\cdot \e, \label{exp_log_constr1}\\
                    ~~&~ \sum_{i=1}^I \theta_i \exp(-\beta z_i )\leq 1.  \label{exp_log_constr2} 
\end{align}  
Let $\z^*$ and $\x_j^*$ for $j=1,\ldots, J$ be the optimal solution to this problem. Due to convexity, the solution can be characterized by the following optimality conditions:
\begin{align}
&x^*_{i,j} = \frac {\nu_j^* \pi_{i,j}}{\xi_i^*},~~i=1,\ldots, I, ~~j=1,\ldots,J,\notag\\
&\eta^* \cdot \theta_i \cdot \exp(-\beta z_i^* ) = \xi_i^*,~~i=1,\ldots,I,\label{exp_log_z}
\end{align}
where $\pmb{\xi}^*=(\xi_1^*, \ldots, \xi_I^*)$ and $\eta^*$ are the Lagrange multipliers for constraints (\ref{exp_log_constr1}) and (\ref{exp_log_constr2}), respectively. It is straightforward to verify that $\xi_i^>0$ for $i=1,\ldots,I$ and $\eta^>0$, as otherwise, the problem would have no solution. This implies that constraint (\ref{exp_log_constr2}) is binding, i.e., $\sum_{i=1}^I \theta_i\exp(-\beta z_i^*) = 1$, which further leads to the following relationships:
\begin{align*}
\eta^* = \sum_{i=1}^i \xi_i^*~~\textrm{and}~~
z_i^* = \frac{1}{\beta}\ln \left(\theta_i\Big/ \frac{\xi_i^*}{\sum_{k=1}^I \xi_k^*} \right).
\end{align*}
Substituting $z^*_i$ and $x^*_{i,j}$ to the equality version of constraint (\ref{exp_log_constr2}) gives equation (\ref{log-exp-xi}) for multiplier $\pmb{\xi}^*$. On the other hand, using Eq. (\ref{def_U_price}), we can compute the limiting price as 
\begin{align*}
p_i^* = \frac{ \theta_i \exp(-\beta z_i^*)}{ \sum \limits_{i=1}^I \theta_i \exp(-\beta z_i^*)}= 
\frac{\xi_i^*}{\sum \limits_{k=1}^I \xi_k^*},
\end{align*}
where the last equality is from Eq. (\ref{exp_log_z}). 
\qed

Although Proposition \ref{prop:exp-log} characterizes the limiting price, the weighting parameter is still unknown. We can use the heuristic weight $\h{\pmb{\nu}}$ given in (\ref{def_crra_weight}) as an approximation for the true weighting parameter $\pmb{\nu}^*$. Since traders have Logarithmic utility ($\gamma=0$), these weighting parameters become $\h{\nu}_j = w{j,0}$ for $j=1,\ldots,J$. By replacing $\nu_j^*$ with $\h{\nu}_j = w_{j,0}$ in (\ref{log-exp-price}) and (\ref{log-exp-xi}), we obtain the heuristic price $\h{p}_i$ as follows:
\begin{align}
\h{p}_i = \frac{\h{\xi}_i}{\sum \limits_{k=1}^I \h{\xi}_k},~~i=1,\ldots, I, \label{exp-log-heristic-price}
\end{align}
where $\h{\xi}_i$ is the solution of the following equations (i.e., the equations (\ref{log_exp_xi}))
\begin{align}
\h{\xi}_i = \frac{\sum \limits_{j=1}^J w_{j,0} \cdot \pi_{i,j} + \frac{1}{\beta} \h{\xi}_i \log \left( \theta_i\Big/\frac{\h{\xi}_i}{\sum_{k=1}^I \h{\xi}_k} \right)}{\sum \limits_{j=1}^J w_{j,0} },\quad i=1,2,...,I.\label{log_exp_xi2}
\end{align}

\begin{figure}[h!]
\centering
\begin{subfigure}{0.49\textwidth}
    \includegraphics[width = \linewidth]{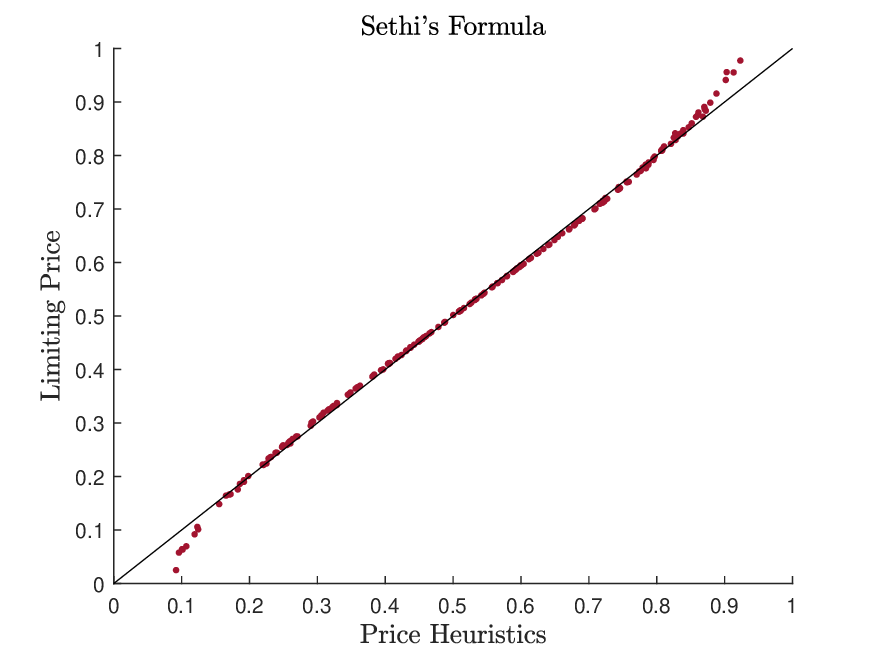}
    \caption{\small Sethi's heuristic price $\breve{p}$}
    \label{fig:Sethi_heuristics}
\end{subfigure}
\begin{subfigure}{0.49\textwidth}
    \includegraphics[width = \linewidth]{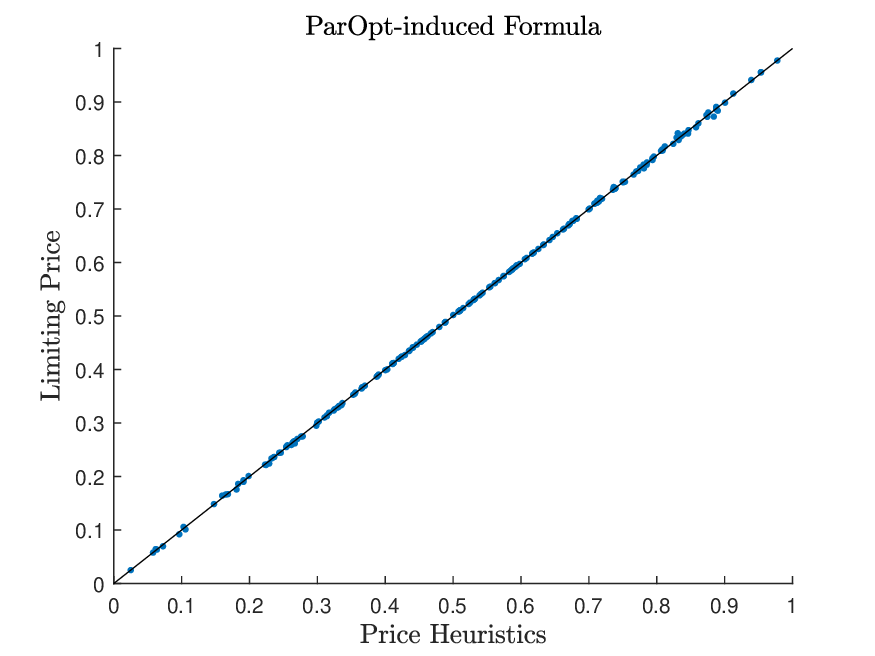}
    \caption{\small Pareto Optimality-induced price $\h{p}$}
    \label{fig:ParOpt_heuristics}
\end{subfigure}
\caption{Comparison of the heuristic prices}\label{heuristic_comparison}
\end{figure}

Finally, we present an evaluation of the price formula (\ref{exp-log-heristic-price}). As a benchmark, we adopt the heuristic price formula given in \cite{sethi2016belief}, which is represented using our notation as:
\begin{align*}
\breve{p}_i = \frac{\sum \limits_{j=1}^J w_{j,0}\cdot\pi_{i,j} }{\sum \limits_{j=1}^J w_{j,0}},~~i=1,\ldots,I.
\end{align*}
We replicate the experiment presented in Section 7 of \cite{sethi2016belief}, where the beliefs of $J=5$ traders are randomly sampled from a beta distribution, and the market maker's liquidity parameter is $\beta=0.05$. Figure \ref{heuristic_comparison} presents the comparison results. In these figures, the y-axis represents the samples of the true limiting price generated from the simulation, and the x-axis represents the heuristic price.

Figure \ref{fig:Sethi_heuristics} shows the performance of Sethi's heuristic price $\breve{p}_i$. As we can see, Sethi's formula exhibits a systematic bias, where the limiting price tends to be higher (lower) than the heuristic price when the heuristic price is high (low). On the other hand, as shown in Figure \ref{fig:ParOpt_heuristics}, our price formula (\ref{exp-log-heristic-price}), which we call the Pareto-Optimal induced price, almost perfectly aligns with the true limiting price. 

\subsection{Price Formula for Market with HARA Utility}\label{apdx:HARA}

In this section, we briefly discuss how to generalize the price formula derived from the CRRA-utility-based market to a more general utility family, namely, the utility with \textit{hyperbolic absolute risk aversion} (HARA utility). We adopt similar notations in Section \ref{sec:CRRA} and define the HARA utility-based MU functions as follows. Given wealth vectors $\y \in \mR^I$ and $\x_j\in \mR^I$, we assume the market maker's MU function and the $j$-th trader's MU function are
\begin{align}
U(\y)  \triangleq\frac{1- \gamma_m}{\gamma_m} \sum_{i=1}^I \theta_i
\left(\frac{a_m y_i}{1-\gamma_m} + b_m \right)^{\gamma_m}~
\textrm{and}~~V_j(\x_j) \triangleq\frac{1- \gamma}{\gamma} \sum_{i=1}^I
\pi_{i,j} \left( \frac{a_j x_{i,j}}{1- \gamma} + b_j
\right)^{\gamma}, \label{eq:hara_UV}
\end{align}
respectively, where $(\theta_1,\ldots,\theta_I)\in \Theta_I$ and $(\pi_{1,j},\ldots, \pi_{I,j})\in \Theta_I$ are the subjective beliefs, $\gamma_m$ and $\gamma$ are the risk aversion parameters
of the market maker and the $j$-th trader, respectively; $a_m>0$, $b_m\ge0$, $a_j> 0$ and $b_j\ge 0$ for $j=1,2, \ldots, J$ are the parameters associated with the HARA utility functions. Clearly, CRRA utility is a special case of HARA utility. For examplem, if we set $b_m=0$ and $a_m= (1-\gamma_m)^{\frac{\gamma_m-1}{\gamma_m}}$ in $U(\y)$ it becomes the CRRA utility.

The above definitions of HARA utility-based MU functions satisfy the convergence conditions in Theorem \ref{thm_Uconv}. Thus, for a given trading sequence, the market state converges to a Pareto optimal point. We may solve problem $\P_{\po}(\pmb{\nu})$ with a set of weighting parameters $\pmb{\nu}^*=(\nu_1^*, \ldots, \nu_J^*)$ for the Pareto optimal wealth allocation, i.e., we have the problem:
\begin{align}
\P_{\po}(\pmb{\nu}^* ):~&~ \max_{\y,\x_1,\ldots, \x_J}~~~
\frac{\gamma }{1-\gamma} \sum_{j=1}^J \left(\nu_j^* \cdot \sum_{i=1}^I \pi_{i,j} \left( \frac{a_j x_{i,j}}{1- \gamma} + b_j
\right)^{\gamma} \right)\notag\\
\textrm{Subject to:}~&~ \y+ \sum_{j=1}^J \x_j ={w}_0^{\all}\cdot \e, \label{hara_c1}\\
                   ~&~  
\frac{\gamma_m}{1-\gamma_m} \sum_{i=1}^I \theta_i \left(\frac{a_m y_i}{1-\gamma_m} + b_m \right)^{\gamma_m} \geq 
U(W_0) \triangleq \frac{\gamma_m}{1-\gamma_m} \left( \frac{a_m W_0}{1-\gamma_m} + b_m \right)^{\gamma_m}. \label{hara_c2}
\end{align}

Similar to the CRRA utility, we have the optimality conditions:
\begin{align}
\nu_j^*\pi_{i,j} a_j \left( \frac{ a_j x_{i,j}}{1-\gamma} +b_j \right)^{\gamma-1} =\xi,~~i=1,\ldots,I;j=1,\ldots,J,\label{hara_eq1}\\
\eta \theta_i a_m \left( \frac{ a_m y_{i}}{1-\gamma_m} + b_m \right)^{\gamma_m-1} = \xi,~~i=1,\ldots,I, \label{hara_eq2}
\end{align}
where $\pmb{\xi}=(\xi_1,\xi_2,\ldots,\xi_I)$ and $\eta \geq 0$ are the Lagrange multipliers for the two contratints. Using (\ref{hara_eq2}) in Eq. (\ref{hara_c2}) gives
\begin{align}
\eta =\frac{\left(\frac{a_m W_0}{1-\gamma_m} +b_m \right)^{1-\gamma_m} }{ \left( \sum \limits_{k=1}^I \theta_k
\left( a_m\theta_k/\xi_k \right)^{\frac{\gamma_m}{1-\gamma_m}} \right)^{\frac{1-\gamma_m}{\gamma_m}} }.\label{hara_eta}
\end{align}
Combining Eqs. (\ref{hara_eq1}), (\ref{hara_eq2}) and (\ref{hara_eta}), we have the sytem of equations for $\pmb{\xi}$:
\begin{align}
&(1-\gamma)\sum_{j=1}^J a_j^{-1}\left( \left( \frac{ \nu_j^* \pi_{i,j} a_j }{ \xi_i} \right)^{\frac{1}{1-\gamma}} -b_j \right)\notag\\
&~~~~~~~~~~~~+\frac{1-\gamma_m}{a_m}\left( \left(\frac{\theta_i a_m}{\xi_i}\right)^{\frac{1}{1-\gamma_m}} \cdot \Big(\frac{a_m W_0 }{1-\gamma_m} + b_m\Big) \cdot \mathcal{L}(\pmb{\theta}, \pmb{\xi})   -b_m \right) = w_0^{\all},~~~i=1,\ldots, I,
\label{hara_xi_equation}
\end{align}
where $\mathcal{L}(\pmb{\theta}, \pmb{\xi}) = \sum_{k=1}^I \theta_k(\theta_k/\xi_k)^{\frac{\gamma_m}{1-\gamma_m}}$. Similar to the case of CRRA utility, the above equations admit a unique solution (see Appedix \ref{apdx:sec:crra_result}), which we denote as $\xi^*=(\xi_1^*, \ldots, \xi_I^*)$. Then, the limiting price can be computed as
\begin{align}
p_i^* =\xi^*_i \Big/ \sum \limits_{k=1}^I \xi_k^*,~~i=1,\ldots, I. \label{hara_price}
\end{align}
In the market with a random trading sequence, we propose to use the following heuristic weights as an approximation, 
\begin{align}
\hat{\nu}_j = \frac{1}{a_j}\Big(\frac{a_j
w_{j,0}}{1-\gamma}+ b_j\Big)^{1-\gamma}, \quad \mbox { for } j=1,2,
\ldots, J.\label{def_omega_conjeture}
\end{align}
Similar to the CRRA utility-based price formula, we may use the heuristic weight $\hat{\nu}_j$ to replace $\nu_j^*$ in Eq. (\ref{hara_xi_equation}). Then, solving these equations gives the solution $\hat{\pmb{\xi}}=(\hat{\xi}_1, \ldots, \hat{\xi}_I)$. The associated approximate limiting price is $\hat{p}_i = \hat{\xi}_i /\sum_{k=1}^I \hat{\xi}_k$. In an unreported experiment, we find that the above approximation scheme accurately predicts the limiting price.

\subsection{Trading Decision for Risk-Neutral Traders}\label{apdx:risk-neutral}
We consider the case when the market maker adopts the exponential utility (given in (\ref{eq:exp_U_Vj})) and all the traders are risk neutral. Using the similar notations in Section \ref{sse:exponential}, in $t$-th round of trading, the $j$-th trader interacts with the market. The market states are $\y_t=(y_{1,t},\ldots, y_{I,t})$ and $j$-the trader's wealth is $\x_{j,t}=(x_{1,j,t},\ldots, x_{I,j,t})$. The $j$-th trader's decision problem is as follows,
\begin{align}
\max_{\z}~~&~\sum_{i=1}^I \pi_{i,j} ( x_{i,j,t}+ z_i )\notag\\
\text{Subject to}~&~~~\sum_{i=1}^I \theta_i(1-e^{-\beta (y_i-z_i) }) \geq 1-e^{-\beta W_0}.\label{risk-neutral-constr}
\end{align}
Checking the optimality condition, we get the optimal solution,   
\begin{align*}
z_i^{*} = y_i - \frac{1}{\beta}\ln\left( \frac{\lambda \beta \theta_i}{ \pi_{i,j}} \right),~~i=1,\ldots,I,
\end{align*}
where $\lambda\geq 0$ is the Lagrange multiplier. It can be verified that $\lambda \not=0$. Thus the constraint (\ref{risk-neutral-constr}) is binding for $z_i^*$, which gives the solution of $\lambda^*= \frac{1}{\beta} e^{\beta W_0}$. After the trading, the market state is updated to $y_{i,t+1} = y_{i,t}-z^*_i = \frac{1}{\beta}\ln\left( \frac{\lambda^* \beta \theta_i}{ \pi_{i,j}} \right)$ for $i=1,\ldots,I$. The price is updated to
\begin{align*}
p_{i,t+1} = \frac{\theta_i e^{-\beta (y_i -z_i^*) }}{\sum \limits_{k=1}^I \theta_k e^{-\beta(y_k-z_k^*)}} = \pi_{i,j},
\end{align*}
for $i=1,\ldots, I$. That is to say, after the $j$-th trader interacts with the market, the price is modified exactly according to the $j$-th trader's belief.

\section{Properties of MU functions and Examples}\label{apdx:sec:MU_function}

\subsection{Examples of MU Functions}\label{apdx:ssec:MU_examples}
One prominent feature of our framework is that it provides more freedom in designing multivariate utility function. The following example shows how various desirable objectives can be made explicit and combined with each other using our MU framework.  We then provide some examples of MU functions which
go beyond the conventional framework of \textit{Expected Utility} or the \textit{Risk measure}-based MU function. 
Let $I$ represent the number of securities, $(\theta_1,\ldots, \theta_I)\in \Theta$ denote the market maker's belief, and $\y=(y_1,\ldots, y_I)$ the wealth vector.

\begin{itemize}
\item We may consider the following MU function,
\begin{align}
U(\y) &=
 -\frac{1}{\beta} \ln\left( \sum_{i=1}^I \theta_i e^{-\beta y_i}
\right)+\eta\sum_{i=1}^I \theta_i\ln\big(y_i+B\big), \label{util_entr_risk}
\end{align}
where $\eta\ge0$ is the parameter. The first term of $U(\y)$ is the so called \textit{Entropic Risk Measure} \cite{follmer2011stochastic} and the second term represents the classical expected utility. It's worth noting that the second term establishes an explicit lower bound of $-B$ for the market maker's terminal wealth. In fact, if an explicit worst-case loss is desired, one can always add a penalty term $\frac{1}{\gamma}\sum_{i=1}^I \theta_i (y_i+B)^{\gamma}$ with $\gamma\le 0$ to the original multivariate utility function $U(\y)$.

\item We can construct the following MU function,
\begin{align}\label{util_min_E}
U(\y) = \min_{i=1,\ldots, I}~\{y_i\}  
+ \frac{\eta}{\gamma} \big(\sum_{i=1}^I \theta_i y_i +B\big)^\gamma,
\end{align}
where $\eta\ge 0$ and $\gamma \le 1$ are the paramter. In this MU function, the first term represents the worst-case terminal wealth for the market maker, and it is a concave function because it is the point-wise minimum of $\y$. The second term establishes a specific lower bound of $-B$ for the expectation, $\sum_{i=1}^I \theta_i y_i$. It's important to note that as the parameter $\gamma$ approaches $1$, the second term approximates the expectation of the wealth (with a shift of $B$). Consequently, a market maker employing this MU function (\ref{util_min_E}) may seek to strike a balance between minimizing worst-case losses and maximizing expectations.

\item Using the MU framework, we can not only design the functional form of the MU function but also customize the domain of the MU function. Thus, it is possible to include some ``hard constraint'' on the market state in MU function. Specifically, consider the function $U(\y)=\sum_{i=1}^I \theta_i u(y_i) $ where  $u(\cdot)$ is an conventional univariate utility function. We then define the domain of this MU function as 
\begin{align*}
\dom(U)\triangleq \{\y:~\y \ge \b{L}\}\cap\{\y:~\rho(\y) \le \rho_0\}\cap\{\y:~\sum_{i=1}^I \theta_i y_i\ge-B\},
\end{align*}
where $\rho_0(\cdot)$ can be viewed as some risk measure. As for the domain of the MU function, the first term bounds the worst-case loss, the second term bounds the risk, and the last term bounds the expectation of the wealth.
\end{itemize}

\subsection{Equivalence Between Various Mechanisms}\label{apdx:ssec:equivalence_mechanism}

In this section, we explicitly assume the domain of $U$ is
polyhedral: $\dom(U)\triangleq \{\y:~\b{a}_k^{\top}\y\ge b_k,~\b{a}_k>\0,~k\in \mathcal{K}\}$ where $\mathcal{K}$ is some index set. The proof for the case $\dom(U) = \mR^I$ is similar.

A \textit{scoring rule} $\b{S}$ is a reward mechanism that prompts participants to report their subjective probability belief and pays them according to their predicting accuracy. Specifically, if someone reports $\p\in \Theta_I$, then he/she gets paid $\b{S}(\p)_i$ if outcome $i$ happens. A scoring rule is \textit{proper} if truthfully reporting belief is a utility
maximizing strategy for a myopic, risk-neutral participant. Additionally, it is \textit{strictly proper} if truthful reporting is the \textit{only} utility-maximizing strategy. \cite{hanson2003combinatorial} invented a market implementation for scoring rules by transforming the belief reporting process into a sequential procedure.

Following a similar rationale with the Market Equivalence Theorem
given by \cite{chen2007utility}, the following proposition shows how
to retrieve a proper scoring rule from a multivariate utility
function. We first introduce the following definition:
\begin{definition}
Consider a mapping $\b{S}(\cdot)$ from a point to a set. If for any
$\p_n\to\p^*$, $\b{s}_n\in\b{S}(\p_n)$, it has
$\lim_{n\to\infty}\b{s}_n\in\b{S}(\p^*)$, then we say $\b{S}(\cdot)$
is upper-hemicontinuous.
\end{definition}

\begin{proposition}\label{prop:b1}
If $U(\cdot):\mR^I\rightarrow \mR$ is a concave function and
$U$-based pricing rule has bounded loss, for $\p\in\Theta_I$, we
introduce $\b{S}$ as,
\begin{align}\label{eq:gum_msr_equiv}
\b{S}(\p)\triangleq - {\arg\min}&~~\Big\{\p^{\top}\y~|~U(\y)\ge
U(W_0\cdot \e),~\y \in \dom(U)\Big\},
\end{align}
where $\dom(U)=\{\y:~\b{a}_k^{\top}\y\ge b_k,~\b{a}_k > \0,~ k\in \mathcal{K}\}$,
then $\b{S}(\p)$ is an upper-hemicontinuous correspondence on
$\Theta_I$ and also a proper scoring rule. Furthermore,
$\b{S}(\cdot)$ is continuous within the interior of $\Theta_I$ when
$U(\cdot)$ is strictly concave.
\end{proposition}

{\noindent\bf Proof of Proposition \ref{prop:b1}.} For convenience,
we denote by $\mc{L}$$=$$\big\{\y~|~U(\y)\ge
U(W_0\cdot \e),~\b{a}_k^{\top}\y\ge b_k,~k \in \mathcal{K}\big\}$ the constraint set
of (\ref{eq:gum_msr_equiv}). Since the market maker has bounded
loss, then there exists a $\underline{\y}\in\mR^I$ such that
$\y\ge\underline{\y},~\forall \y\in\mc{L}$. This suggests that
$\inf_{\y\in\mc{L}}\{\e_i^{\top}\y\}$ has a solution (possibly a set of
extended reals). for all $i$. Then for any $\p\in\Theta_I$ we have
$\inf_{\y\in\mc{L}}\{\p^{\top}\y\}\ge \sum_i p_i
\inf_{\y\in\mc{L}}\{\e_i^{\top}\y\}$, which implies
(\ref{eq:gum_msr_equiv}) has a solution for any $\p\in\Theta_I$.
Specifically, for $\p\in\tm{interior}(\Theta_I)$ the intersection of
recession cone of $\p^{\top}\y$ and  $\mc{L}$ is  $\{\0\}$, therefore the
solution set is compact \cite{bertsekas2003convex}, and the solution
set is a unique element when $U$ is strictly concave, hence
$\b{S}:\tm{interior}(\Theta_I)\to\mR^I$ is a function.

For the upper-hemicontinuity of $\b{S}$, suppose there exists
$\p_n\to\h{\p}$, $\b{s}_n\in\b{S}(\p_n)$ such that:
$\lim_{n\to\infty}\b{s}_n=\h{\b{s}}\not\in\b{S}(\h{\p})$. Here we
    consider extended reals, i.e.,
some elements of $\b{s}_n,\h{\b{s}}$ may be $+\infty$. Then we apply
the Bolzano-Weistrass theorem on extended reals $\b{s}_n$ to derive
a convergence subsequence that converges to some  $\h{\b{s}}$. By
continuity of $U(\cdot)$, it has $U(-\h{\b{s}}) \ge U(W_0 \cdot \e)$ and $
-\b{a}_k^{\top}\h{\b{s}}\ge b_k$ for all $k$.
This means $\h{\b{s}}$ is a feasible solution for
(\ref{eq:gum_msr_equiv}) at point $\hat{\p}$. {Due to the optimality
    of $-\b{S}(\h{\p})$ at $\h{\p}$, there exists an $\overline{\b{s}}\in
    \b{S}(\hat{\p})$ such that $-\h{\p}^{\top} \overline{\b{s}} < -
    \h{\p}^{\top}\h{\b{s}}$.
    Again by continuity, monotonicity of $U(\cdot)$ and the fact that
    $\b{a}_k>\0,~k\in \mathcal{K}$,  we can subtract a small $\epsilon \e$ from
    $\overline{\b{s}}$ to get another $\tilde{\b{s}}$ such that,
    \begin{align*}
        \begin{cases}
            -\h{\p}^{\top} \tilde{\b{s}}< -\h{\p}^{\top}\h{\b{s}},\\
            U(-\tilde{\b{s}}) > U(W_0\cdot \e),\\
            -\b{a}_k^{\top}\tilde{\b{s}}> b_k,~~k \in \mathcal{K}.
        \end{cases}
    \end{align*}
    Since we have $\p_n\to\h{\p}$, $\b{s}_n\to\h{\b{s}}$, there exists
    an $N$ large enough such that the following system of inequalities
    hold,
    \begin{align*}
        \begin{cases}
            -\p_N^{\top} \tilde{\b{s}}< -\p_N^{\top}\b{s}_N, \\
            U(-\tilde{\b{s}}) > U(W_0\cdot \e),\\
            -\b{a}_k^{\top}\tilde{\b{s}}> b_k,~~k\in \mathcal{K}.
        \end{cases}
\end{align*}}
Such a fact means $\tilde{\b{s}}$ is a feasible solution of
(\ref{eq:gum_msr_equiv}) at $\p_N$, and its corresponding objective
value is smaller than $\b{s}_N$ which is a contradiction. Thus, when
$\p_n\to\h{\p}$, $\b{s}_n\in\b{S}(\p_n)$ implies
$\lim_{n\to\infty}\b{s}_n=\h{\b{s}}\in\b{S}(\h{\p})$, i.e.,
$\b{S}(\cdot)$ is upper-hemicontinuous. Furthermore,
upper-hemicontinuity is equivalent with continuity when
$\b{S}(\cdot)$ is a function, therefore $\b{S}$ is continuous on
$\tm{interior}(\Theta_I)$ for a strictly concave MU function
$U(\cdot)$.

As for properness, when a risk-neutral trader is facing a scoring
rule $\b{S}$, if we assume his/her true probability estimate is
$\p$, then his/her decision problem can be expressed as,
\begin{equation}
   \begin{aligned}
\max_{\b{r}\in\Theta_I} ~~&~~\p^{\top}\b{s}\\
       \textrm{Subject to}~&~~\b{s}\in\b{S}(\b{r}),\\
           ~~&~~\b{S}(\b{r})=-\underset{\y\in\mc{L}}{\arg\max}~-\b{r}^{\top}\y,
   \end{aligned}
\end{equation}
where $\mc{L}=\big\{\y:~U(\y)\ge U(W_0\cdot \e),~~\b{a}_k^{\top}\y\ge
b_k,~\forall k \big\}$. From the structure of such problem, it has
\begin{align*}
    \p^{\top}\b{s}_1\ge\p^{\top}\b{s}_2,\quad\forall \b{r}\ne\p,\b{s}_1\in\b{S}(\p),\b{s}_2\in\b{S}(\b{r}),
\end{align*}
which means truthfully reporting $\p$ is a utility-maximizing
strategy. $\hfill\Box$

\begin{remark}
   From the above discussion one can also easily verify that the MU-induced
   scoring rule is strictly proper within the interior of $\Theta_I$ when the
   underlying utility is strictly concave.
\end{remark}

The next proposition shows how to induce an MU with bounded loss
from a proper scoring rule.
\begin{proposition}\label{prop:b2}
    Let $\b{S}:\Theta_I\to R$ be a proper scoring rule and $\b{S}$ is continuous
    on $\Theta_I$, let
     \begin{align*}
         \mc{L}\triangleq &-\underset{\p\in\Theta_I}{\cap}\big\{\b{s}:~ \p^{\top}\b{s}\le \p^{\top}\b{S}(\p)    \big\}\\
         \mc{D}\triangleq &~~~\big\{\b{S}(\p):~\p\in\Theta_I  \big\}
    \end{align*}
    then $\mc{L}$ is a closed and convex set and $-\mc{D}$ lie on
    the boundary of $\mc{L}$. Define
    \begin{align}
        U(\y)\triangleq -\inf\{m:~\y+m\cdot\e\in\mc{L}\}=\underset{\p\in\Theta_I}{\inf}~\big\{
        \p^{\top}\big(\b{S}(\p)+\y \big) \big\}\label{defn:msr_U}
    \end{align}
    then $U:\mR^I\to\mR$ is a weakly monotone\footnote{A multivariate utility $U$ is said to be weakly monotone if
    $\y\ge\h{\y}\implies U(\y)\ge U(\h{\y})$,  $y_i>\h{y}_i,\forall i\implies U(\y)>U(\h{\y})$ and for any $i$ and any $\y\in\mR^I$
there exists some  $\beta>0$ such that  $U(\y-\beta\e_i)<U(\y)$.}
multivariate utility function, while the $U$-based MU suffers
bounded loss and satisfies
\begin{align}\label{eq:MSRtoGUM}
-\b{S}(\b{r}) \in\underset{\y}{\arg\min}~\big\{~\b{r}^{\top}\y
~|~U(\y)\ge 0 \big\}.
\end{align}
\end{proposition}

{\noindent\bf Proof of Proposition \ref{prop:b2}.} Clearly,
$\big\{\b{s}~|~\p^{\top}\b{s}\le \p^{\top}\b{S}(\p) \big\}$ defines a closed
halfspace for all $\p\in\Theta_I$. Since closed halfspace is
natively convex, $\mc{L}$ can be regarded as the intersection of
closed convex sets. Thus, it is also closed and convex. Also, for
$\p\in\Theta_I$, it is easy to see that $-\mc{D}$ lies on the
boundary of $\mc{L}$.


Note that (\ref{defn:msr_U}) can be expressed as,
\begin{align*}
  U(\y)\triangleq -\min_{m}&~~~\big\{~m~|~-\p^{\top}(\y+m\cdot\e)\le\p^{\top}\b{S}(\p),\quad\forall \p\in\Theta_I\big\}.
\end{align*}
The constraint set $-\p^{\top}(\y+m\cdot\e)\le\p^{\top}\b{S}(\p)$ for all
$\p\in\Theta_I$ can be equivalently written as:
\begin{align*}
    -\p^{\top}\y-m\le\p^{\top}\b{S}(\p),~~\forall \p\in\Theta_I\implies m\ge -\p^{\top}\big(\b{S}(\p)+\y\big),~~\forall \p\in\Theta_I.
\end{align*}
Therefore MU function $U(\y)$ is equivalent to
\begin{align}
    U(\y)=  -\underset{\p\in\Theta_I}{\sup}~\big\{  -\p^{\top}\big(\b{S}(\p)+\y \big) \big\}=\underset{\p\in\Theta_I}{\inf}~\big\{  \p^{\top}\big(\b{S}(\p)+\y \big) \big\}.
\end{align}

At last, we check whether $U(\cdot)$ meets Assumption
\ref{assumption:U_mild}.
\begin{itemize}
\item Domain condition: Since $\b{S}(\cdot)$ is continuous on $\Theta_I$, $U(\cdot)$ is actually
  defined as the negative of minimum value of a continuous function over a closed and
        compact set, then Weistrass Theorem says such minimum value is attainable and
        well-defined, therefore $U(\cdot)$ is well defined on whole $\mR^I$.
\item Continuity condition: Suppose $\y_n\to \overline{\y}$ but $U(\y_n)$ does not
        approach $U(\overline{\y})$. Then, it must have one of the following cases:
\begin{enumerate}
  \item There exists a $\delta>0$ such that $\exists
            N>Z$, $U(\y_N)>U(\overline{\y})+\delta,~~\forall Z>0$;

  \item There exists a $\delta>0$ such that $\exists
    N>Z,~U(\y_N)+\delta<U(\overline{\y})$, $\forall Z>0$.

\end{enumerate}

Let $\p_N^*\in\underset{\p\in\Theta_I}{\arg\min}~\big\{
\p^{\top}\big(\b{S}(\p)+\y_N
    \big) \big\}$ and $\overline{\p}^*\in\underset{\p\in\Theta_I}{\arg\min}~\big\{
\p^{\top}\big(\b{S}(\p)+\overline{\y} \big) \big\}$. Then for case (1),
by the continuity of $\b{S}$ and $\y_n\to\overline{\y}$, we know
there exists an $N$ sufficiently large such that
\begin{align*}
U(\y_N)={\p_N^*}^{\top}\big(\b{S}(\p_N^*)+\y_N \big) >
{\overline{\p}^*}^{\top}\big(\b{S}(\overline{\p}^*)+\y_N \big),
\end{align*}
which is contradictory to the optimality of $\p_N^*$.

As for case (2), there also exists an $N$ sufficiently large such
that,
\begin{align*}
U(\overline{\y})={\overline{\p}^*}^{\top}\big(\b{S}(\overline{\p}^*)+\overline{\y}
    \big) > {\p_N^*}^{\top}\big(\b{S}(\p_N^*)+\overline{\y} \big)
\end{align*}
which is contradictory to the optimality of $\overline{\p}^*$.
Therefore, it must have $\y_n\to \overline{\y}\implies U(\y_n)\to
U(\overline{\y})$.

\item Bounded loss condition: Since $\Theta_I$ is a closed and bounded set while $\b{S}$ is continuous on $\Theta_I$,
therefore, $\mc{D}=\big\{\b{S}(\p):~\p\in\Theta_I\big\}$ is also a
closed and bounded set. By the structure of $U$ we know $U(\y)\ge 0
\iff \y\in\mc{L}$. Since $\e_i\in\Theta_I,~\forall i=1,\ldots,I$
then for any
        $\y$ satisfying $U(\y)\ge 0$, it has:
\begin{align*}
\e_i^{\top}\y\ge-\e_i^{\top}\b{S}(\e_i),~~\forall i=1,\ldots,I,~\implies
\inf_{U(\y)\ge 0} y_i\ge -\e_i\b{S}(\e_i),~~\forall i=1,\ldots,I.
\end{align*}
Since $\mc{D}$ is bounded, the left-hand side of the above
    inequality is also bounded, therefore the worst-case loss of the market
    maker is bounded.

\item Concavity for $\y_1,\y_2\in\mR^I$: Note that the following inequality holds,
        \begin{align*}
            U\big(\alpha \y_1 + (1-\alpha)\y_2\big)&=\underset{\p\in\Theta_I}{\inf}~\big\{  \p^{\top}\big(\b{S}(\p)+ \alpha\y_1+(1-\alpha)\y_2 \big) \big\}\\
                                                   &\ge\alpha\underset{\p\in\Theta_I}{\inf}~\big\{  \p^{\top}\big(\b{S}(\p)+ \y_1 \big) \big\} +(1-\alpha)\underset{\p\in\Theta_I}{\inf}~\big\{  \p^{\top}\big(\b{S}(\p)+ \y_2 \big) \big\}\\
                                                   &= \alpha U(\y_1)+(1-\alpha) U(\y_2).
        \end{align*}

\item Weak monotonicity: The following inequalities holds true,
  \begin{align*}
  U(\y+\bo{\delta})&=\underset{\p\in\Theta_I}{\inf}~\big\{  \p^{\top}\big(\b{S}(\p)+ \y +\bo{\delta}\big) \big\}\ge
            \underset{\p\in\Theta_I}{\inf}~\big\{ \p^{\top}\big(\b{S}(\p)+ \y \big) \big\}=U(\y),~~\forall \bo{\delta}\ge \0,\\
  U(\y+\bo{\delta})&=\underset{\p\in\Theta_I}{\inf}~\big\{
            \p^{\top}\big(\b{S}(\p)+ \y +\bo{\delta}\big) \big\}>
            \underset{\p\in\Theta_I}{\inf}~\big\{  \p^{\top}\big(\b{S}(\p)+ \y \big)
            \big\}=U(\y),\\
            &~~~~~~~~~~~~\forall \bo{\delta} \mbox{ such that } \delta_i>0,~\forall i.
\end{align*}
Also for every $i$ and any $\y\in\mR^I$ it has:
\begin{align*}
             U(\y-\beta\e_i) = \underset{\p\in\Theta_I}{\inf}~\big\{  \p^{\top}\big(\b{S}(\p)+ \y -\beta\e_i\big) \big\}\le
             \e_i^{\top}\b{S}(\e_i)+\e_i^{\top}\y -\beta.
\end{align*}
Since $\e_i^{\top}\b{S}(\e_i)+y_i$ is bounded for  $\y\in\mR^I$ and
$\beta$ is arbitrary, for some $\beta>0$ it has
$U(\y-\beta\e_i)<U(\y)$ for all $i$ and  $\y\in\mR^I$.
\end{itemize}

As for (\ref{eq:MSRtoGUM}), first when $\y=-\b{S}(\b{r})$, by
properness of $\b{S}$ we have $\p^{\top}\b{S}(\p)-\p^{\top}\b{S}(\b{r})\ge
0,~~\forall \p\in\Theta_I$ therefore:
\begin{align*}
    U\big(-\b{S}(\b{r})\big) =\inf_{\p\in\Theta_I}\big\{  \p^{\top}\b{S}(\p)-\p^{\top}\b{S}(\b{r}) \big\}\ge 0
\end{align*}
which means $-\b{S}(\b{r})$ is a feasible solution of
(\ref{eq:MSRtoGUM}).

On the other hand, suppose there exists a $\tilde{\y}\in\mR^I$ such
that: $\b{r}^{\top}\tilde{\y}<-\b{r}^{\top}\b{S}(\b{r})$,
which means
\begin{align*}
    \b{r}^{\top}\b{S}(\b{r})+\b{r}^{\top}\tilde{\y}<0\implies U(\tilde{\y})=\inf_{\p\in\Theta_I}\big\{  \p^{\top}\b{S}(\p)+\p^{\top}\tilde{\y} \big\}<0
\end{align*}
that is to say any $\tilde{\y}$ with objective function value lower
than $-\b{r}^{\top}\b{S}(\b{r})$ is not a feasible solution of
(\ref{eq:MSRtoGUM}) therefore we must have
$-\b{S}(\b{r})\in\underset{U(\y)\ge 0}{\arg\min}~\b{r}^{\top}\y$.

Cost function-based mechanism is a prediction market making method
proposed by \cite{chen2007utility}. It is now one of the most
widely-used market making mechanism in prediction market literature
due to its implementation simplicity. A cost function
$C:\mR^I\to\mR$ should satisfy the following properties:
\begin{itemize}
    \item $C(\cdot)$ is continuous and defined on whole  $\mR^I$;
    \item Convexity: $C(\alpha\q_1+(1-\alpha)\q_2)\le \alpha
        C(\q_1)+(1-\alpha)C(\q_2)$;
    \item Monotonicity: for any $\q,\hat{\q}$, if $\q \ge \hat{\q}$, it has
        $C(\b{q})\ge C(\hat{\q})$ and the inequality is held strictly if $\q >
        \hat{\q}$;
    \item Translation invariance: $C(\q + t \cdot \e)=C(\q)+t,~\forall t\in\mR$.
\end{itemize} 
Then given current outstanding securities $\q$, a new order
$\Delta\q$ is charged with  $C(\q+\Delta\q)-C(\q)$.

One can see that a $\rho(\y)\triangleq  C(-\y)$ defines a risk
measure, therefore  $V(\y)\triangleq -C(-\y)$ is always a
multivariate utility function. The following proposition formally
establishes their equivalence.
\begin{proposition}\label{prop:b3}
For any given $\q\in\mR^I$ any multivariate-utility $U(\cdot)$, let
\begin{align*}
    C_U(\q)\triangleq  \min&~~ W\\
    s.t. &~~ U(W\cdot\e-\q)\ge U(W_0\cdot\e),\\
         &~~ W\cdot\e-\q\in\dom(U),
\end{align*}
then $C_U(\cdot)$ is a well-defined cost function, and $C_U(\q_t+\Delta\q)-C_U(\q_t)$ is exactly the price charge of $\Delta\q$ under $U$-based MU when outstanding securities are  $\q_t$. Conversely, for any cost function $C(\cdot)$, let
\begin{align}
    U_C(\y)=-C(-\y),\label{eq:c2gum}
\end{align}
then $U_C(\y)$ is a qualified multivariate utility as per Assumption \ref{assumption:U_mild}. Furthermore, denote by $W_t$ the amount of cash the cost function-based market maker collected by time $t$, then such a cost function-based market maker is equivalent to $U_C( W_t \cdot \e-\q_t) = U_C( W_0\cdot\e)$.
\end{proposition}
 {\noindent\bf Proof of Proposition \ref{prop:b3}.} MU$\rightarrow$ Cost function: That $C_U$ is a function follows directly from uniqueness of solution of $\P_{\tm{Cmin}}$; the translation invariance property is obvious. We now focus on the continuity and convexity of $C_U$:
\begin{itemize}
    \item Continuity: Suppose there exists a sequence $\q_n\to \overline{\q}$ such that $\lim_{n\to\infty}C_U(\q_n)=\h{C}\ne C_U(\overline{\q})$
        ($\h{C}$ may be $+\infty$). If $\h{C}>C_U(\overline{\q})$
        , then $U(\h{C} \cdot \e-\overline{\q})>U(C_U(\overline{\q})\cdot \e-\overline{\q})\ge
        U(W_0 \cdot \e)$, $\b{a}_k^{\top}(\h{C} \cdot \e-\overline{\q})>b_k,~\forall
        k$. By continuity of $U$ we know there exists an $N$ sufficiently large
        such that:
        $U(C_U(\q_N) \cdot \e-\q_N)>U(W_0 \cdot \e)$, $\b{a}_k^{\top}(C_U(\q_N)\e-\q_N)>b_k,~\forall
        k$. Now we can achieve a smaller objective value by slightly reducing
        $C_U(\q_N)$, which contradicts with the optimality of
        $C_U(\q_N)$. Conversely, if $\h{C}<C_U(\overline{\q})$, then by
        continuity of $U$ we have $U(\h{C}\cdot \e-\overline{\q})\ge
        U(W_0\cdot \e)$, $\b{a}_k^{\top}(\h{C}\cdot \e-\overline{\q})\ge b_k,~\forall k$
        , which again contradicts with the optimality of $C_U(\overline{\q})$.
        Therefore $\q_n\to
        \overline{\q}\implies C_U(\q_n)\to C_U(\overline{\q})$, that is to
        say $C_U$ is continuous.
    \item Convexity: By the concavity of $U$ and the convexity of $\dom(U)$ we
        know that for any $\q_1,\q_2$:
        \begin{align*}
                &U\Big(\big(\alpha C_U(\q_1)+(1-\alpha) C_U(\q_2)\big)\cdot\b{e} - \big(\alpha  \q_1+(1-\alpha) \q_2\big)\Big)\\
                    &=U\Big( \alpha\big( C_U(\q_1)\b{e} - \q_1 \big) +(1-\alpha)\big( C_U(\q_2)\b{e} - \q_2 \big) \Big)\\
                        &\ge \alpha U\Big(C_U(\q_1)\b{e} - \q_1\Big) + (1-\alpha)U\Big(C_U(\q_2)\b{e} - \q_2\Big)\\
                            &\ge U(W_0\cdot \e).
        \end{align*}
        Similarly, it has
        \begin{align*}
                \big(\alpha C_U(\q_1)+(1-\alpha) C_U(\q_2)\big)\cdot\b{e} - \big(\alpha  \q_1+(1-\alpha) \q_2) \in \dom(U),
            \end{align*}
            which means $\alpha C_U(\q_1)+(1-\alpha) C_U(\q_2)$ is a feasible
            solution of (\ref{eq:c2gum}) at $\alpha  \q_1+(1-\alpha)
            \q_2$, therefore by the optimality of $C_U(\cdot)$ we have:
            \begin{align*}
                    C_U(\alpha  \q_1+(1-\alpha) \q_2)\le \alpha C_U(\q_1)+(1-\alpha) C_U(\q_2).
                \end{align*}
\end{itemize}
The price equivalence between $C_U$ and (\ref{eq:c2gum}) is obvious.\hfill\qed

Cost function $\rightarrow$ MU: Clearly $U_C$ inherits concavity,
continuity and monotonicity from $C$, therefore is a qualified risk
preference function per Assumption \ref{assumption:U_mild}. On the
other hand, assume the cost function-based market maker has $W_t$
cash by time $t$, while the aggregate outstanding securities are
$\q_t$ then the price charge for an incoming new order is $\Delta
W_t = C(\q_t+\Delta\q_t) - C(\q_t)$.
By translation invariance of $C$ it has
 \begin{align*}
     C(-(W_t+\Delta W_t)\cdot \e+(\q_t+\Delta \q_t)) =& -W_0-\sum_{l=1}^{t-1} \big(C(\q_{l+1})-C(\q_l)\big) \\
        &- \big( C(\q_t+\Delta\q_t) + C(\q_t) \big) - C(\q_t+\Delta\q_t)\\
     =&-W_0=C(-W_0\cdot \e),
\end{align*}
which further implies $U_C(W_t\cdot \e-\q_t)=U_C(W_0\cdot \e)$ for $t$. \hfill\qed

\subsection{Additional Properties of MU Function}\label{apdx:MU_function_property}

\begin{itemize}
\item \textbf{No-arbitrage for Surebet}: Under Assumption \ref{assumption:U_mild}, the market making mechanism induced by 
multivariate utility function $U(\cdot)$ is well defined and has the following properties.

\begin{proposition}\label{thm_Uniq}
If the market maker adopts MU function $U(\cdot)$ to price any order $\Delta \q$ by solving problem $\P_{\cmin}(\Delta \q)$, it has following properties:
\begin{description}
\item[(i)] It can appropriately price any finite order $\Delta \q$ with finite price $\Delta w^*$ and the price $\Delta w^*$ is unique;
\item[(ii)] The $k\in \mR$ units of surebet order $\Delta \q = k \cdot \e$ is always charged with $k$ dollars.\footnote{As all outcomes of the prediction market are mutually exclusive and exhaustive, the surebet means that the order covers all the states. The surebet order is in form $\Delta q= k \cdot \e$ for some $k$.}
\end{description}
\end{proposition}

\begin{proof}{Proof.}
Given market state $(\q, w)$ and incoming order $\{\Delta \q\}$, we denote the
constraint set of $\P_{\textrm{cmin}}$ at this state by
$\mathcal{C}(w,\q,\Delta\q )$. By the structure of the problem it is not hard
to see that the only common direction of recession of set
$\mathcal{C}(w,\q,\Delta\q)$ and objective function is $\0$, therefore
$\P_{\cmin}$ admits a solution for any $\{\q, w\}$ and $\Delta \q$, and the
solution is clearly unique. As for the price of surebet $\Delta q= k\e$, we
consider two cases. In the first case, the market state $\{\q, w\}$ resulted
from previous trading satisfies $U(w\cdot  \e - \q)=U(W_0\cdot\e)$. If we solve problem
$\P_{\cmin}$ for the surebet order $k\cdot \e$ in current round of trading, the
constraint becomes $U((w+\Delta w)\cdot \e - (\q + k \cdot \e))\geq U(W_0\cdot\e)$, which
implies $\Delta w^* =k$. In the second case, if the previous trading gives
$U(w\cdot \e - \q) > U(W_0\cdot \e)$, we may still have the solution $\Delta w^*=k$ for the
order $k\cdot \e$. Otherwise, the previous trading may not achieve its optimal price. 
\end{proof}

\item \textbf{Bounded Worst-case Loss}: Although a market maker is designed to provide liquidity, it is certainly not designed to be infinitely exploited. To avoid such catastrophe, one might wish to put a cap on the worst-case loss. Obviously MU function based market maker suffers bounded worst-case loss if and only if the utility level set $\mathcal{L}\triangleq \{\y \in\mR^I~|~U(\y)\ge U(W_0\cdot \e)\}$ is lower bounded by some vector $\udl{\y} \in \mR^I$, i.e., it
has $\udl{\y}\leq \y$ for all $\y \in \mathcal{L}$. The utility level set $\mathcal{L}$ actually describes all the
acceptable terminal wealth $\y$ of the market maker. Indeed, each element of $\udl{\y}=\big(\udl{y}_1, \udl{y}_2,\ldots,
\udl{y}_n\big)$  can be characterized as follows,
\begin{align}
 \P_i~&~\min_{\y} \{y_i~|~\y=(y_1,\cdots, y_I),~U(\y)\geq U(W_0\cdot\e)\}. \label{prob_LBi}
 \end{align}
Worst-case loss bound can be found by solving subproblems $\P_i$ for all $i=1,\cdots, I$. Obviously, the worst-case loss
has a natural bound when the domain of utility $U$ itself is bounded from below, e.g.,  $U(\y)=\sum_i \pi_i
(y_i+B_i)^\gamma,~\gamma<1$. If such a domain bound design is not desired, then we need to find to other means to bound the loss. One viable way is to resort to transaction restrictions. Specifically, in real world traders are usually
constrained to have non-negative wealth position when true state is revealed (It is known as the \textit{no-bankruptcy restriction}). Under this case, bounded loss can be easily guaranteed. Besides the \textit{no-bankruptcy restriction}, we provide the following results for the bounded worst-case loss.

\begin{proposition}\label{prop_lb}
If multivariate utility function $U(\cdot)$ satisfies Assumption \ref{assumption:U_mild} and $\sup_{\y} U(\y)=B<+\infty$ with
\begin{align}
\lim \limits_{t\rightarrow \infty }U(t (\e - \e_i) )<B,~~\textrm{for}~i=1,\cdots, I, \label{prop_lb_cond}
\end{align}
then the MU-function based market maker has bounded loss.
\end{proposition}

\begin{proof}
    The multivariate utility function is bounded above means that, there exist some finite number
    $B$, such that $U(\y) < B$ for any $\y\in \dom(U)$. Under the condition given in
    this proposition, we assume the market maker has unbounded loss. Without loss of
    generality, we assume the first element of $\y$ may go to $-\infty$ and $y\in
    \mathcal{L}$. This assumption is equivalent to say there exists vector
    $\y(\tau)\triangleq (-\tau, g(\tau), g(\tau),\cdots, g(\tau)) \in \mR^I$ with
    $\y(\tau)
    \in \mathcal{L}$ when $\tau\rightarrow \infty$, where $\tau>0$ and $g(\tau)$ is any function satisfying
    $g(\tau)>\tau$. Clearly, it has $\y(\tau)= g(\tau )( \e-\e_i )- \tau  \e_i $. Since $U(\cdot)$ is a concave function, it has
    \begin{align}
    &\lim_{\tau \rightarrow\infty} U\big( g(\tau)(\e -\e_i)- \tau \e_i\big)
    \le \lim_{\tau \rightarrow \infty}\Big( U\big( g(\tau)(\e-\e_i)\big)
    -\tau \nabla U( g(\tau )(\e-\e_i))^{\top} \e_i \Big)  \label{inf1}\\
    &\lim_{\tau \rightarrow \infty}U\big( g(\tau )(\e- \e_i)+\sqrt{t}\e_i \big)
    \le \lim_{\tau \rightarrow \infty}\Big( U(g(\tau)(\b{e}-\b{e}_i))
    +\sqrt{\tau} \nabla U\big( g(\tau)(\b{e}-\b{e}_i) \big)^{\top} \e_i \Big) \label{inf2}
    \end{align}
    The left-hand side of inequality (\ref{inf2}) tends to the infimum $B$. According to (\ref{prop_lb_cond}), it has
    $\lim_{\tau\rightarrow\infty}U( g(\tau)(\e-\e_i))<B$. Thus, there exists some $\delta>0$ satisfying $\delta = B - \lim
    \limits_{\tau \rightarrow\infty}U(g(\tau)(\e-\e_i)) > 0$. Also, inequality (\ref{inf2}) becomes
    \begin{align}
        \lim_{t\rightarrow\infty}\sqrt{t}\cdot\nabla U\big(g(t)(\e - \e_i)\big)^{\top} \e_i\ge\delta
    \end{align}
    Substituting the above inequality in (\ref{inf1}) and noticing $\tau=\sqrt{\tau}\sqrt{\tau}$, it has
    \begin{align}
        \lim_{t\rightarrow\infty} U\big( g(\tau)(\e - \e_i)- \tau \e_i\big)
      &\le \lim_{\tau \rightarrow \infty}\Big( F\big(g(\tau)(\e-\e_i)\big)- \sqrt{\tau} \delta \Big)\notag\\
      &< B -\lim_{\tau \rightarrow \infty}\sqrt{\tau} \delta = -\infty.
    \end{align}
    That is to say, the utility of $U(\y(\tau))$ is infinitely low, thus can not be maintained above $U(W_0\cdot\e)$. It
    contradicts our previous statement $\y(\tau)\in \mathcal{L}$, which also implies our assumption that $\mathcal{L}$
    has no lower bound is not true. 
\end{proof}

\item \textbf{Acceptance of Limit Order:} The MU-function based pricing rule can be easily modified to accept the limit order. We refer to $\p$ as the `instantaneous' price since the actual price charge for any non-trivial order
$\Delta \b{q}$ is not $\p^{\top} \Delta \q$ but rather the amount $\Delta w^*$
calculated by solving problem $\P_{\cmin}(\Delta \q)$. This difference is due to the fact
that the instantaneous price varies continuously along with the state variable
$\y$. Such a phenomenon also persists in other mechanisms like the cost-function
based mechanism. A limit order implementation could be a good rescue for this situation. The limit order could provide much more leeway for someone to express her valuation. \cite{agrawal2011unified} formally characterize a
limit order by a triplet $(\b{a}, \Gamma, p^{\textrm{lim}})$, where $\b{a}$ is a
vector that describes the order by setting each component of $\b{a}$ being either $1$ (if the specified state is desired) or  $0$ (if it is not desired), $\Gamma \in\mR$ denotes the maximum desired units of that bundle and
$p^{\textrm{lim}}\in\mathbb{R}$ is the maximum averaged price the trader is
willing to pay for the bundle. \cite{agrawal2011unified} shows that SCPM market
maker decides the units to grant via solving a convex optimization problem. They
also show that such implementation is a generalized Vickrey-Clarke-Groves(VCG)
scheme(see, e.g., \cite{nisanalgorithmic}) and therefore admits $\textit{myopic
truthful reporting}$ in the sense that traders will not lie about their true
valuation of bid price $p^{\textrm{lim}}$. Similar to the SCPM, our MU function-based framework can be easily modified to incorporate limit order submission.
Specifically, given market state $(\q, w)$, a limit order of the form $(\b{a},
\Gamma, p^{\textrm{lim}})$, the price of this limit order can be computed by
solving the following problem, 
\begin{equation}
\label{P_smax_multivariate utility}
    \begin{aligned}
        \P_{\textrm{Smax}}:~~\max_{0\leq \tau\leq \Gamma,\Delta w}~&~p^{\textrm{lim}} \cdot \tau - \Delta w\\
        \textrm{Subject to:}~&~ U(( w+\Delta w)\cdot\b{e} - (\q + \tau \b{a})) \ge U(W_0\cdot\e), \\
        ~&~(w+\Delta w) \cdot\e - ( \q+ \tau \b{a})\in\dom(U).
    \end{aligned}
\end{equation}
After solving problem $\P_{Smax}$, the trader is granted $\tau^*$ shares and
charged with  $\Delta w^*$, where $(\tau^*,\Delta w^*)$ is the solution of
problem $\P_{Smax}$. The objective function $\tau p^{\textrm{lim}} - \Delta w$
can be understood as follows. The term $\tau p^{\textrm{lim}}$ is the maximum price one is willing to pay for the order, and $\Delta w$ is the price actually charged for such an order. Maximizing $\tau p^{\textrm{lim}} - \Delta w$ is
actually finding the highest possible cost reduction for the trader. Like SCPM, our limit order implementation is a generalized VCG scheme and admits truthful
reporting of the bid price. The proof of this claim is similar to the one of Theorem 1 in \cite{agrawal2011unified}.
Thus, we omit the details.
\end{itemize}

\end{document}